\documentclass[12pt,preprint]{aastex}

\begin{document}

\title{Optical and Near-Infrared Photometry of Distant Galaxy Clusters\footnotemark[1]}

\author{S.A.\ Stanford,\altaffilmark{2,3}} 
\affil{Physics Department, University of California, Davis, CA 95616}
\email{adam@igpp.ucllnl.org}
 
\author{Peter R.\ Eisenhardt,\altaffilmark{2}} 
\affil{Jet Propulsion Laboratory, California Institute of Technology
4800 Oak Grove Drive, Pasadena, CA 91109}
\email{prme@kromos.jpl.nasa.gov}

\author{Mark Dickinson,\altaffilmark{2}}
\affil{Space Telescope Science Institute, 3700 San Martin Drive,
Baltimore, MD 21218}
\email{med@stsci.edu}

\author{B.P.\ Holden,\altaffilmark{3}} 
\affil{Physics Department, University of California, Davis, CA 95616}
\email{bholden@igpp.ucllnl.org}

\and

\author{Roberto De Propris}
\affil{Research School of Astronomy \& Astrophysics, Australian National
University, Weston, ACT 2611, Australia}
\email{propris@mso.anu.edu.au}

\footnotetext[1]{Based on observations made with the NASA/ESA Hubble
Space Telescope, obtained from the Data Archive at the Space
Telescope Science Institute, which is operated by the Association of
Universities for Research in Astronomy, Inc., under NASA contract NAS
5-26555. These observations are associated with proposals \#5790,
\#7872, and \#8269.}

\altaffiltext{2}{Visiting Astronomer,
National Optical Astronomy Observatories, which are operated by the
Association of Universities for Research in Astronomy, Inc. (AURA)
under cooperative agreement with the National Science Foundation.}

\altaffiltext{3}{Institute of
Geophysics and Planetary Physics, Lawrence Livermore National
Laboratory, Livermore, CA 94550}

\begin{abstract}

We present optical and near-infrared photometry of 45 clusters of galaxies
at $0.1 < z < 1.3$.  Galaxy catalogs in each cluster were defined at the
longest wavelenth available, generally the $K$-band, down to approximately
two magnitudes below $M^\ast$.  We include finding chart images of the band
used for catalog definition.  The photometry has been used in previously
published papers to examine the origin and evolution of galaxies in distant
clusters.

\end{abstract}


\section{Introduction}

The identification and study of distant galaxy clusters is of interest
in current astronomical research.  An important use of galaxy clusters
lies in studying the formation and evolution of galaxy populations.
Progress on this front has been made in a large number of recent
studies \citep{aragon93,ellis97,dressler97,bender98,dokkum98a,pahre98,couch98,lubin98,
dokkum00,kelson00,post01}.

Beginning in 1991, we sought to make use of the advent of relatively large
format near-IR detectors to study the galaxy populations of distant
clusters.  We collected imaging data in both the near-IR and the optical on
a large sample of clusters drawn from several samples.  These data were
analyzed and detailed results were presented on the evolution of early-type
galaxies at moderate redshifts in Stanford, Eisenhardt, \& Dickinson (1995,
1998), on the evolution of the $K$-band luminosity function in
\citet{rdp99}, on the Butcher-Oemler effect in $K$-selected galaxy samples
in \citet{sed4}, and on the evolution of early-type galaxies in high
redshift clusters in \citet{sed3}.  So far the photometry for these various
studies has been published for only two clusters, Abell 370 and Abell 851 in
\citet{sed95}.  In this paper, we present the photometry on the other 43
clusters used in our published work on distant clusters.
 
\section{Sample}

The sample of clusters which we observed is listed in
Table~\ref{sample}.  This sample is heterogenous, following no
well-defined criteria.  It consists of many of the best-known
clusters, the reason being that such objects would most likely have
imaging data available in the $HST$ archive.  The clusters were taken
from such sources as the EMSS catalog \citep{gioia90,henry92}, the
Gunn-Hoessel-Oke sample \citep{gho}, and the Abell catalog.  Where
possible, we have included the X-ray luminosity for each cluster in
Table~\ref{sample} using values published in the literature, assuming
$H_0=65$ km s$^{-1}$ Mpc$^{-1}$, $\Omega_m = 0.3$, and $\Omega_\Lambda
= 0.7$.

\section{Observations}

Nearly all of the clusters were observed following these
prescriptions: the depth of the imaging at 5$\sigma$ in a $\sim$2
arcsec aperture reaches two magnitudes fainter than present-day
(unevolved) $M^\ast$; the bands consist of at least two of the three
standard filters in the near-IR, $JHK$, and two more in the optical
which change with redshift so as to bracket the rest frame 4000\AA\
break; and the area covered is approximately 1 Mpc in diameter
centered on the the brightest cluster galaxy.  Unless stated otherwise
we use the following cosmological parameters where necessary: $H_0=65$
km s$^{-1}$ Mpc$^{-1}$, $\Omega_m = 0.3$, and $\Omega_\Lambda = 0.7$.
The limiting magnitude in the IR-band used to define the catalog, and
the field size covered (defined as that encompassed in the longest
wavelength near-IR image for which at least half of the maximum
exposure time was achieved) are listed in Table~\ref{sample}.  For the
vast majority of the clusters we obtained $K$-band data and these were
used to define the galaxy catalogs in each cluster.  For a small
number of clusters, $K$-band data were not obtained and so $H$-band
data were used to define the galaxy catalogs.  The actual bands which
were observed, along with the date and place of each observation, are
listed in Table~\ref{obslog}.  The codes for the 'tel.' (telescope)
columns in this table are as follows: K4 is the 4~m and K2.1 is the
2.1~m telescope, both at Kitt Peak National Observatory; C0.9 is the
0.9~m and C1.5 is the 1.5~m telescope, both at Cerro Tololo
Interamerican Observatory; P5 is the 5~m telescope at Palomar; W3.5 is
the Wisconsin-Indiana-Yale-NOAO 3.5~m telescope; and HST is the Hubble
Space Telescope. In Table~\ref{obslog}, the optical bands are
generically listed as red or blue at the top and the actual band
(dependent on redshift) is given for each cluster.  The details of
these observations have been given previously in \citet{sed95},
\citet{s97} and \citet{sed98}.

The data were reduced following standard procedures, as described in
\citet{sed95} and \citet{sed98}.  The optical data were calibrated
onto the Landolt system using observations of Landolt standard stars,
and the near-IR data onto the CIT system using observations of the
UKIRT faint standards \citep{ukirt,haw01} after suitable
transformations.  The typical rms of these calibrations is 0.02 in
the optical and 0.03 mag in the near-IR.  Images were geometrically
transformed to be coaligned with the longest wavelength image. The
point spread functions of the images at the various wavelengths were
matched by convolution to a common effective angular resolution.

\section{Photometry}

Catalogs were defined for each cluster using the longest wavelength image,
usually the $K$-band but occasionally $H$-band.  A modified version (K.\
Adelberger, private communication) of FOCAS \citep{valdes82} was used for
object detection and photometry.  The details of our use of FOCAS,
including extensive simulations which were run to understand completeness
and errors in the photometry, are given in \citet{sed95} and \citet{sed98}.
All photometry was corrected for Galactic reddening using the interstellar
extinction curve given in \citet{mathis90} with values for $E(B-V)$ taken
from \citet{bh82}.

The photometry is given for each cluster in its own table.  In the
heading of each table the size of the aperture used to define optical
and near-IR colors is given.  The positions for each object are given
in arcsec relative to the brightest central galaxy, with positive
numbers being to the east and north of the central galaxy.  The
location of this central galaxy is given by the coordinates in
Table~\ref{sample}, and made clear in the finding charts shown in the
figures.  The relative positions of the objects as given in the
photometry tables are good to only $2$~arcsec on average.  For the
longest wavelength band, both the total magnitude e.g.\ $Kt$ (as
defined by FOCAS) and the aperture magnitude e.g.\ $Ka$ are given.
The aperture radius was chosen to be equal to the FWHM of the point
spread function.  Uncertainties, which do not include the errors of
the transformation to the standard magnitude system, at the one
$\sigma$ level are given for the photometry, not including the
systematics which are discussed more fully in \citet{sed95} and
\citet{sed98}.  The following codes are used in the photometry tables.
For cases where there are no data for an object, the entry is $-99$,
and when the data at the position of an object are bad, such as
because of a cosmic ray or a bleed trail, the entry is $99$.  A color
which has $0.00$ for the positive error is a lower limit; the negative
error in that case is a 1 $\sigma$ limit.

For the 25 clusters for which we have made use of $HST$ archival WFPC2
images, we include a column in the photometry tables with a
morphological code.  These clusters were analyzed in
\citet{sed95,sed98}, \citet{dokkum01b}, and \citet{sed3}.  The codes
are the following: 0 means the object was outside the WFPC2 field, 1
is an E/S0, 2 is an Sa or Sb, 3 is an Sc or Sd, 4 is Irr, and 5 is
disturbed and/or interacting.  In one case, RXJ 0848+4453, we have
NIC3 imaging as well as WFPC2 imaging and so include two columns of
morphological types using the same codes.

All photometry tables and their associated finding charts are available in full only
electronically from The Astrophysical Journal.

\acknowledgments

The authors thank NOAO for large amounts of telescope time, and the
staffs of the KPNO, CTIO, and Palomar Observatories for their
assistance.  We also thank the referee for a timely report. Support
for SAS came from NASA/LTSA grant NAG5-8430.and was provided by NASA
through grants from the Space Telescope Science Institute, which is
operated by the Association of Universities for Research in Astronomy,
Inc., under NASA contract NAS 5-26555, for proposals \#5790, \#8269,
and \#7872.  The work by SAS and BPH at LLNL was performed under the
auspices of the U.S. Department of Energy under Contract No.\
W-7405-ENG-48.  Portions of this work were carried out by the Jet
Propulsion Laboratory, California Institute of Technology, under a
contract with NASA.

\begin{figure}
\plotone{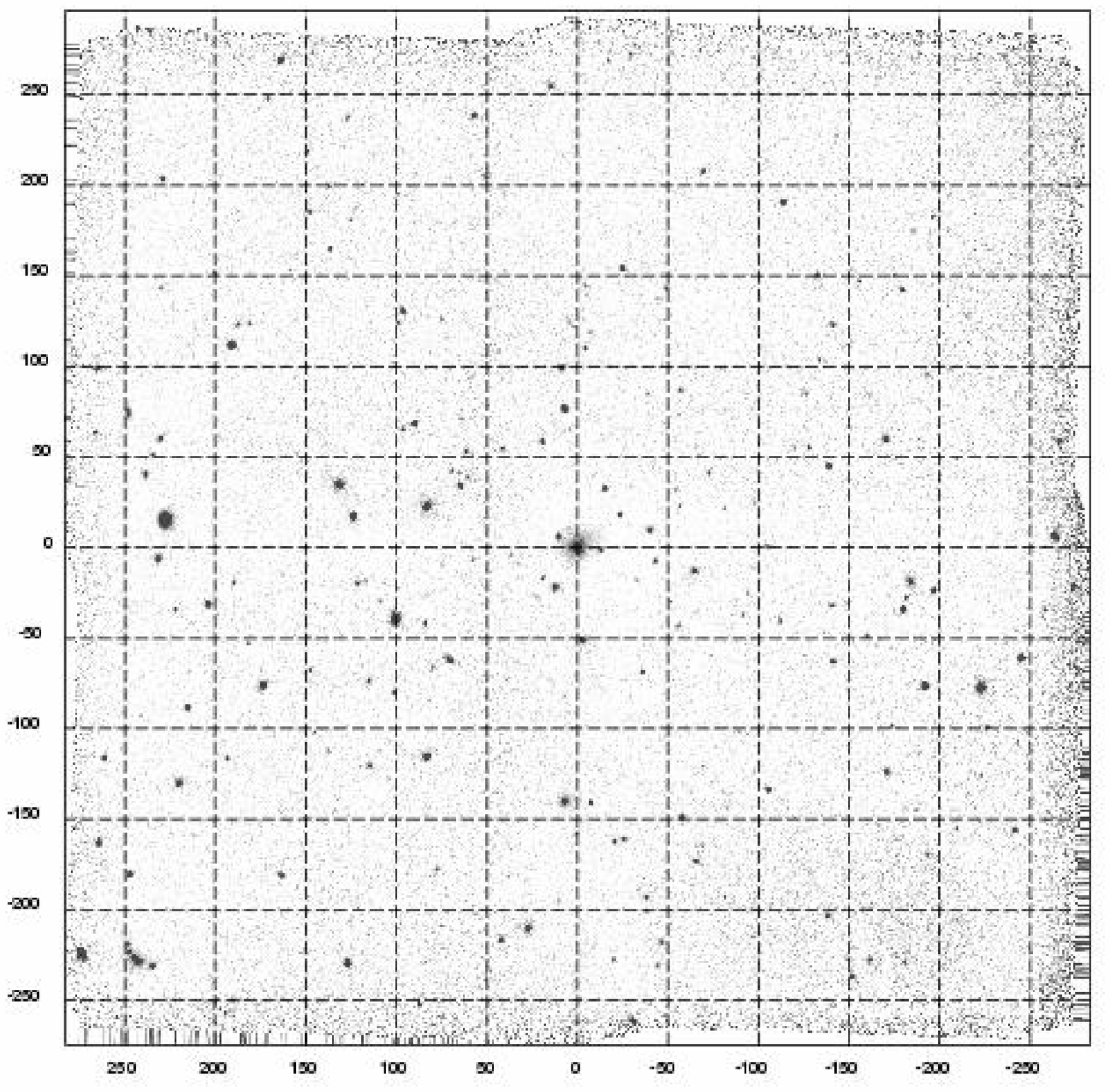}
\caption{$K$-band image of Abell 1146.  The coordinates are in relative arcsec from the
central brightest galaxy.  North is up and East to the left. }
\label{a1146k}
\end{figure}
\clearpage
\begin{figure}[p]
\plotone{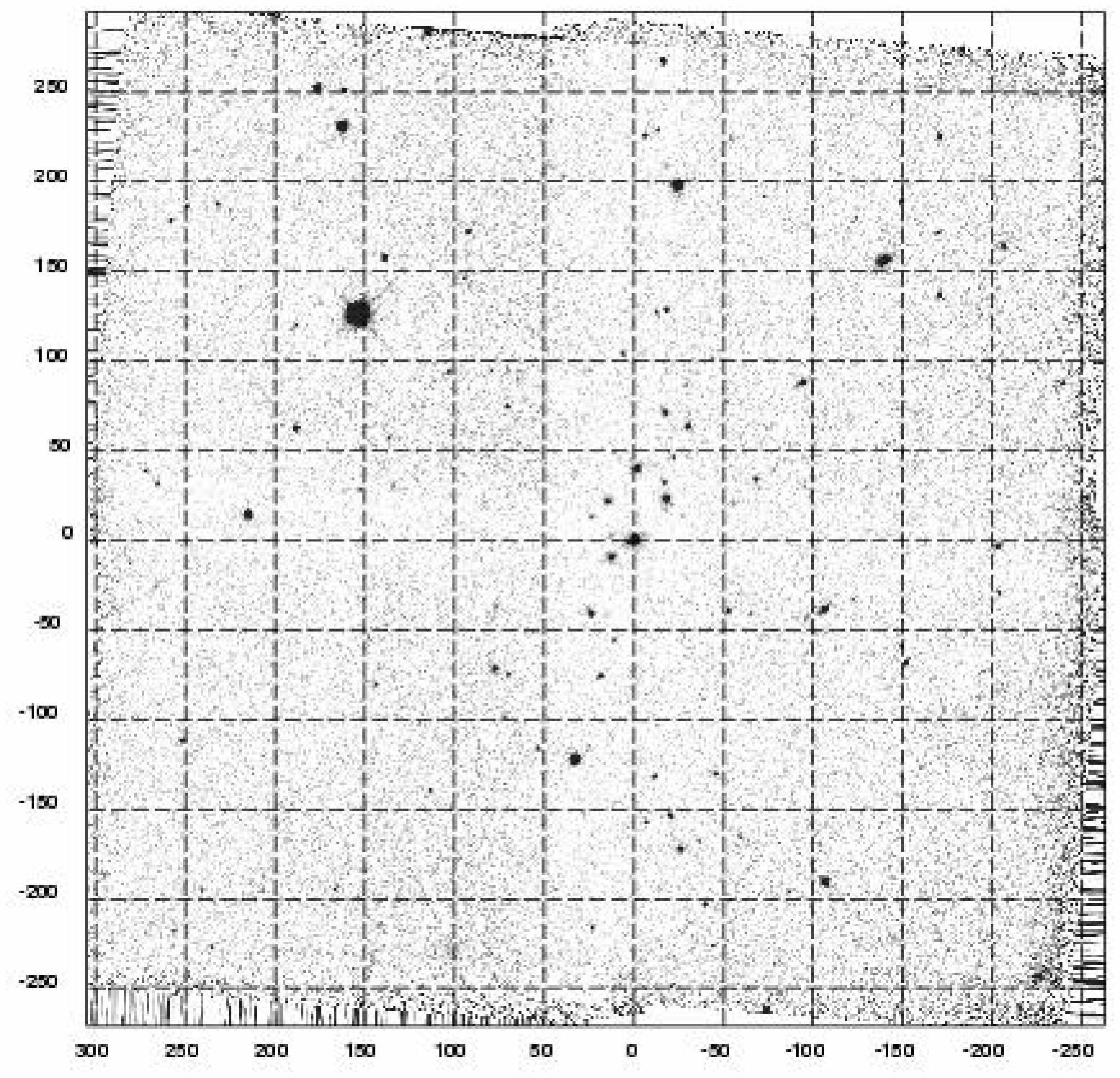}
\caption{$K$-band image of Abell 3305.  The coordinates are in relative arcsec from the
central brightest galaxy.  North is up and East to the left. }
\label{a3305k}
\end{figure}
\clearpage
\begin{figure}[p]
\plotone{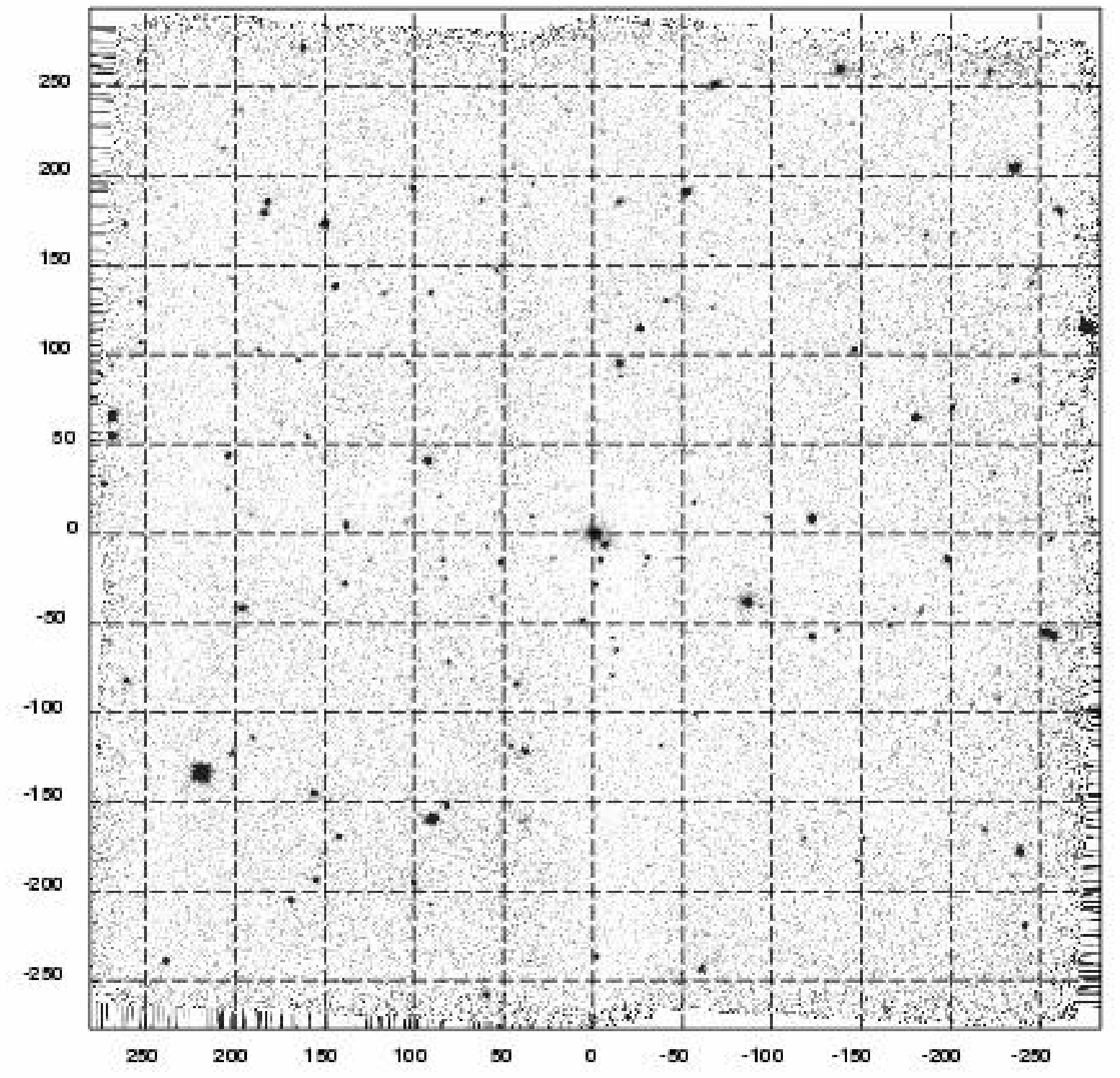}
\caption{$K$-band image of MS0906.5+1110.  The coordinates are in relative arcsec from the
central brightest galaxy.  North is up and East to the left. }
\label{m0906k}
\end{figure}
\clearpage
\begin{figure}[p]
\plotone{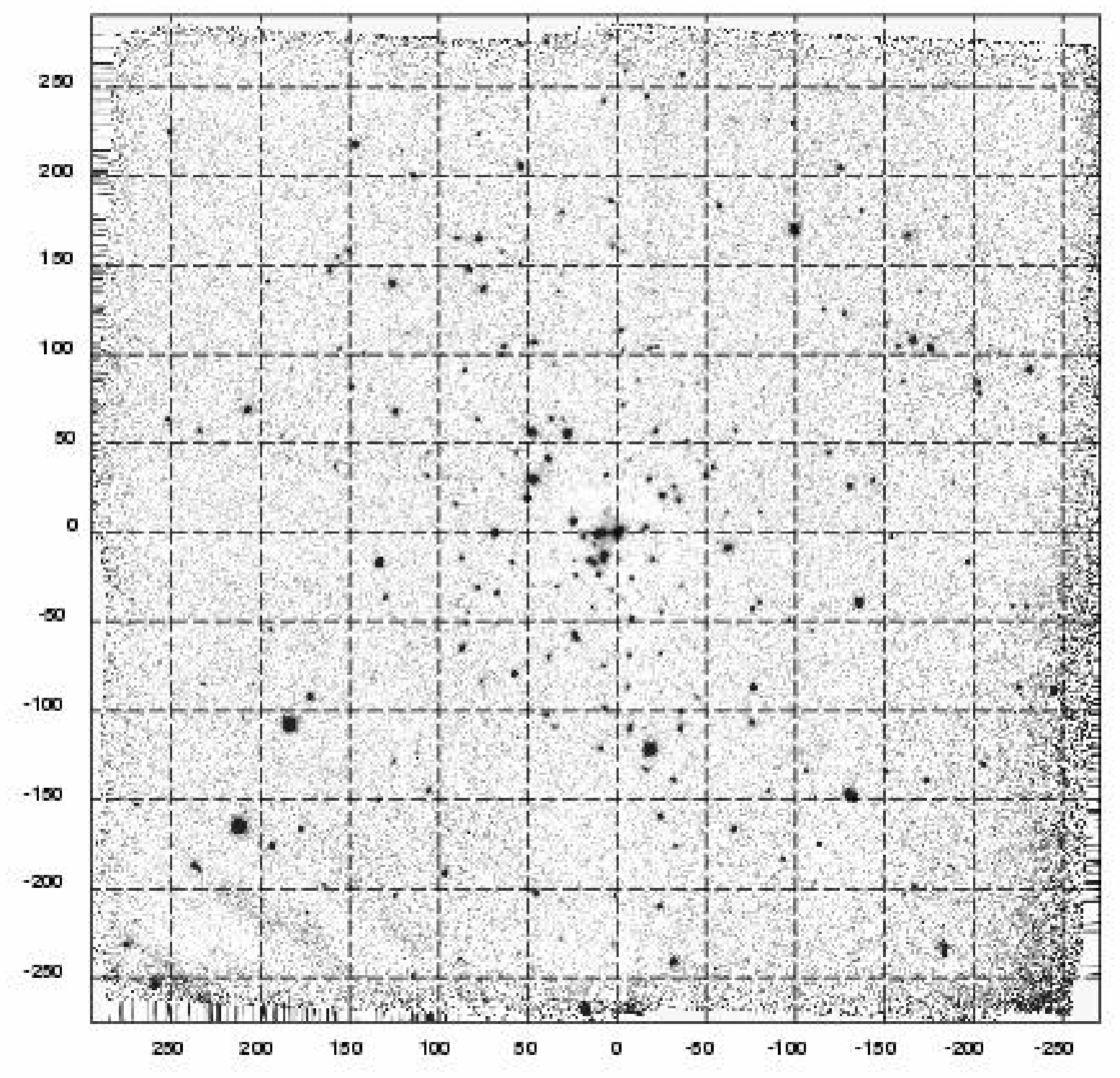}
\caption{$K$-band image of Abell 1689.  The coordinates are in relative arcsec from the
central brightest galaxy.  North is up and East to the left. }
\label{a1689k}
\end{figure}
\clearpage
\begin{figure}[p]
\plotone{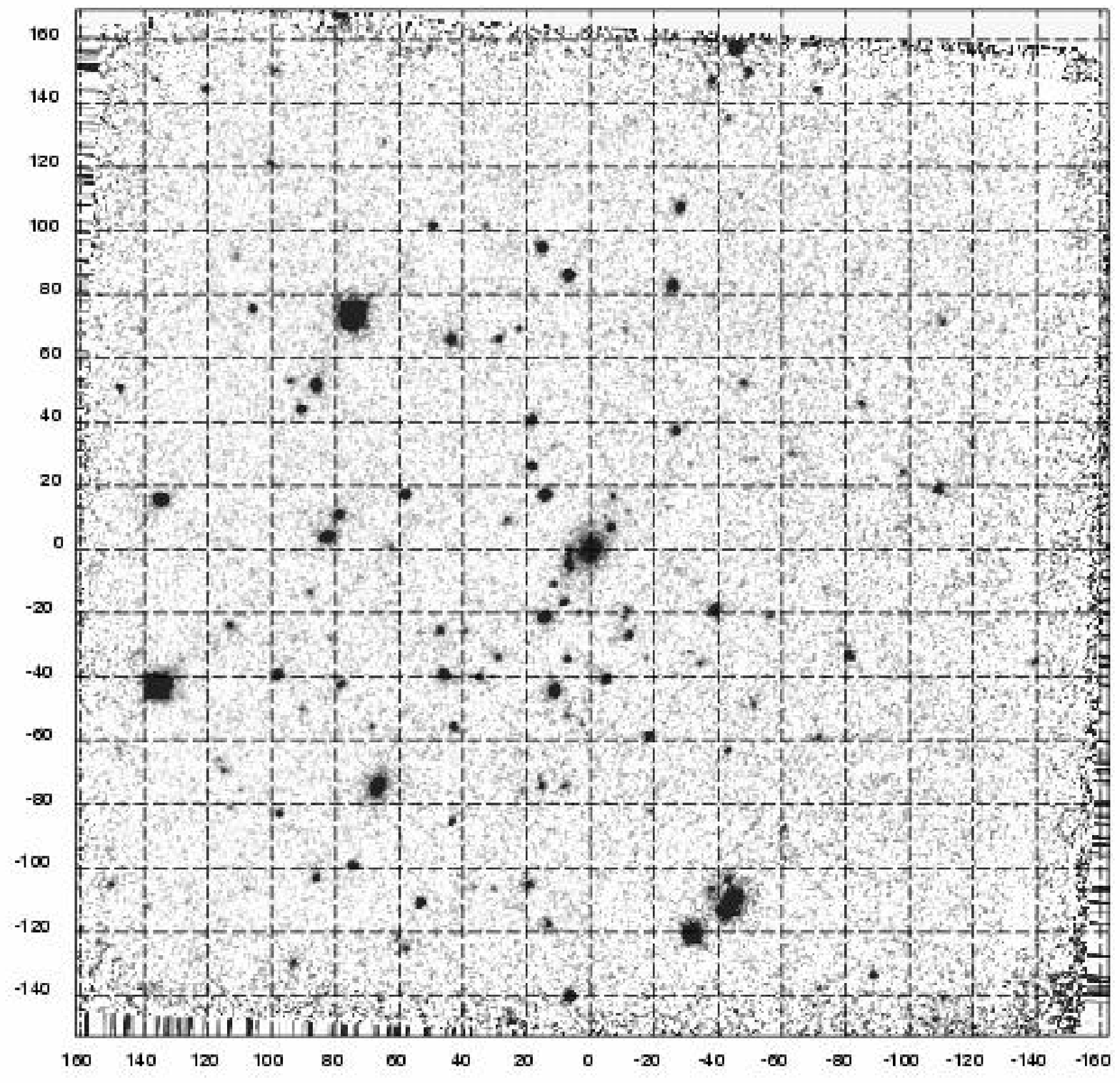}
\caption{$K$-band image of Abell 1942.  The coordinates are in relative arcsec from the
central brightest galaxy.  North is up and East to the left. }
\label{a1942k}
\end{figure}
\clearpage
\begin{figure}[p]
\plotone{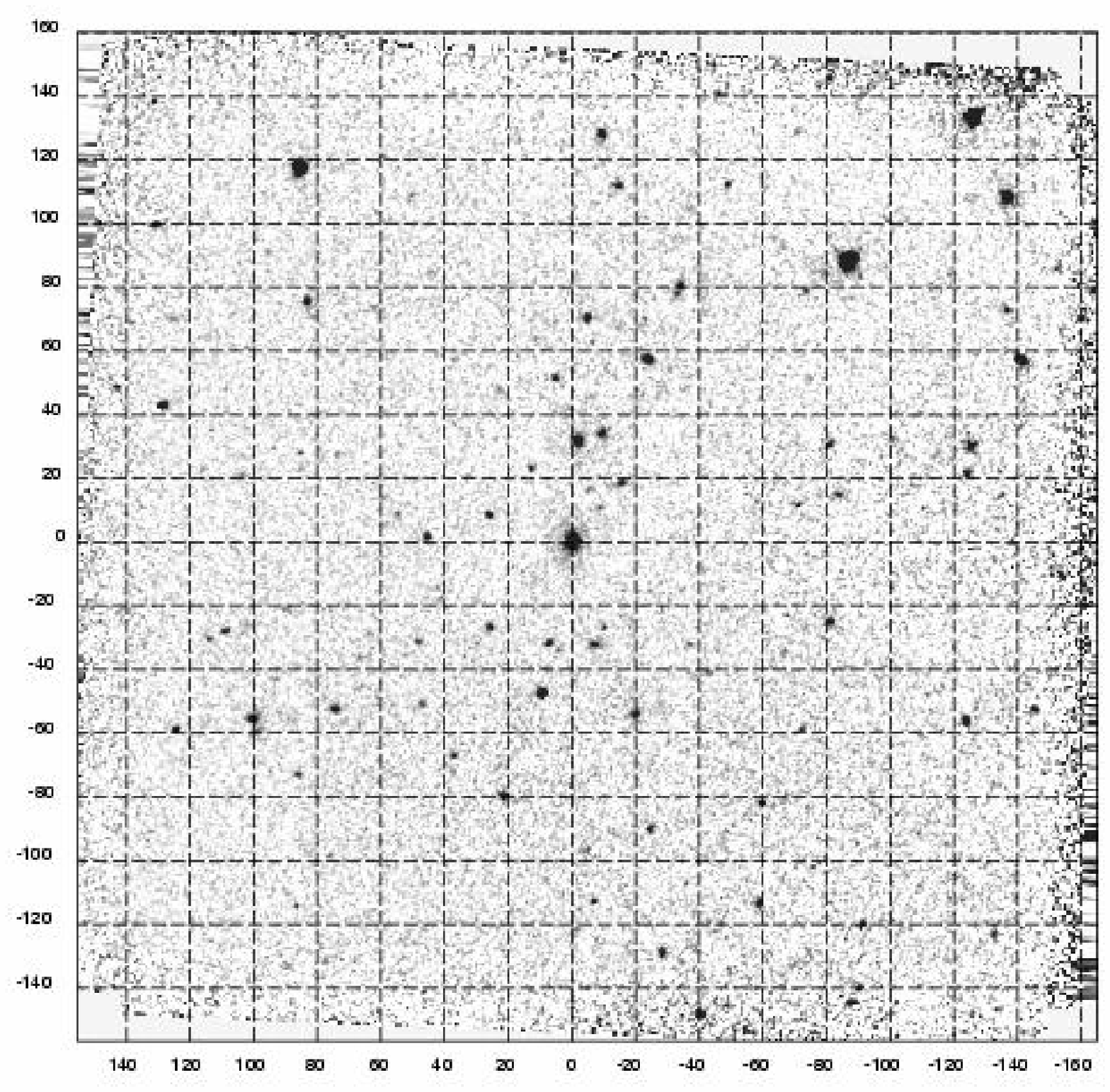}
\caption{$K$-band image of MS 1253.9+0456.  The coordinates are in relative arcsec from the
central brightest galaxy.  North is up and East to the left. }
\label{m1253k}
\end{figure}
\clearpage
\begin{figure}[p]
\plotone{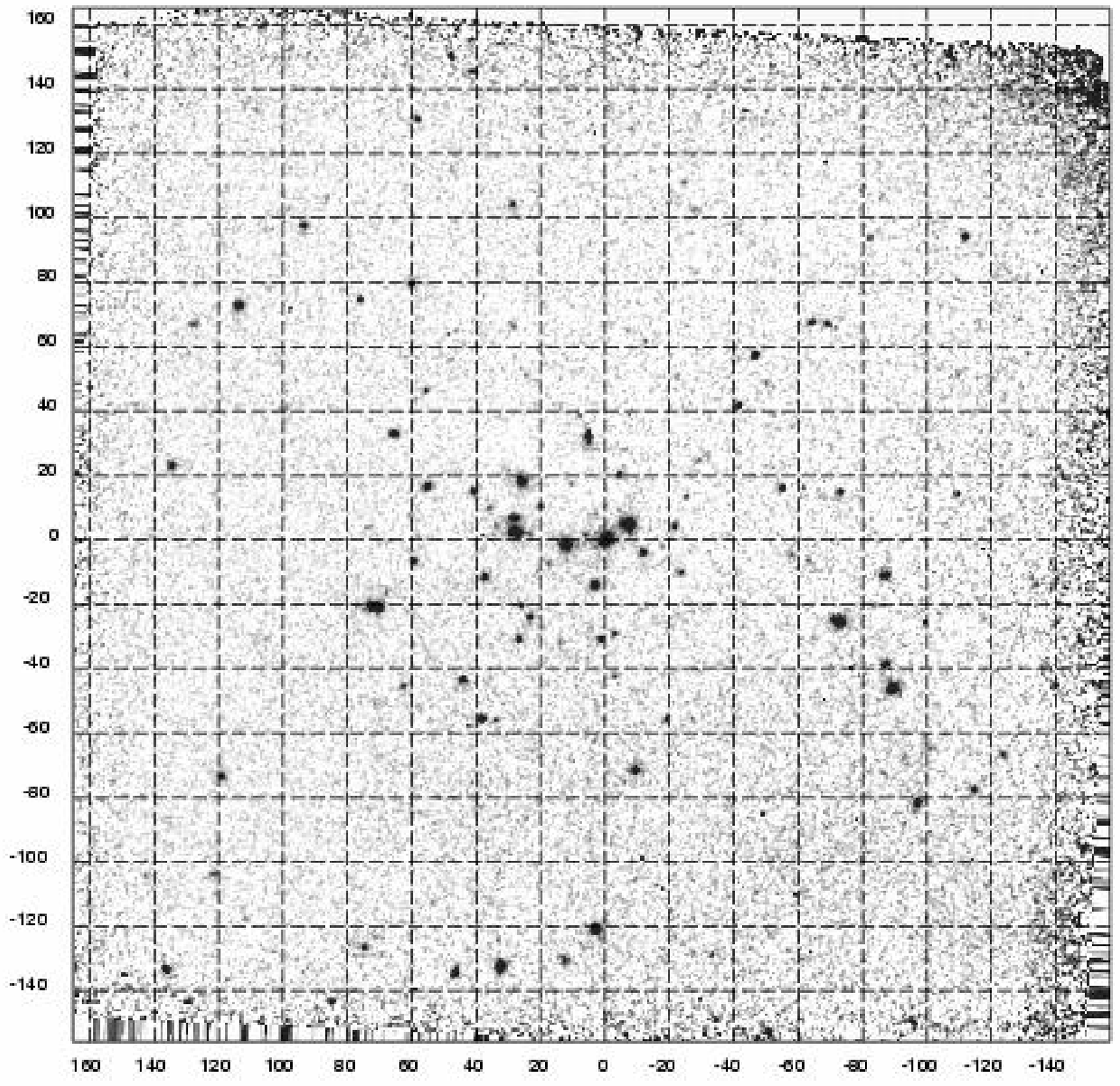}
\caption{$K$-band image of Abell 1525.  The coordinates are in relative arcsec from the
central brightest galaxy.  North is up and East to the left. }
\label{a1525k}
\end{figure}
\clearpage
\begin{figure}[p]
\plotone{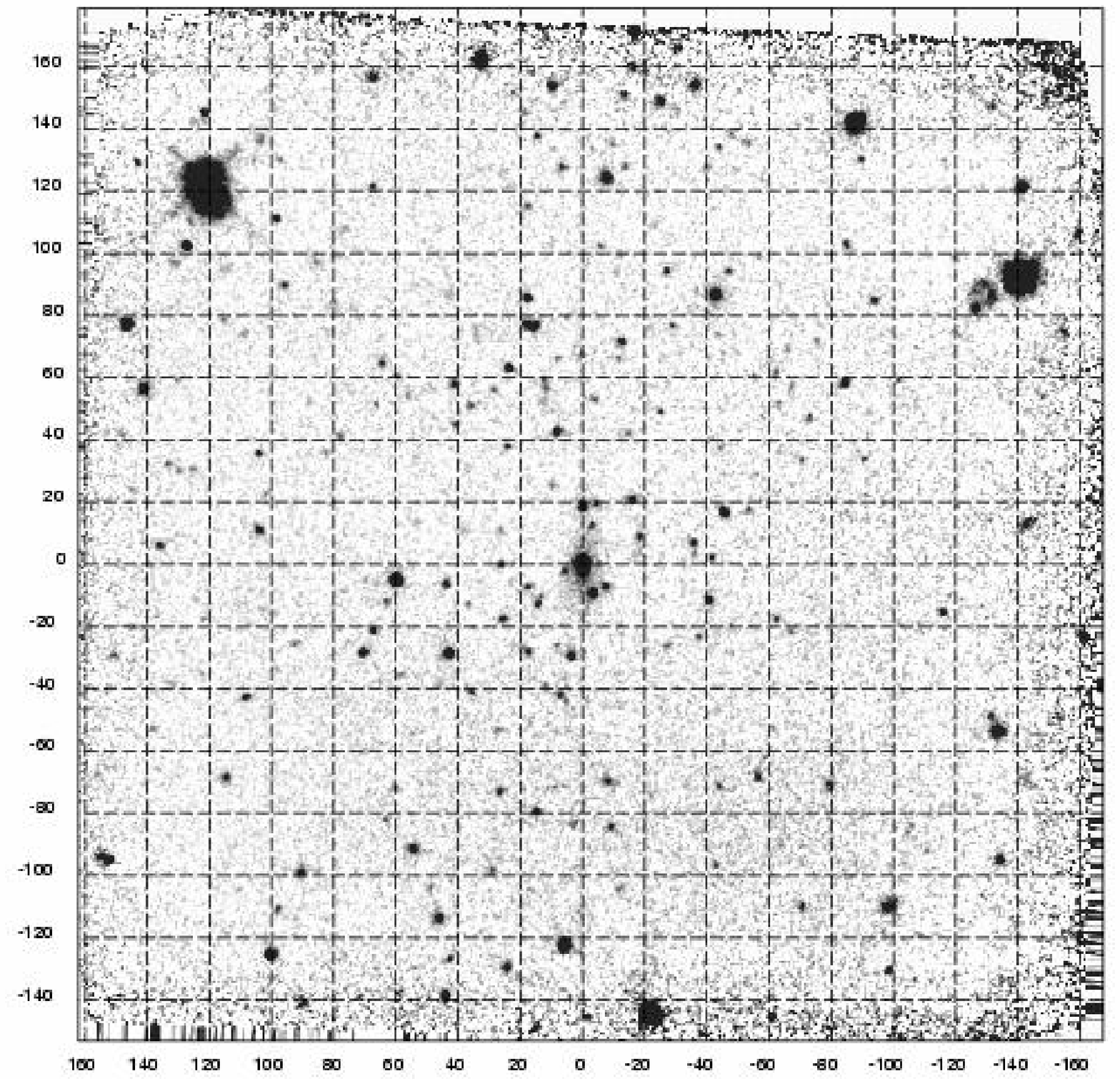}
\caption{$K$-band image of MS 1008.1-1224.  The coordinates are in relative arcsec from the
central brightest galaxy.  North is up and East to the left. }
\label{m1008k}
\end{figure}
\clearpage
\begin{figure}[p]
\plotone{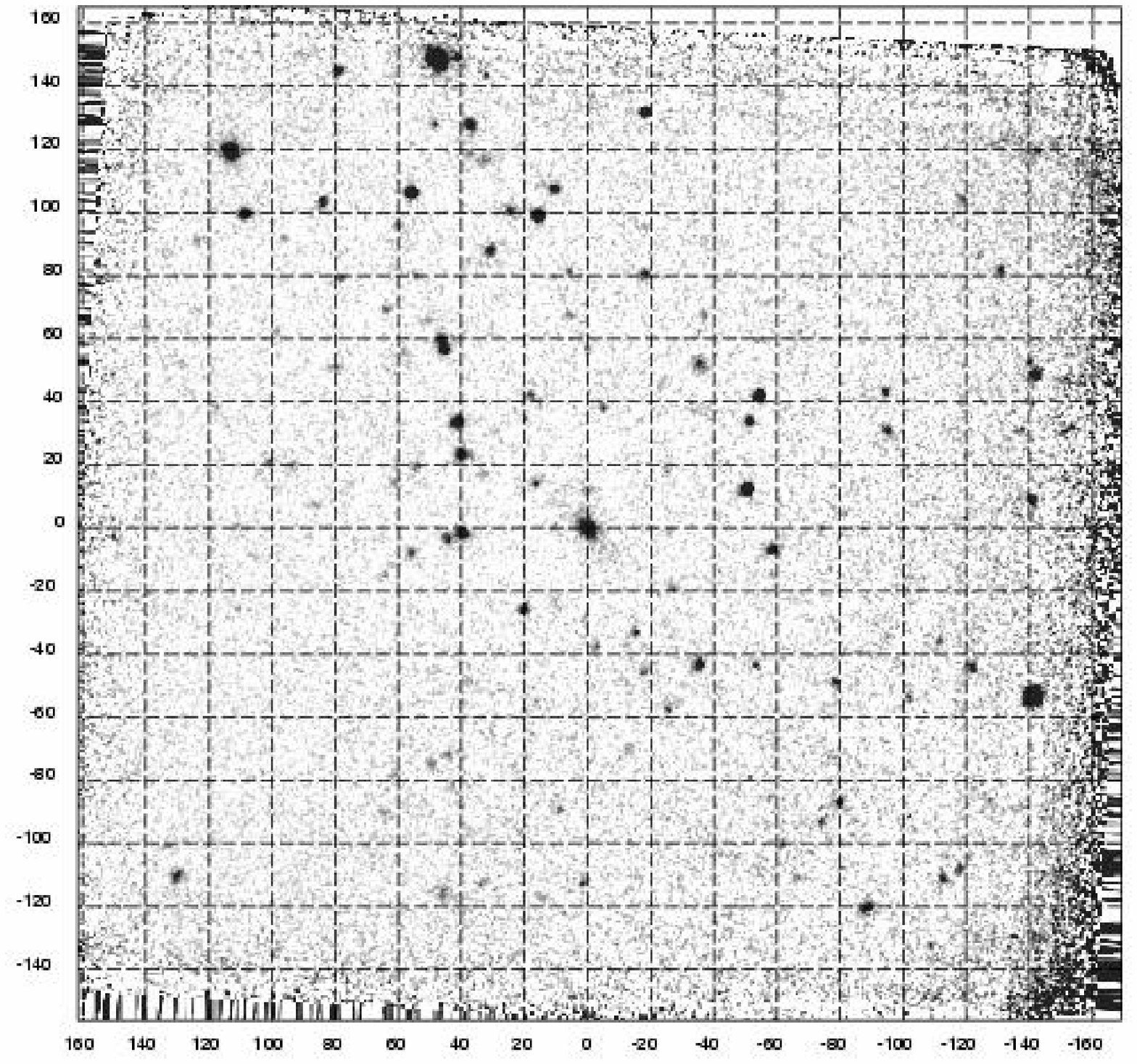}
\caption{$K$-band image of MS 1147.3+1103.  The coordinates are in relative arcsec from the
central brightest galaxy.  North is up and East to the left. }
\label{m1147k}
\end{figure}
\clearpage
\begin{figure}[p]
\plotone{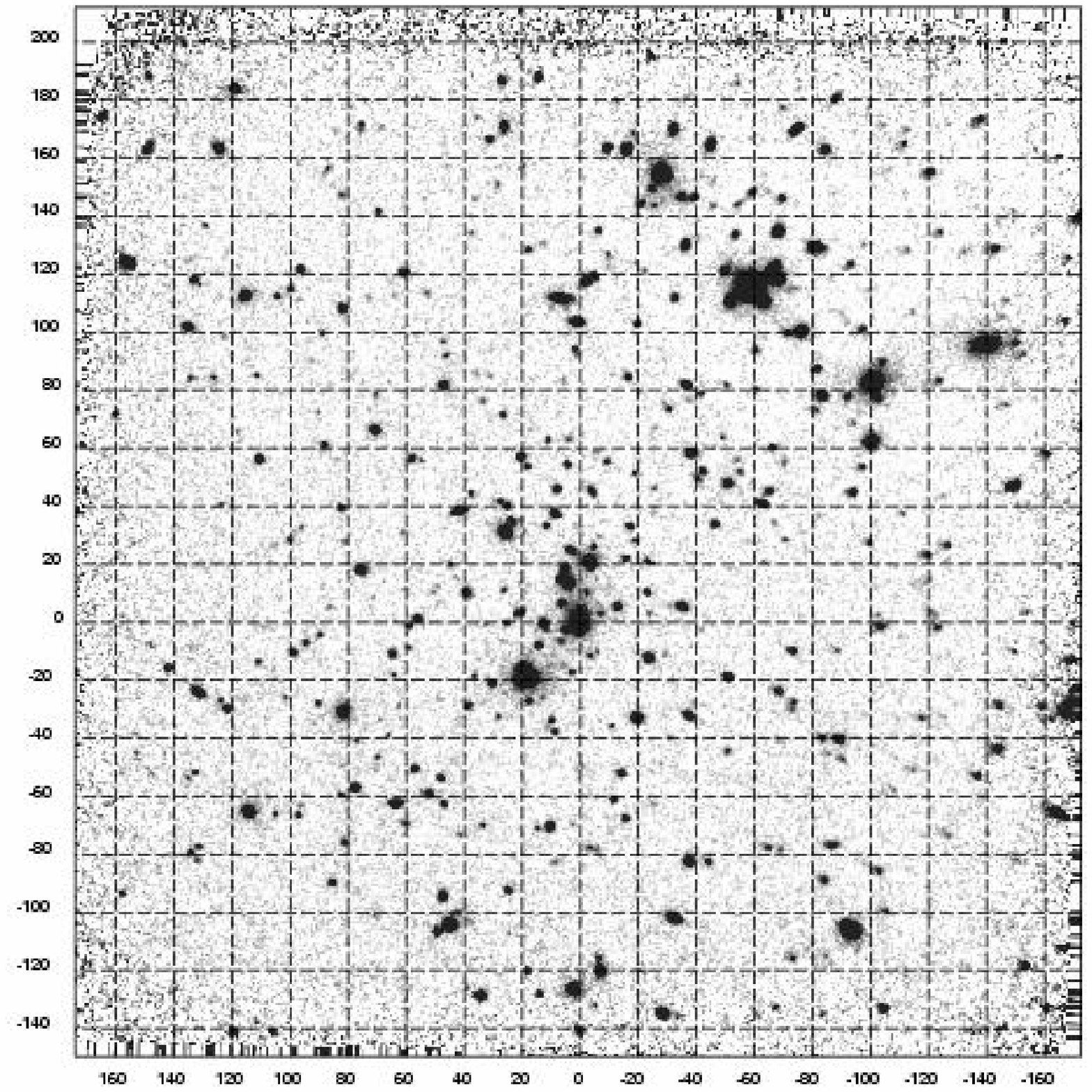}
\caption{$K$-band image of AC~118.  The coordinates are in relative arcsec from the
central brightest galaxy.  North is up and East to the left. }
\label{ac118k}
\end{figure}
\clearpage
\begin{figure}[p]
\plotone{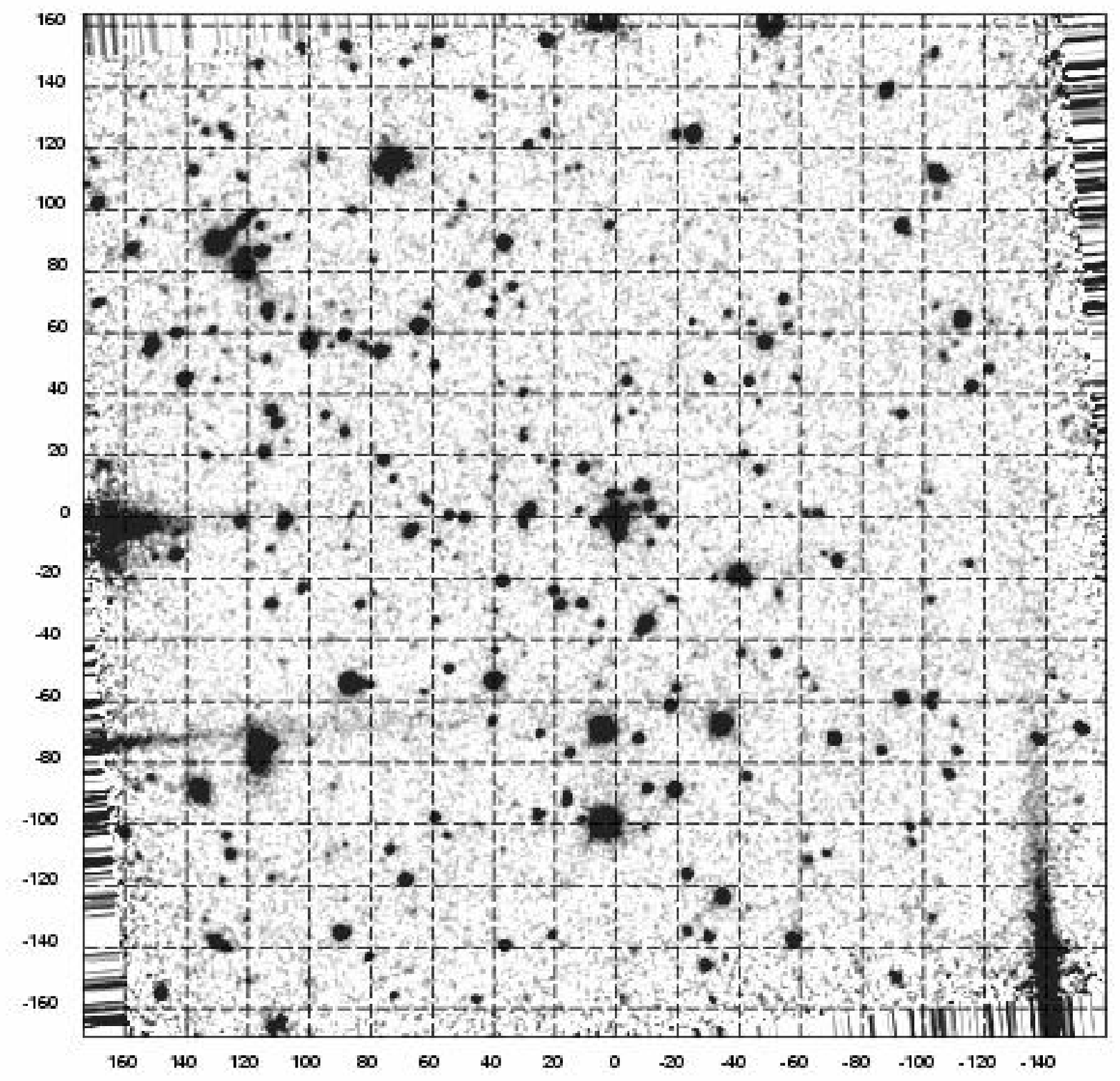}
\caption{$K$-band image of AC~103.  The coordinates are in relative arcsec from the
central brightest galaxy.  North is up and East to the left. }
\label{ac103k}
\end{figure}
\clearpage
\begin{figure}[p]
\plotone{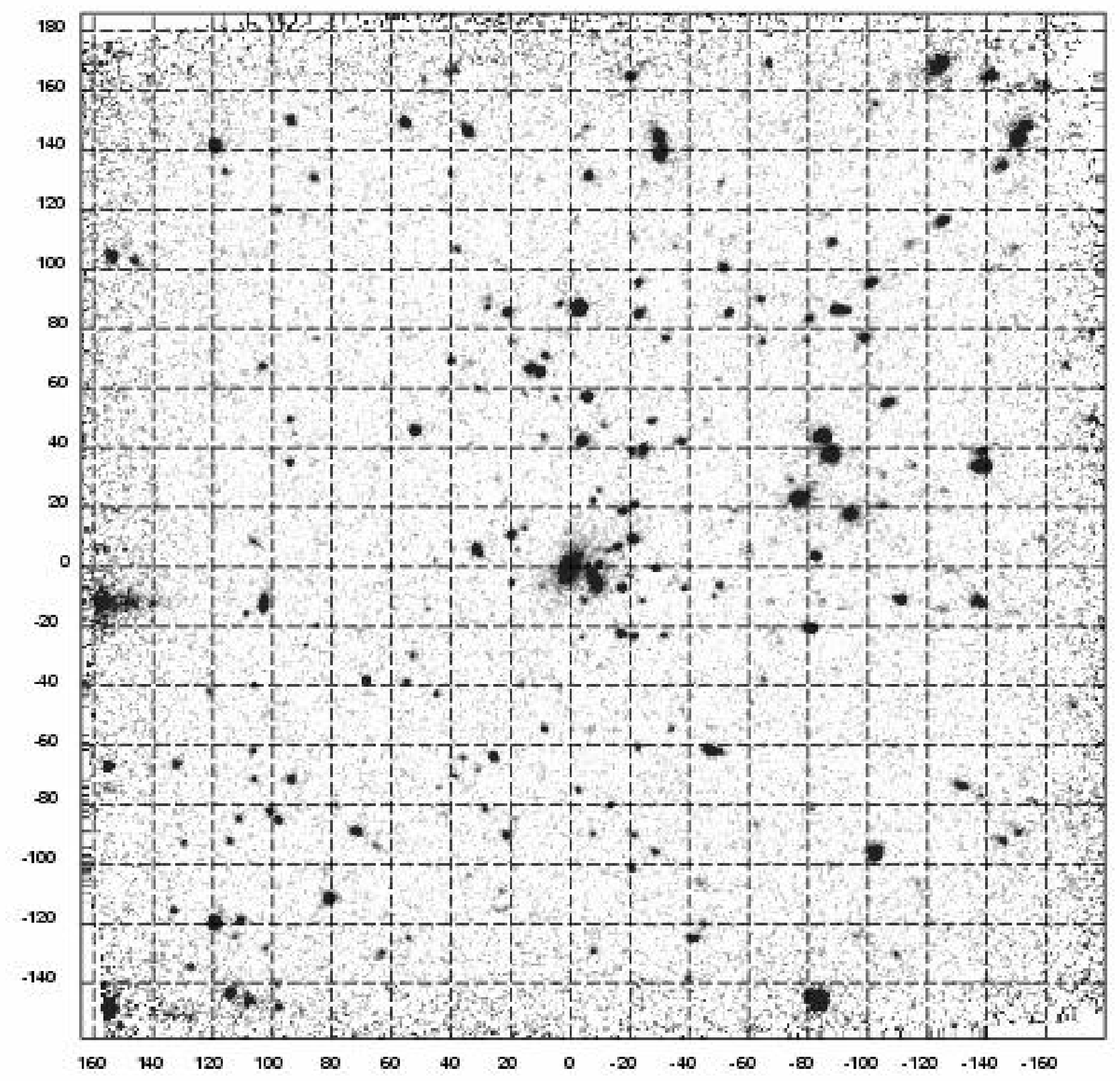}
\caption{$K$-band image of AC~114.  The coordinates are in relative arcsec from the
central brightest galaxy.  North is up and East to the left. }
\label{ac114k}
\end{figure}
\clearpage
\begin{figure}[p]
\plotone{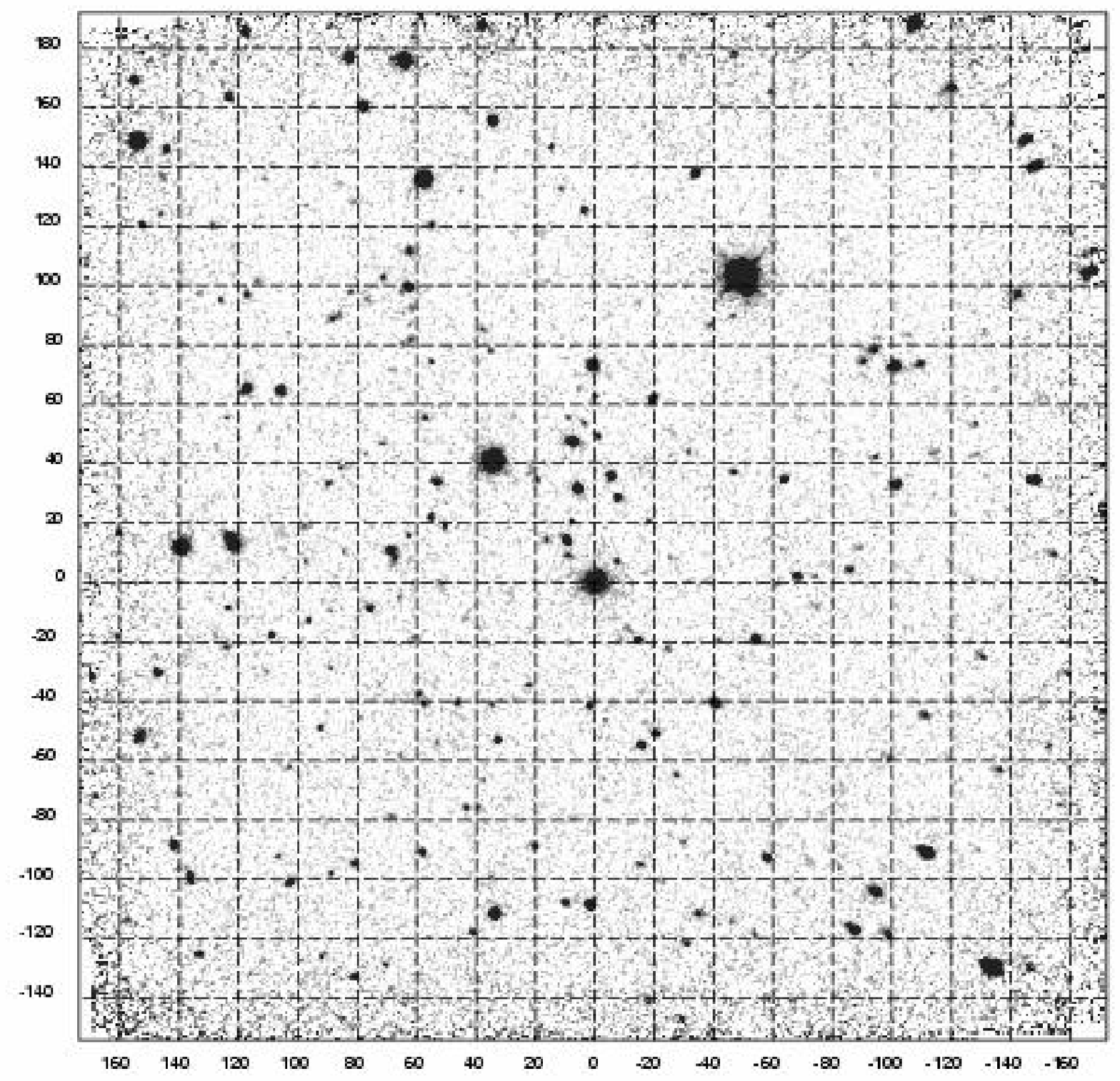}
\caption{$K$-band image of MS 2137.3-2339.  The coordinates are in relative arcsec from the
central brightest galaxy.  North is up and East to the left. }
\label{m2137k}
\end{figure}
\clearpage
\begin{figure}[p]
\plotone{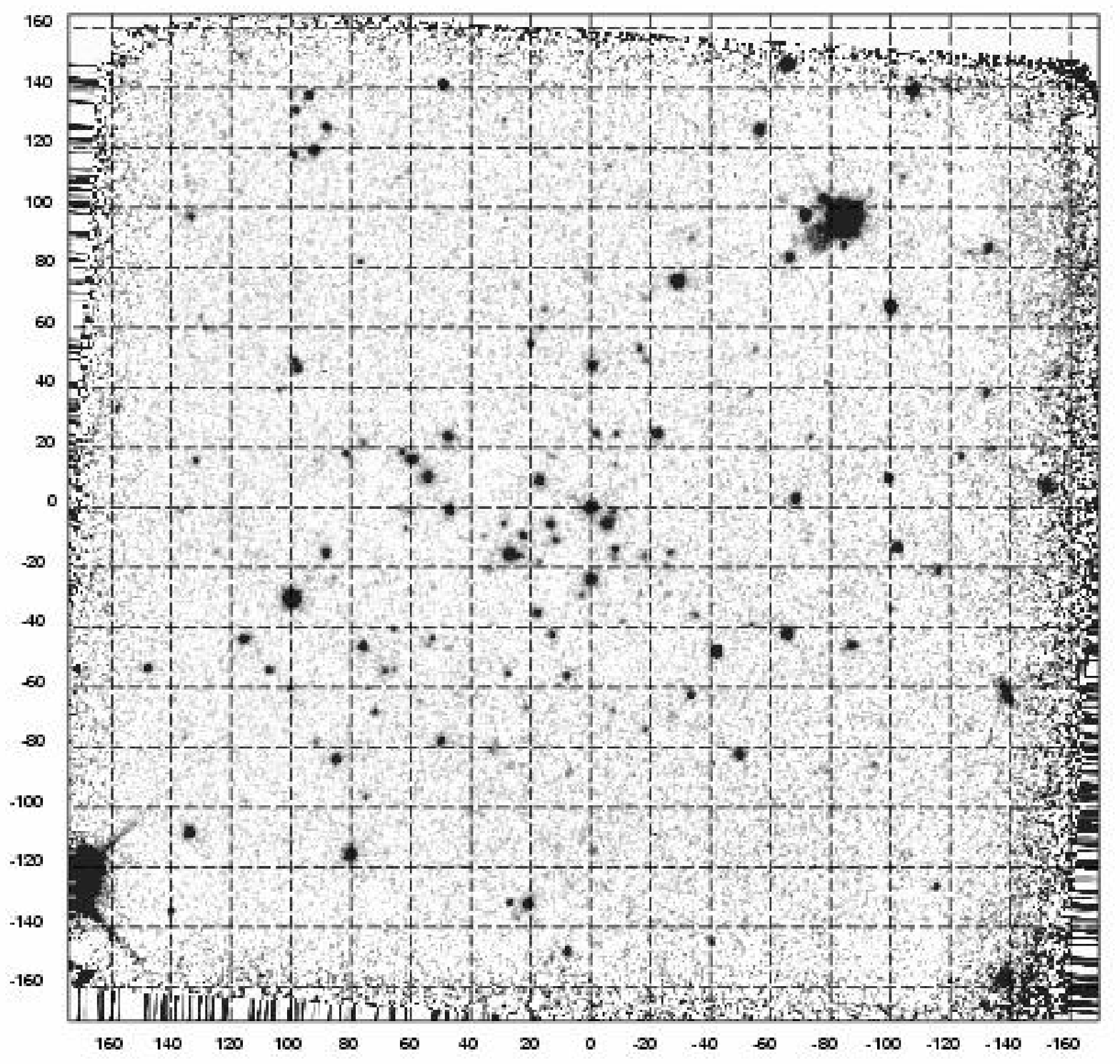}
\caption{$K$-band image of Abell S0506.  The coordinates are in relative arcsec from the
central brightest galaxy.  North is up and East to the left. }
\label{s0506k}
\end{figure}
\clearpage
\begin{figure}[p]
\plotone{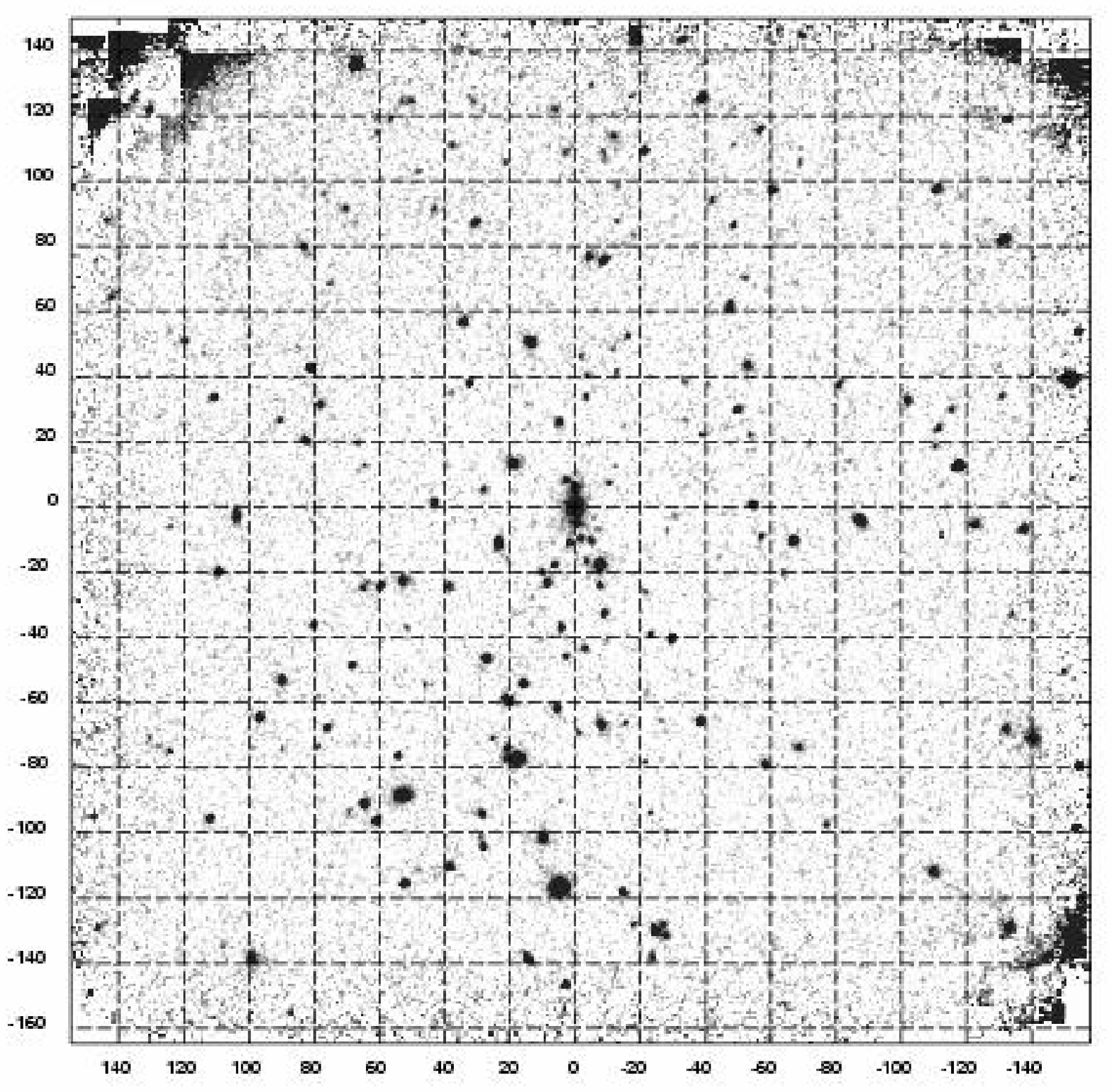}
\caption{$K$-band image of MS 1358.1+6245.  The coordinates are in relative arcsec from the
central brightest galaxy.  North is up and East to the left. }
\label{m1358k}
\end{figure}
\clearpage
\begin{figure}[p]
\plotone{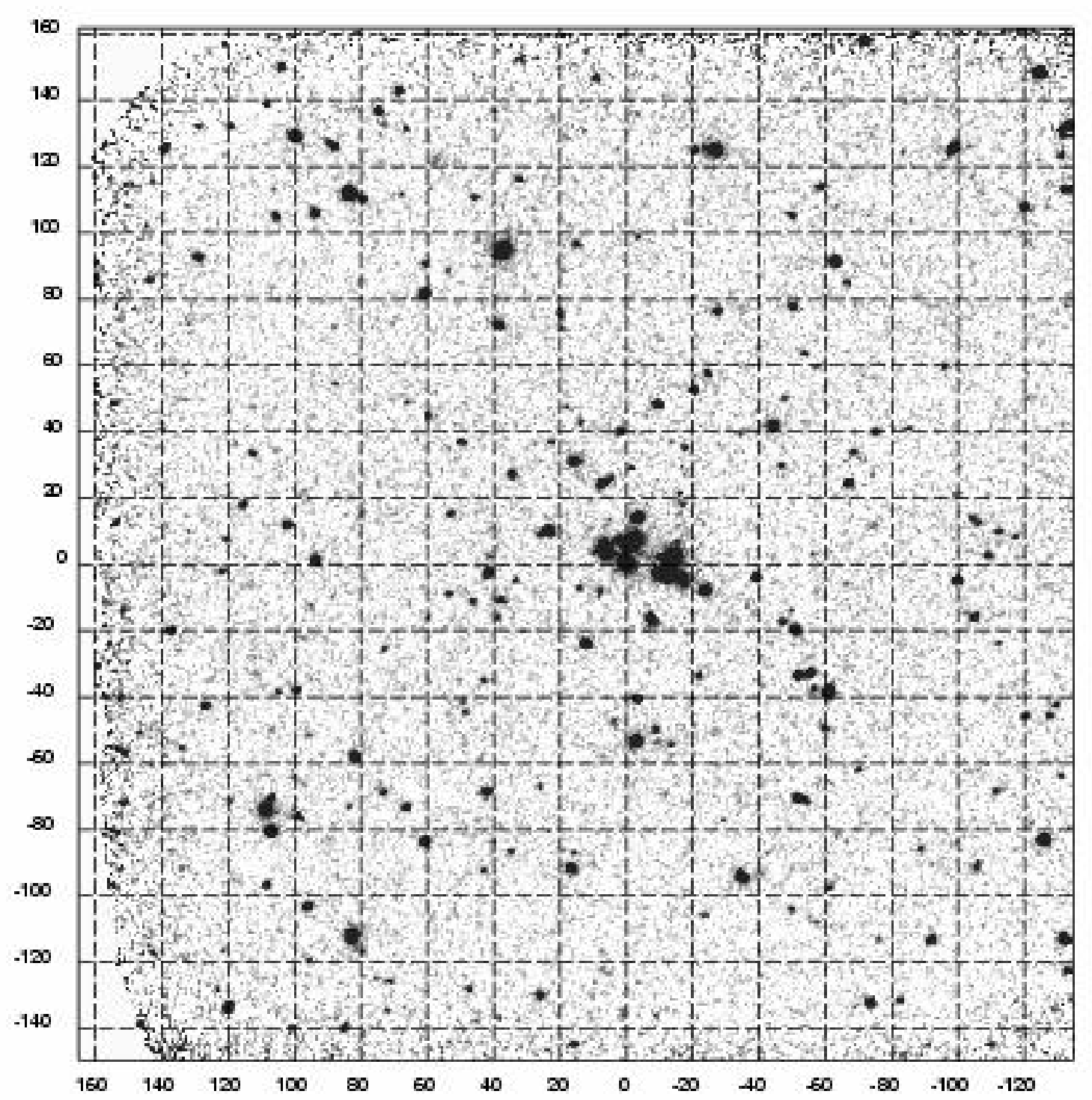}
\caption{$K$-band image of CL~2244-02.  The coordinates are in relative arcsec from the
central brightest galaxy.  North is up and East to the left. }
\label{c2244k}
\end{figure}
\clearpage
\begin{figure}[p]
\plotone{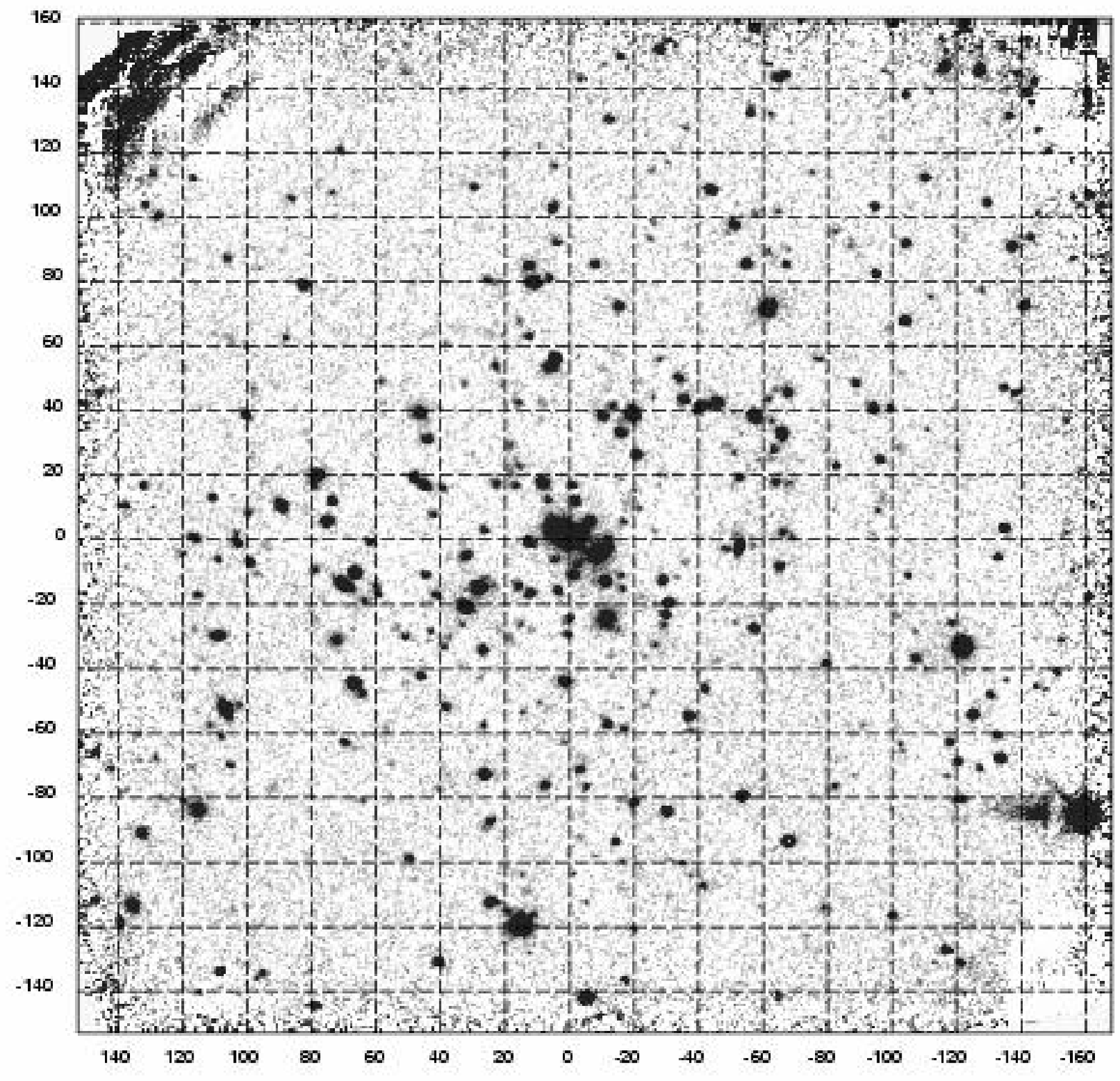}
\caption{$K$-band image of CL~0024+16.  The coordinates are in relative arcsec from the
central brightest galaxy.  North is up and East to the left. }
\label{c0024k}
\end{figure}
\clearpage
\begin{figure}[p]
\plotone{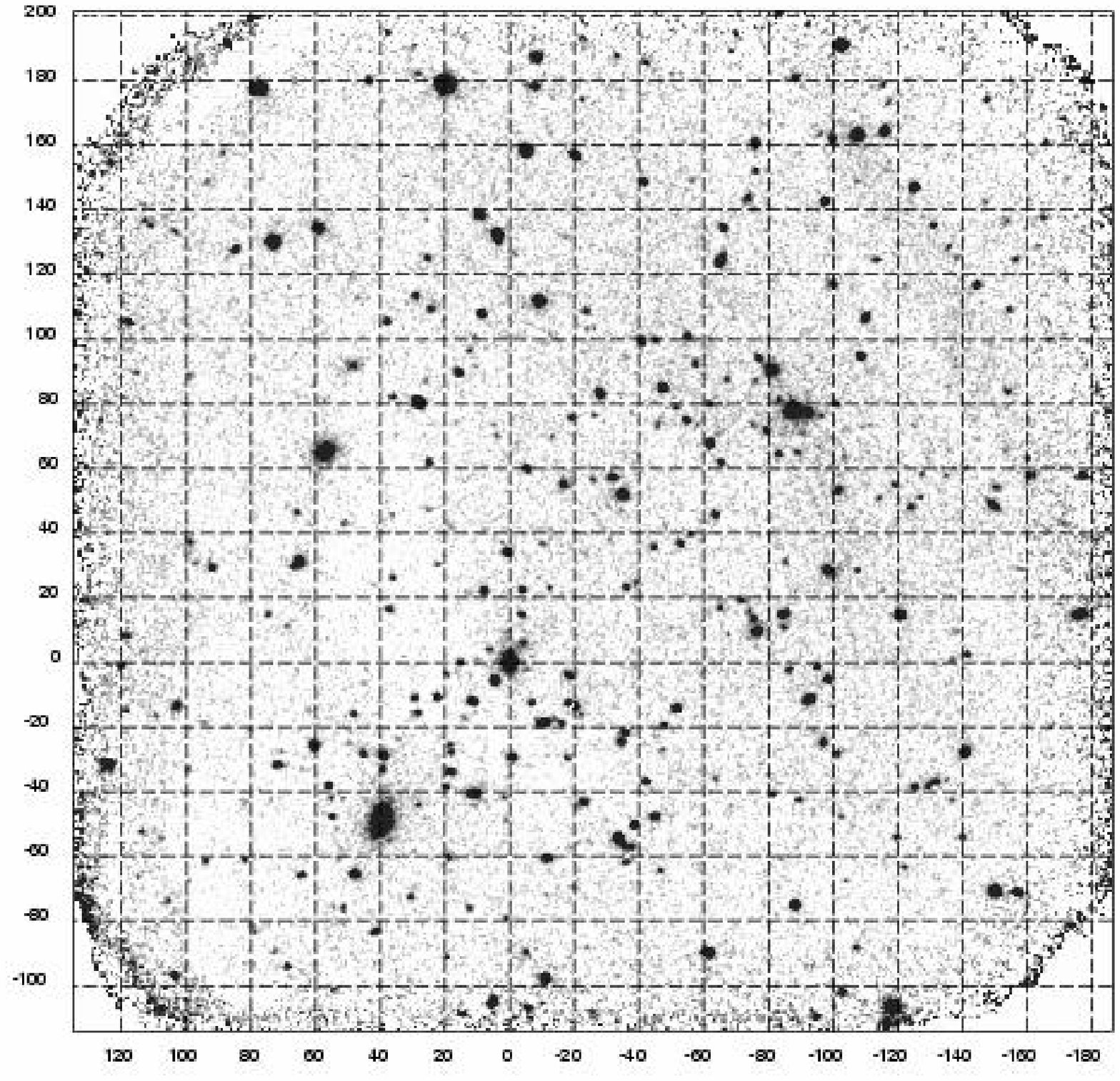}
\caption{$K$-band image of GHO~0303+1706.  The coordinates are in relative arcsec from the
central brightest galaxy.  North is up and East to the left. }
\label{g0303k}
\end{figure}
\clearpage
\begin{figure}[p]
\plotone{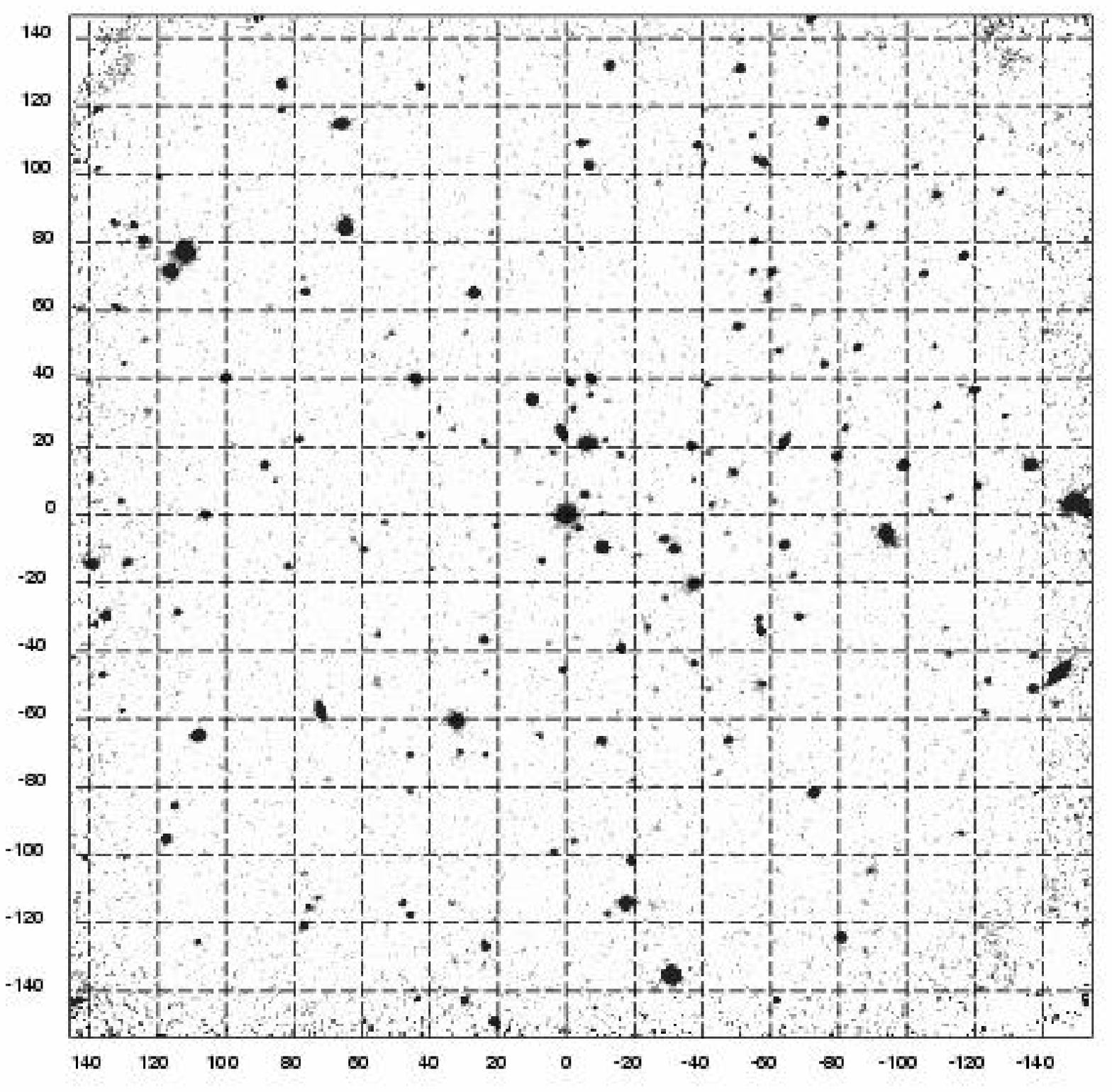}
\caption{$K$-band image of 3C~313.  The coordinates are in relative arcsec from the
central brightest galaxy.  North is up and East to the left. }
\label{3c313k}
\end{figure}
\clearpage
\begin{figure}[p]
\plotone{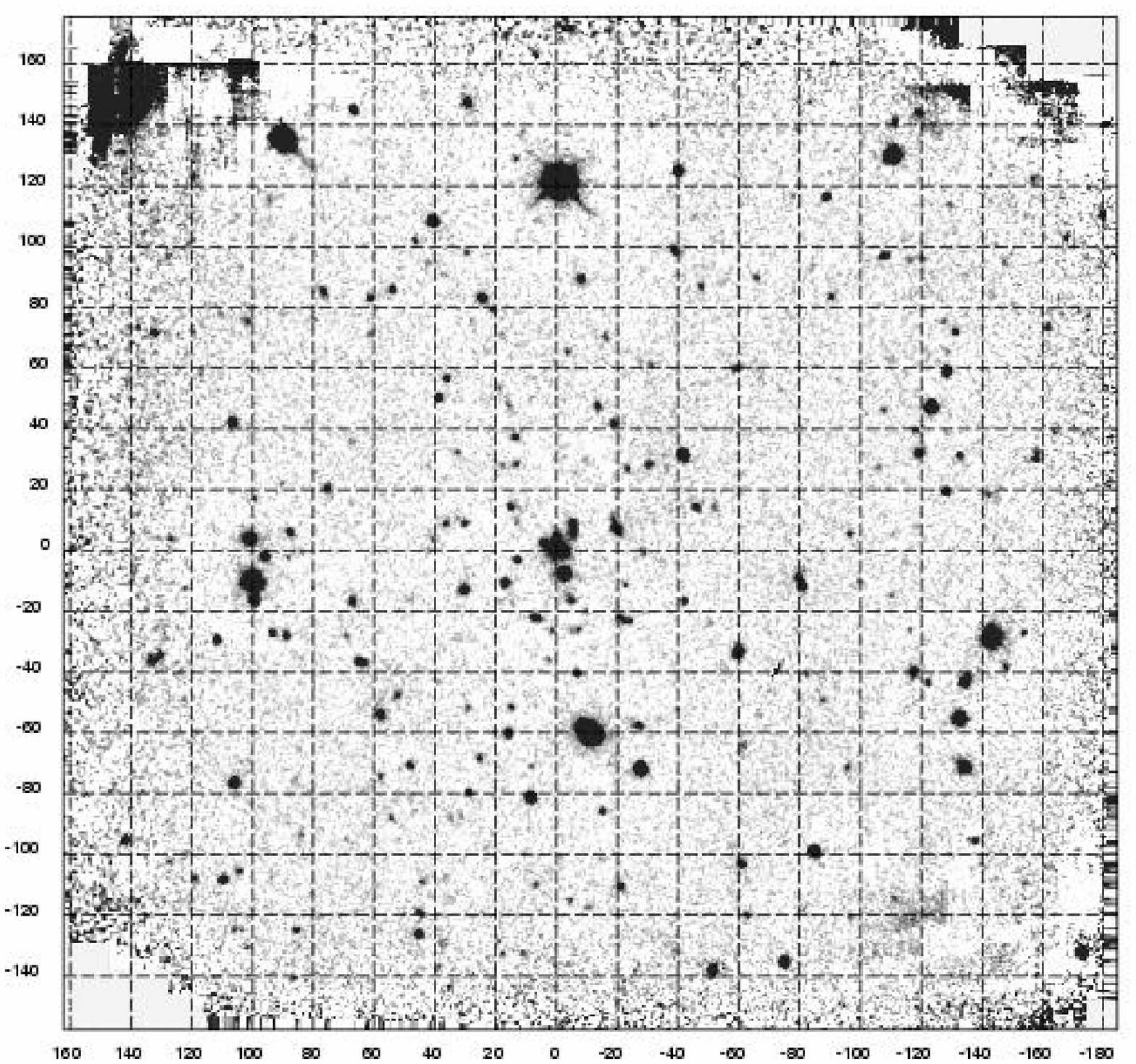}
\caption{$K$-band image of 3C~295.  The coordinates are in relative arcsec from the
central brightest galaxy.  North is up and East to the left. }
\label{3c295k}
\end{figure}
\clearpage
\begin{figure}[p]
\plotone{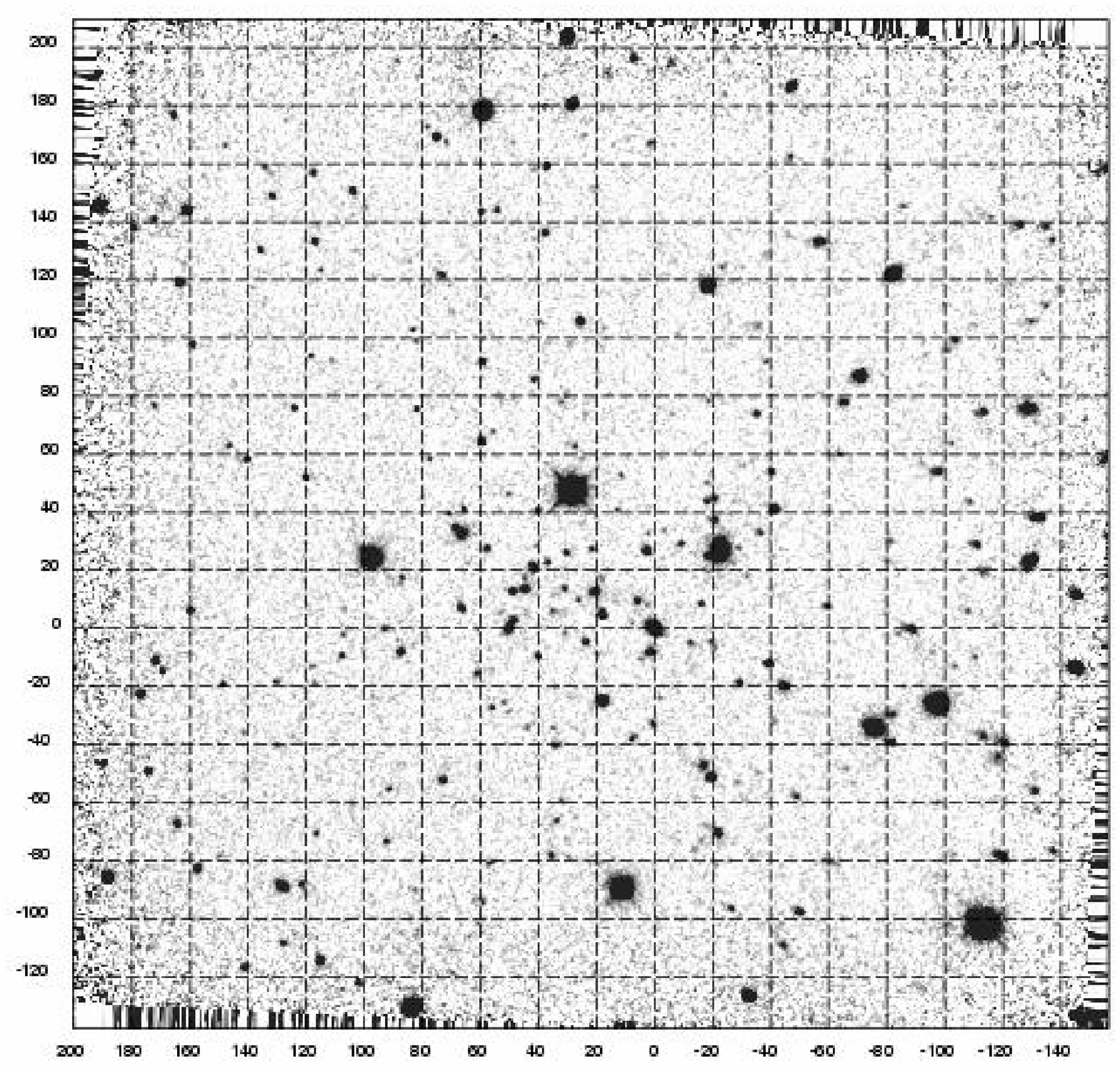}
\caption{$K$-band image of F1557.19TC.  The coordinates are in relative arcsec from the
central brightest galaxy.  North is up and East to the left. }
\label{f1557k}
\end{figure}
\clearpage
\begin{figure}[p]
\plotone{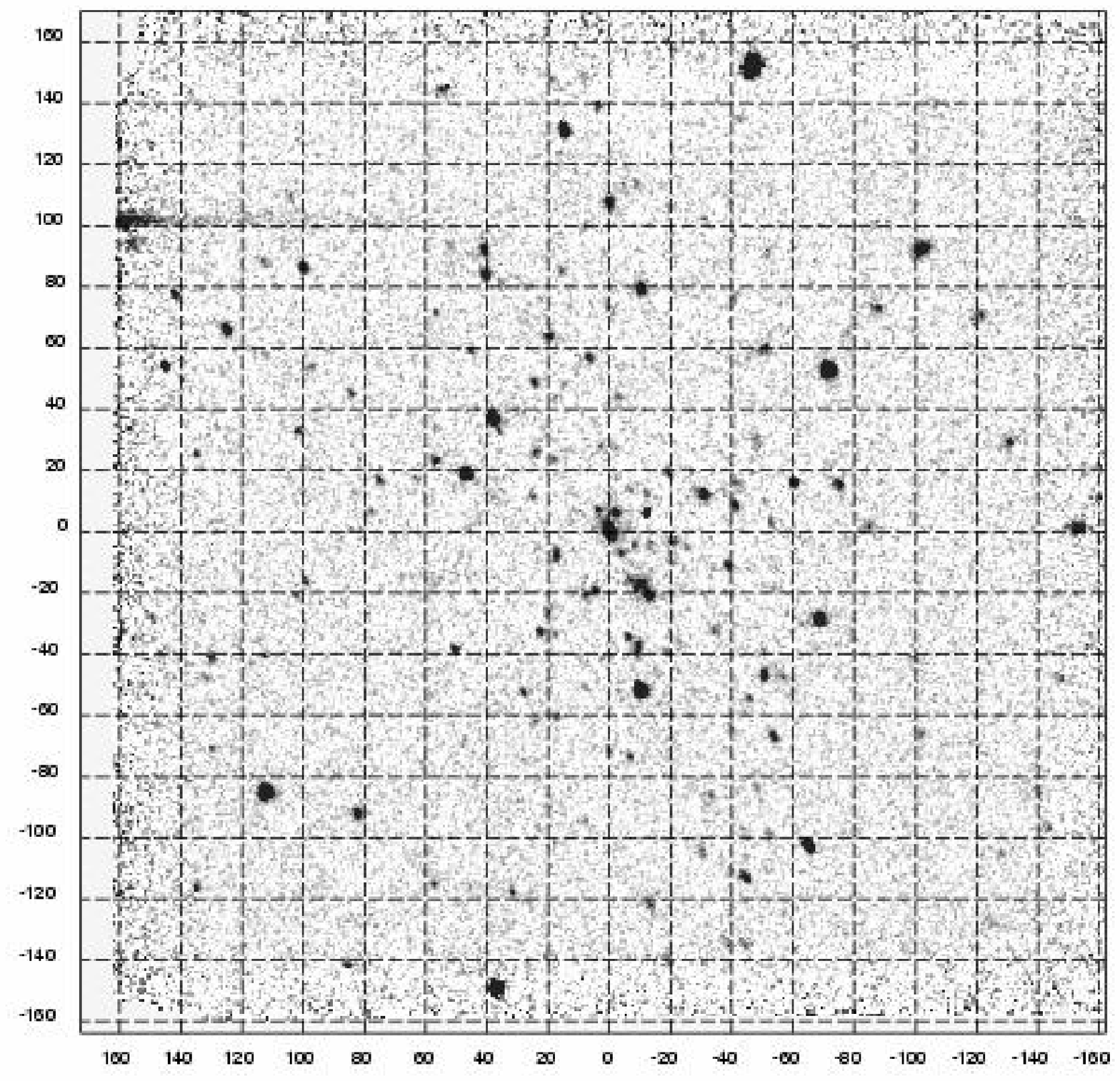}
\caption{$K$-band image of Vidal 14.  The coordinates are in relative arcsec from the
central brightest galaxy.  North is up and East to the left. }
\label{v14k}
\end{figure}
\clearpage
\begin{figure}[p]
\plotone{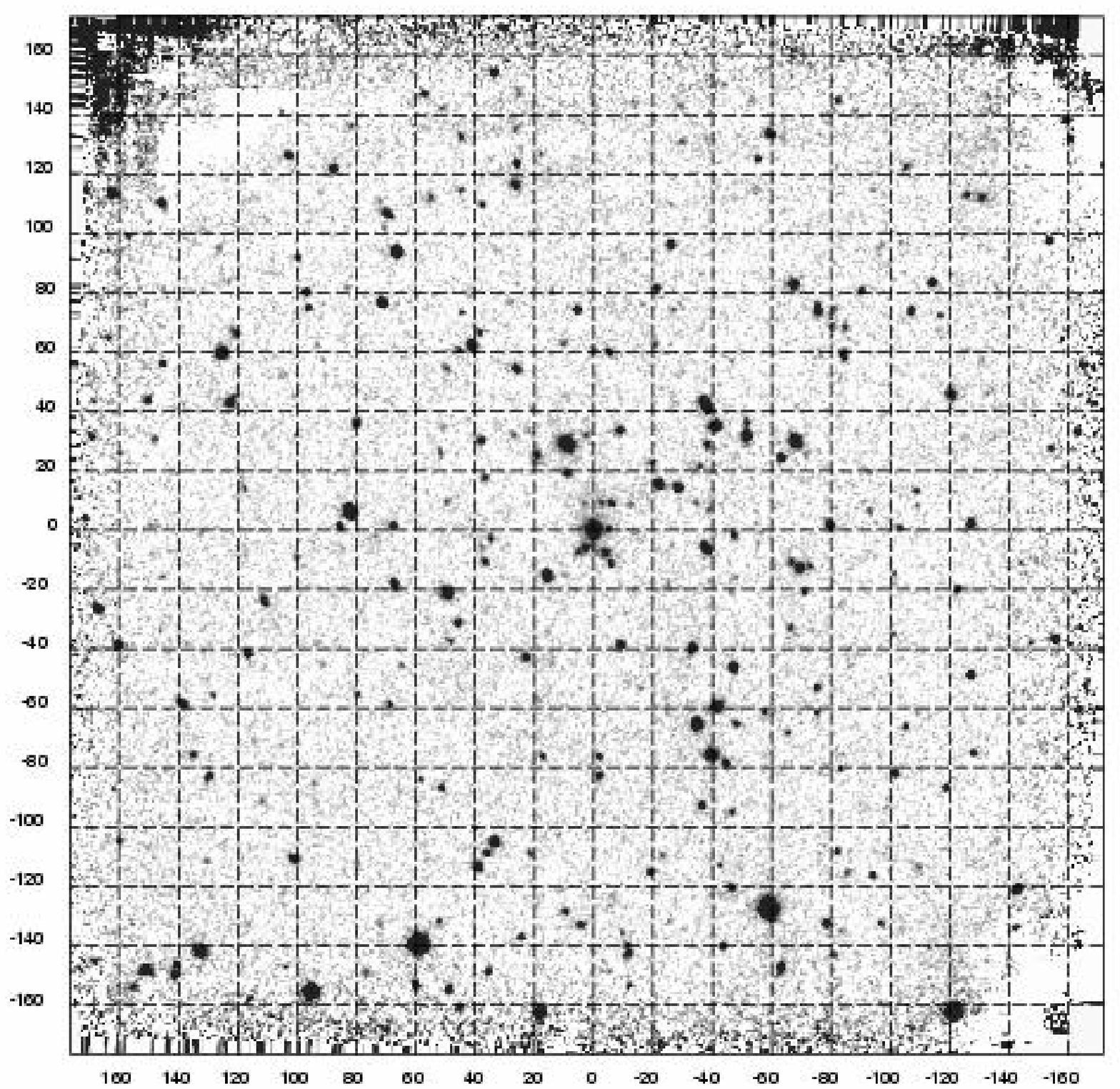}
\caption{$K$-band image of GHO~1601+4253.  The coordinates are in relative arcsec from the
central brightest galaxy.  North is up and East to the left. }
\label{g1601k}
\end{figure}
\clearpage
\begin{figure}
\plotone{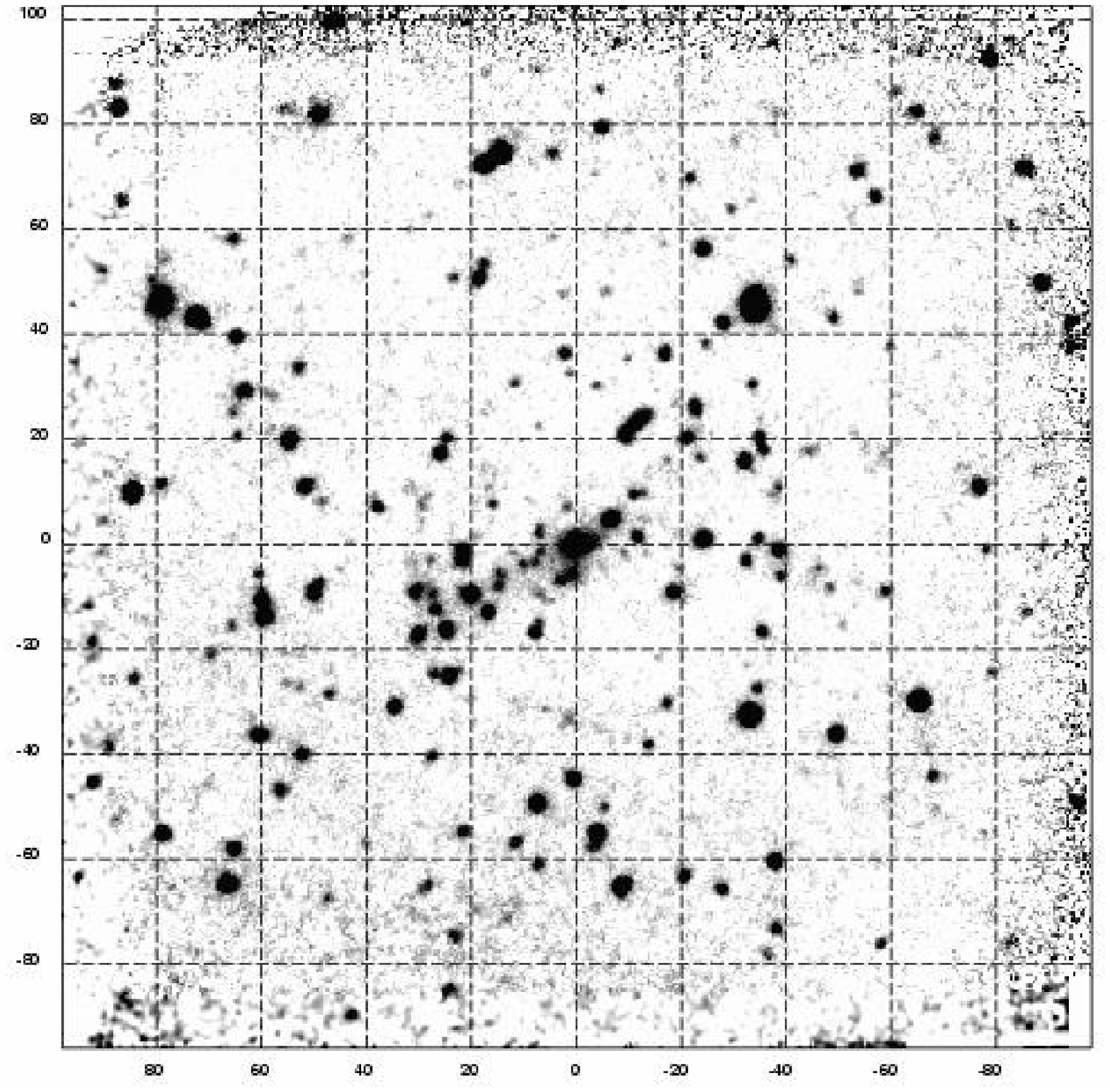}[p]
\caption{$K$-band image of MS 0451.6-0306.  The coordinates are in relative arcsec from the
central brightest galaxy.  North is up and East to the left. }
\label{m0451k}
\end{figure}
\clearpage
\begin{figure}[p]
\plotone{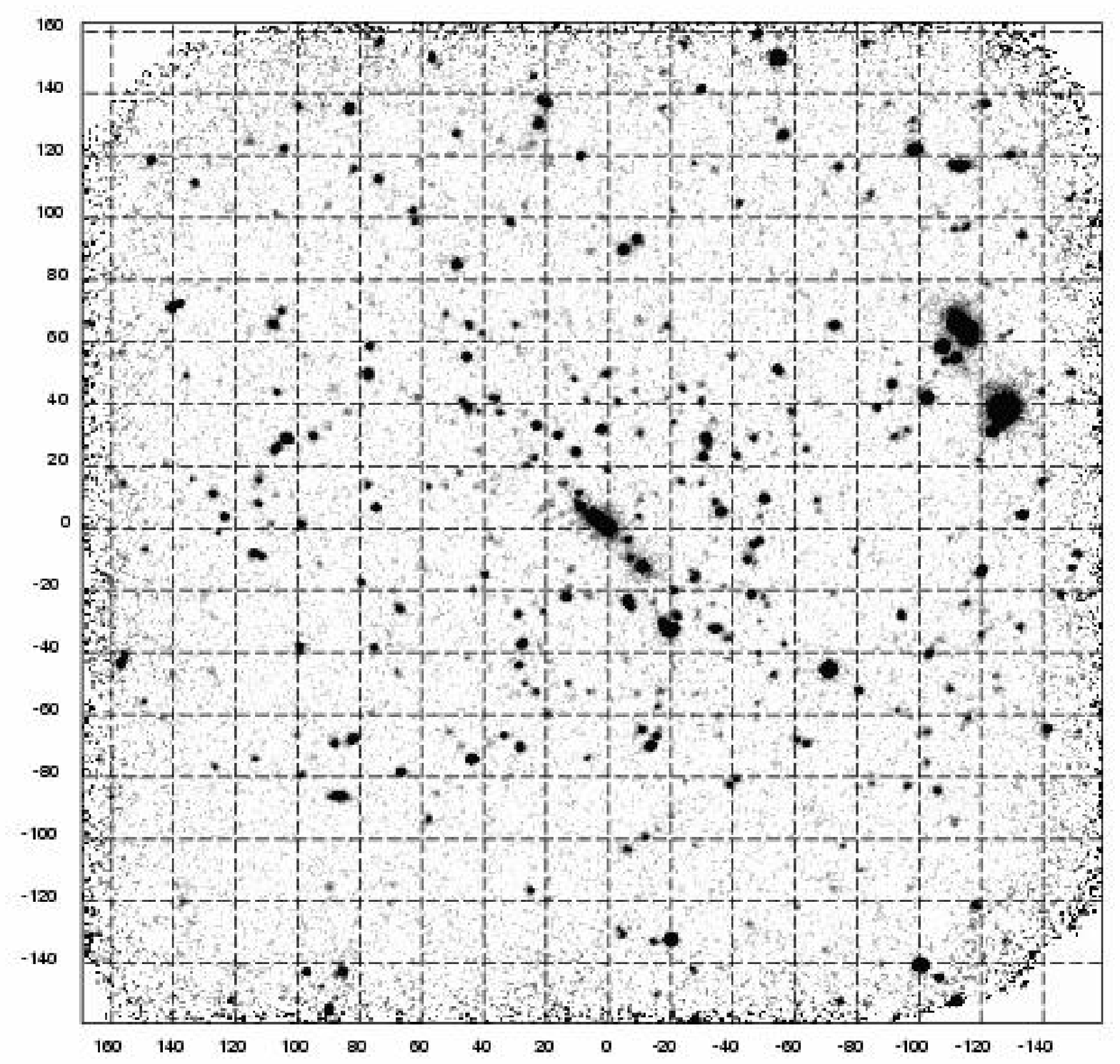}
\caption{$K$-band image of CL~0016+16.  The coordinates are in relative arcsec from the
central brightest galaxy.  North is up and East to the left. }
\label{c0016k}
\end{figure}
\clearpage
\begin{figure}[p]
\plotone{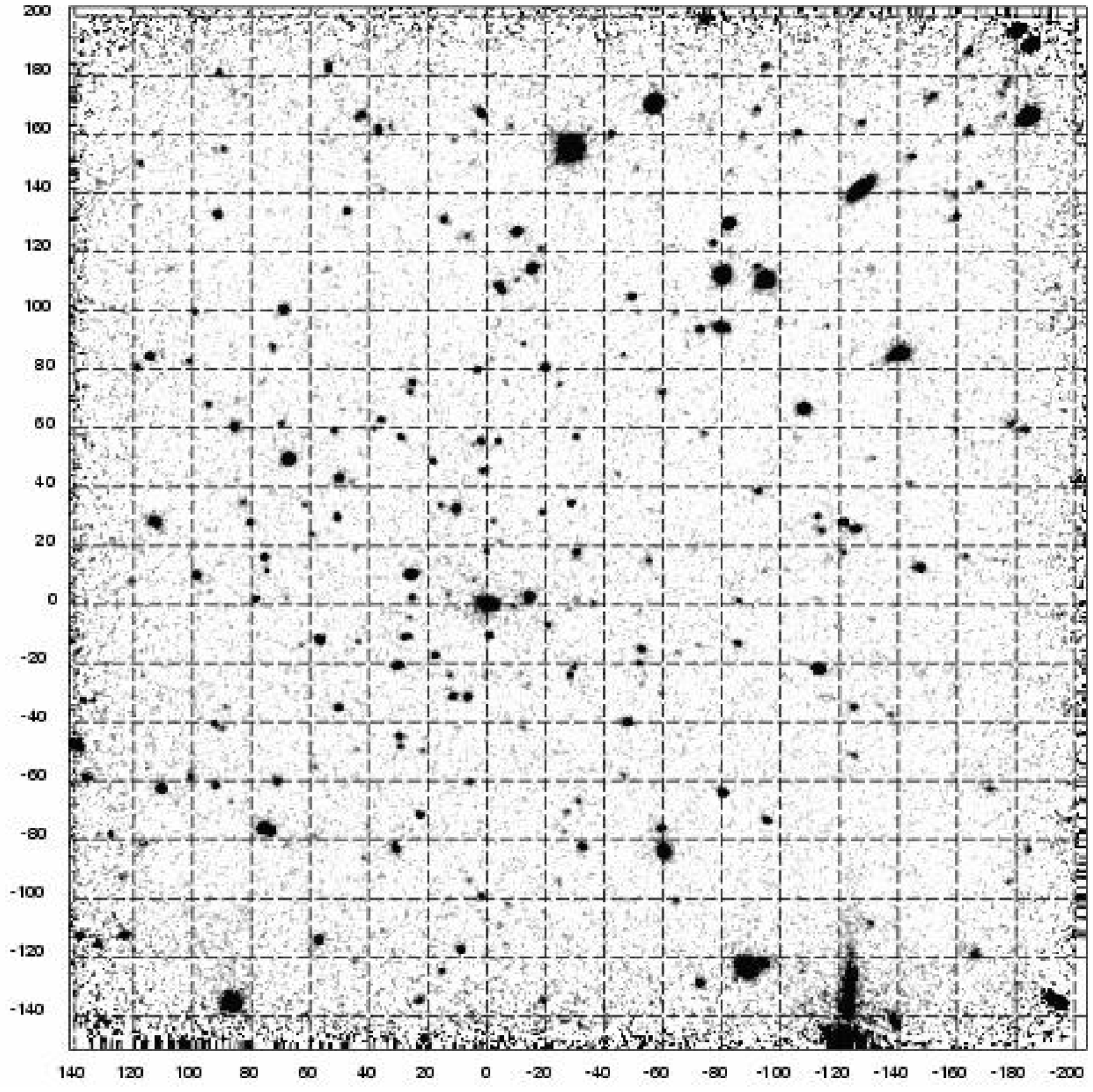}
\caption{$K$-band image of J1888.16CL.  The coordinates are in relative arcsec from the
central brightest galaxy.  North is up and East to the left. }
\label{j1888k}
\end{figure}
\clearpage
\begin{figure}[p]
\plotone{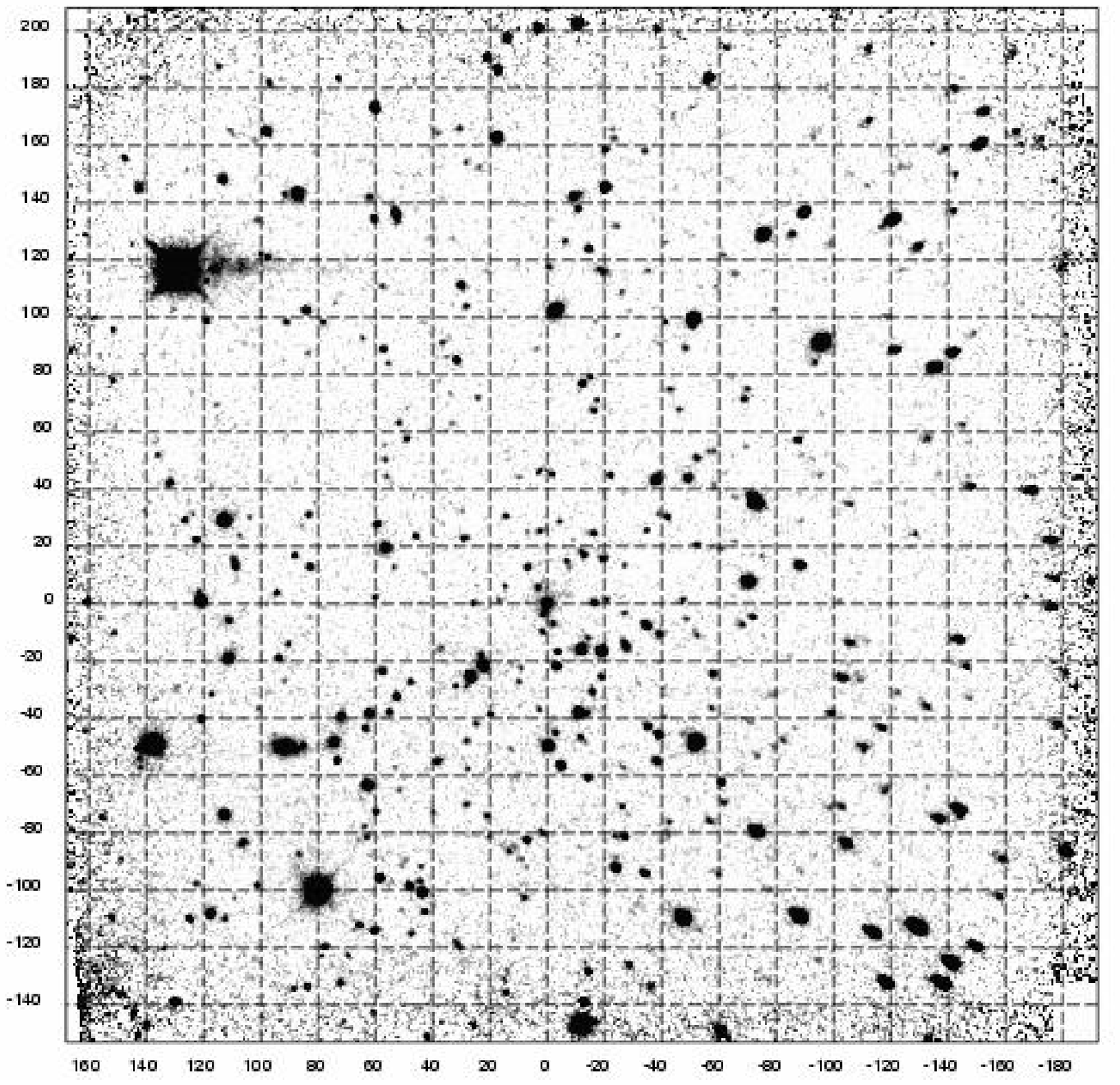}
\caption{$K$-band image of MS 2053.7-0449.  The coordinates are in relative arcsec from the
central brightest galaxy.  North is up and East to the left. }
\label{m2053k}
\end{figure}
\clearpage
\begin{figure}[p]
\plotone{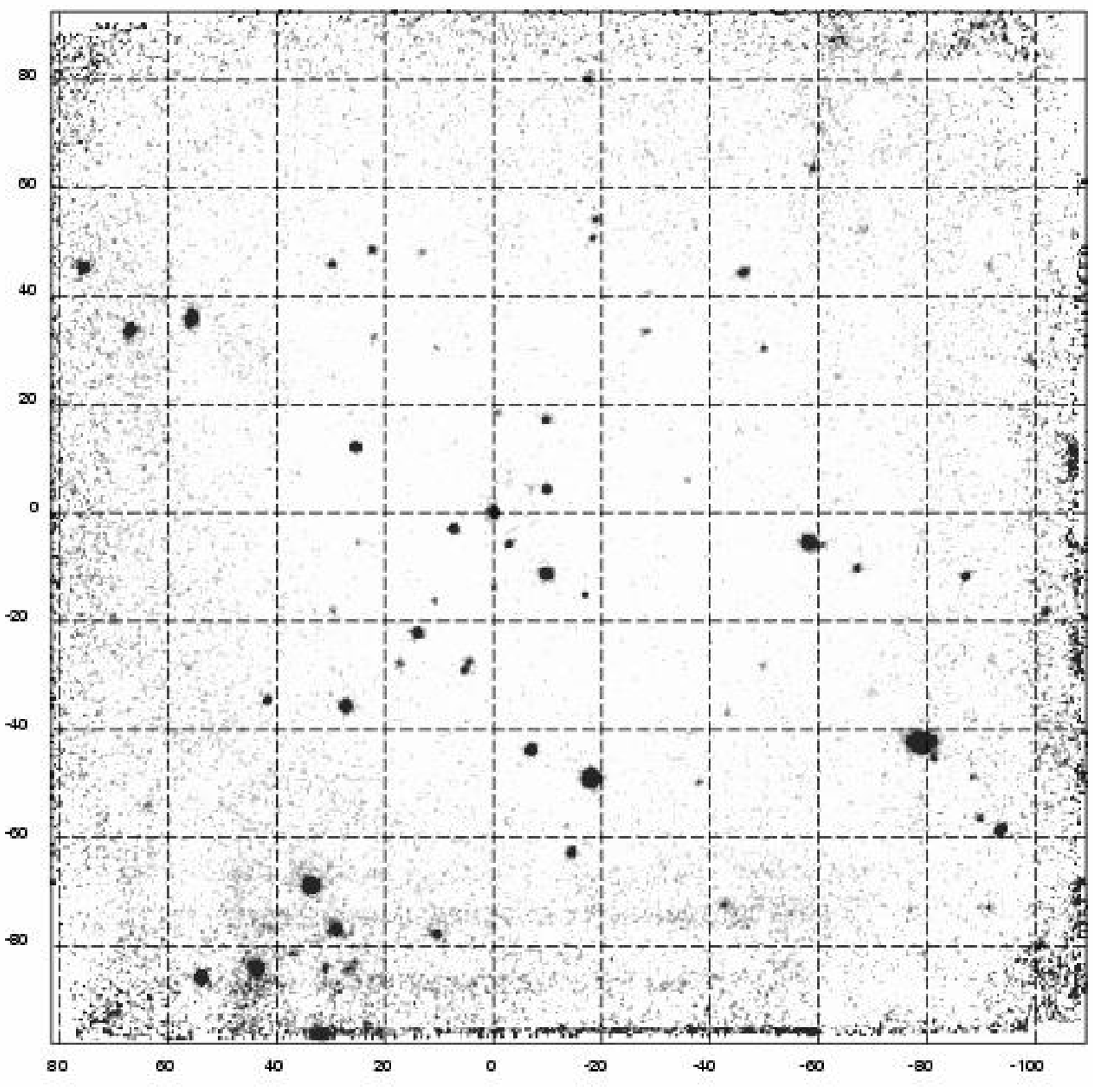}
\caption{$K$-band image of GHO~0317+1521.  The coordinates are in relative arcsec from the
central brightest galaxy.  North is up and East to the left. }
\label{g0317k}
\end{figure}
\clearpage
\begin{figure}[p]
\plotone{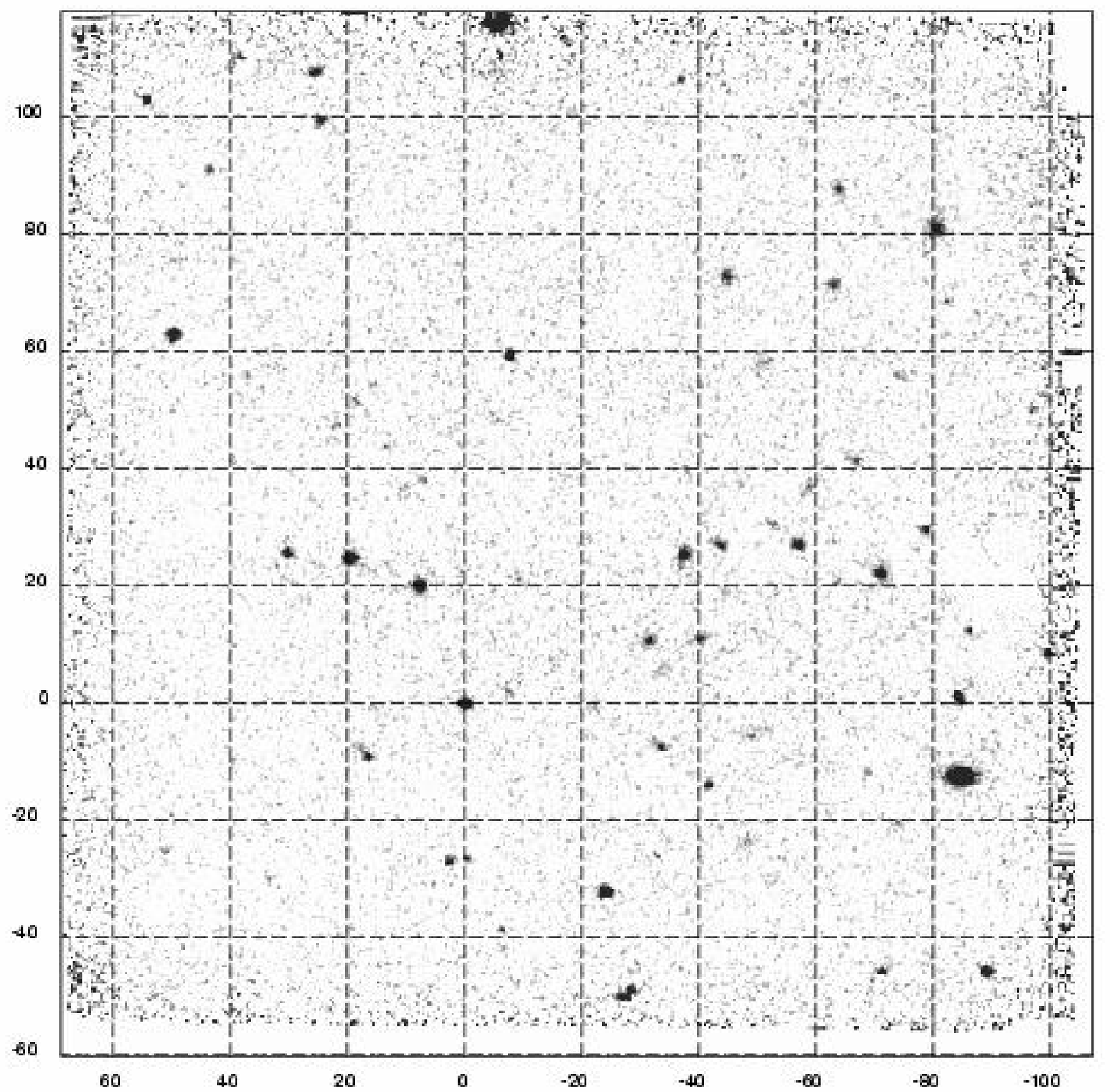}
\caption{$H$-band image of GHO~0229+0035.  The coordinates are in relative arcsec from the
central brightest galaxy.  North is up and East to the left. }
\label{g0229h}
\end{figure}
\clearpage
\begin{figure}[p]
\plotone{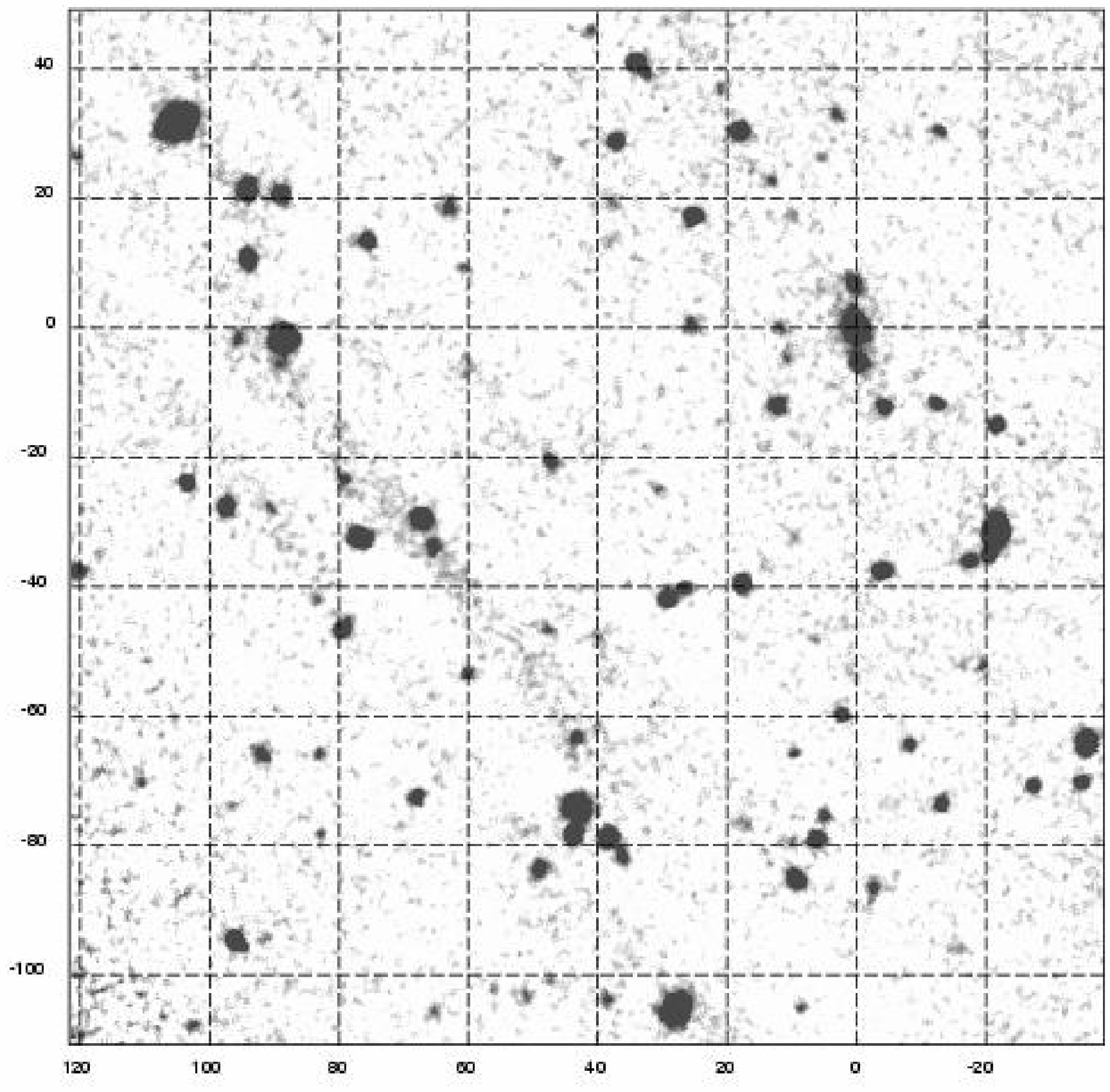}
\caption{$K$-band image of 3C~220.1.  The coordinates are in relative arcsec from the
central brightest galaxy.  North is up and East to the left. }
\label{3c220p1k}
\end{figure}
\clearpage
\begin{figure}[p]
\plotone{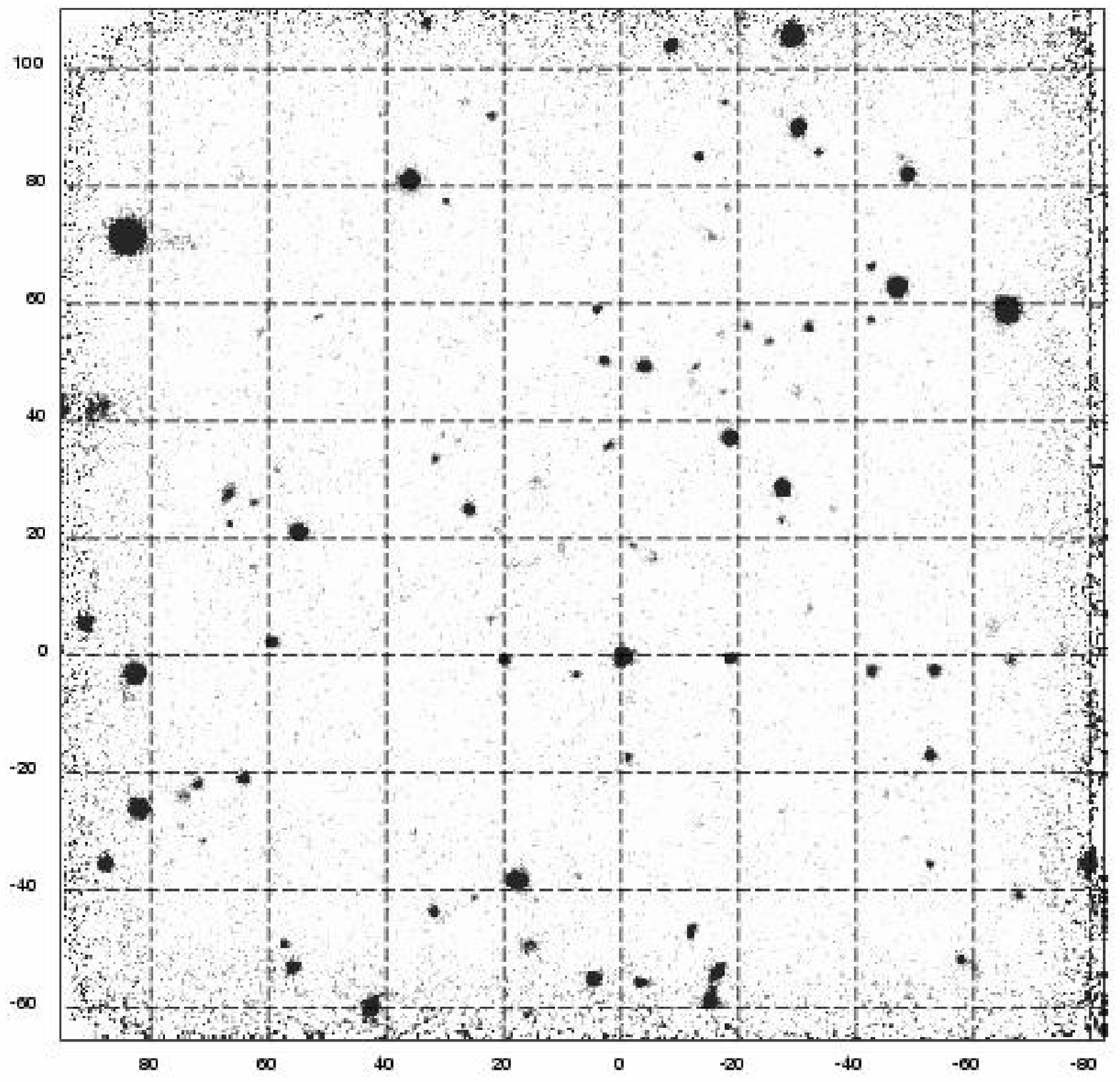}
\caption{$H$-band image of GHO~2155+0321.  The coordinates are in relative arcsec from the
central brightest galaxy.  North is up and East to the left. }
\label{g2155h}
\end{figure}
\clearpage
\begin{figure}[p]
\plotone{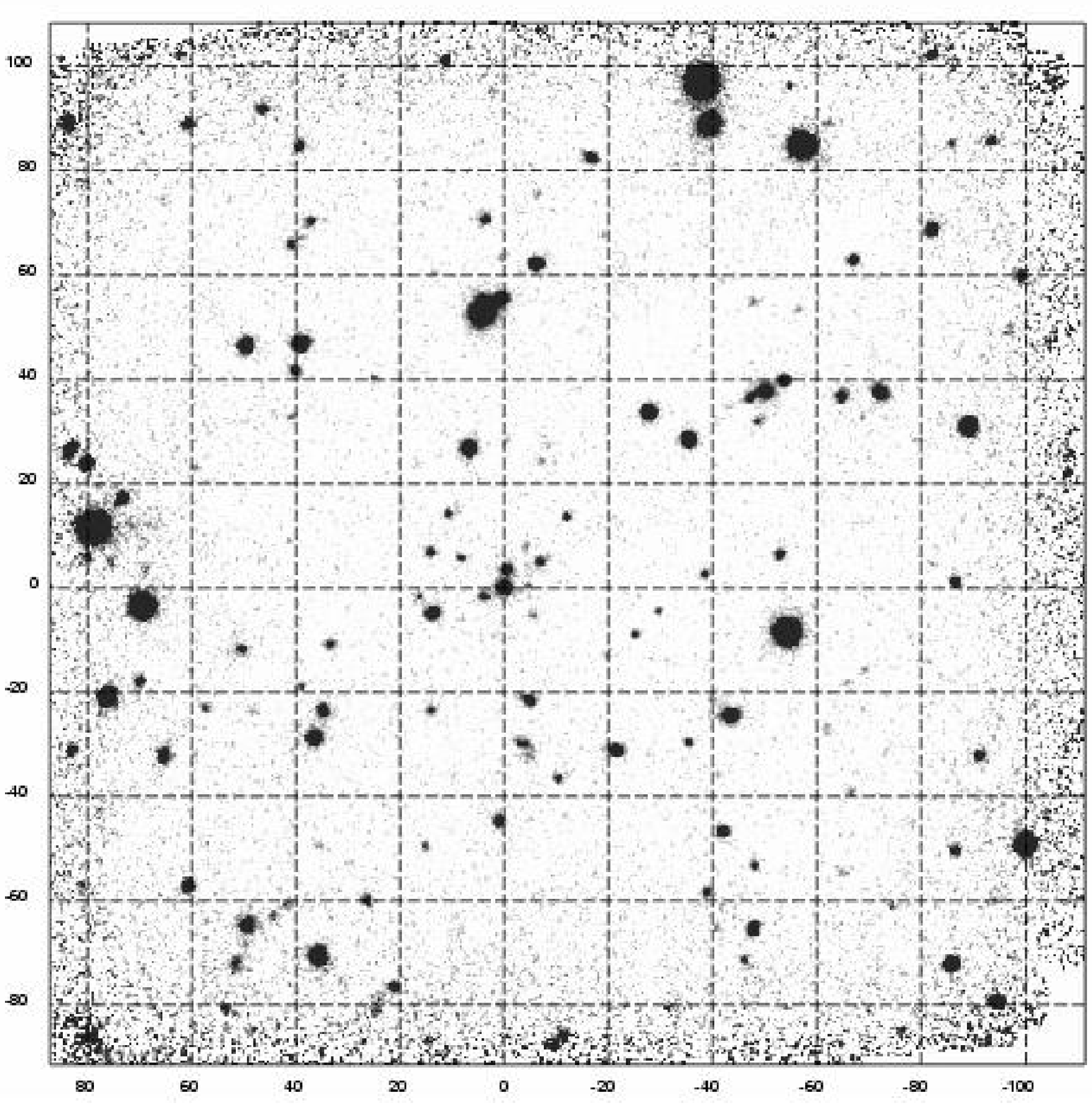}
\caption{$K$-band image of GHO~2201+0258.  The coordinates are in relative arcsec from the
central brightest galaxy.  North is up and East to the left. }
\label{g2201k}
\end{figure}
\clearpage
\begin{figure}[p]
\plotone{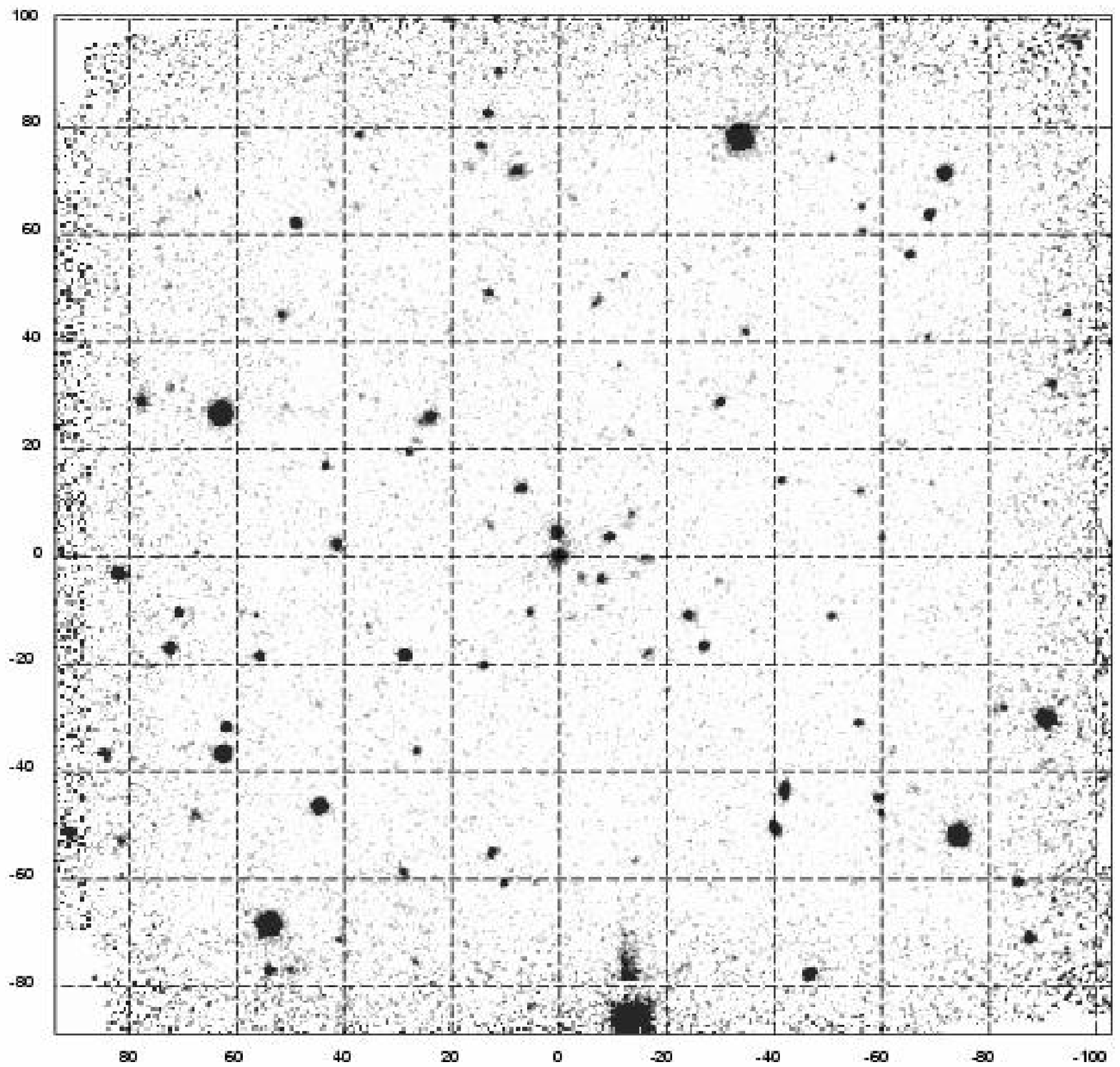}
\caption{$K$-band image of 3C~34.  The coordinates are in relative arcsec from the
central brightest galaxy.  North is up and East to the left. }
\label{3c34k}
\end{figure}
\clearpage
\begin{figure}[p]
\plotone{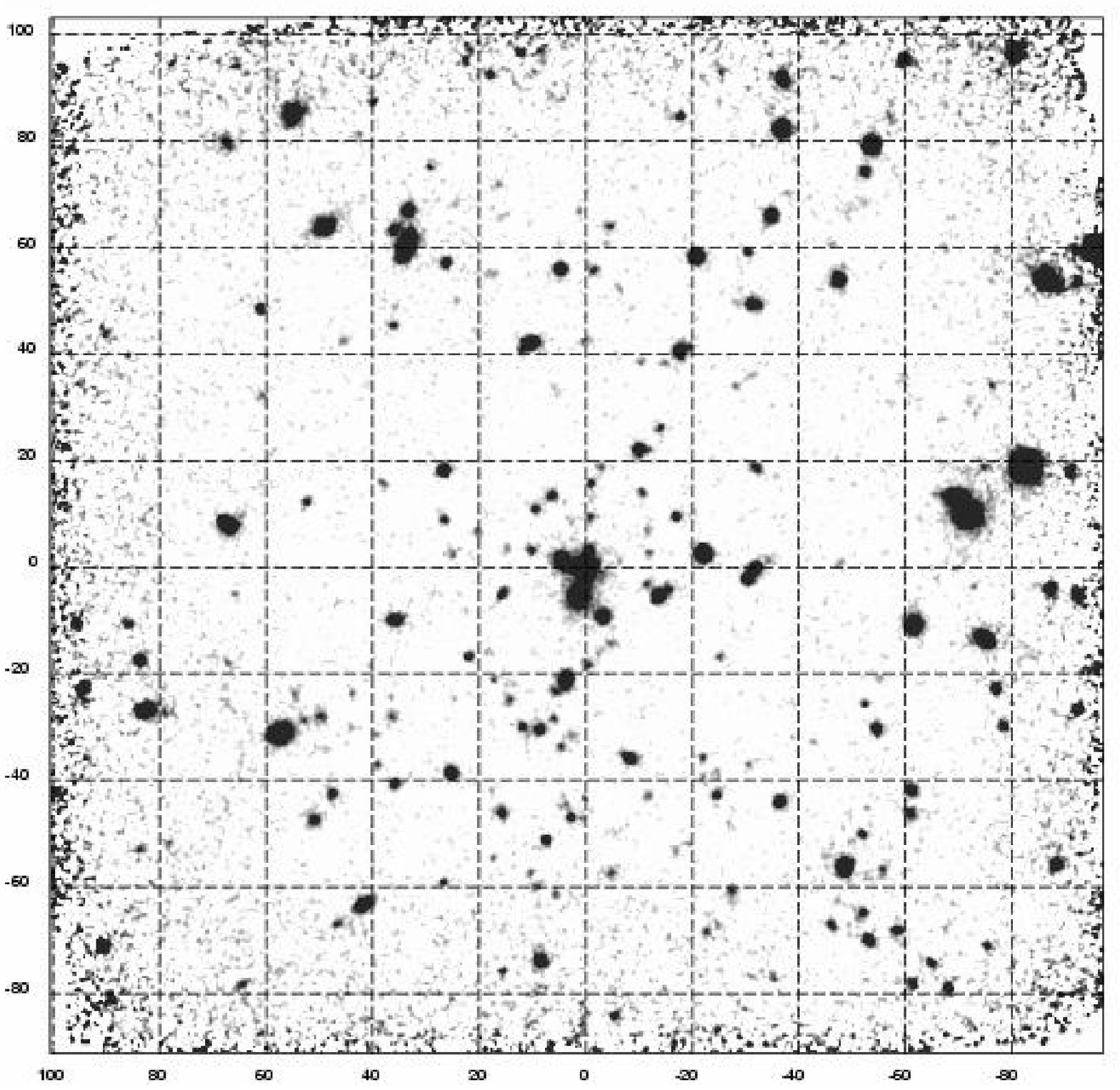}
\caption{$K$-band image of GHO~1322+3027.  The coordinates are in relative arcsec from the
central brightest galaxy.  North is up and East to the left. }
\label{g1322k}
\end{figure}
\clearpage
\begin{figure}[p]
\plotone{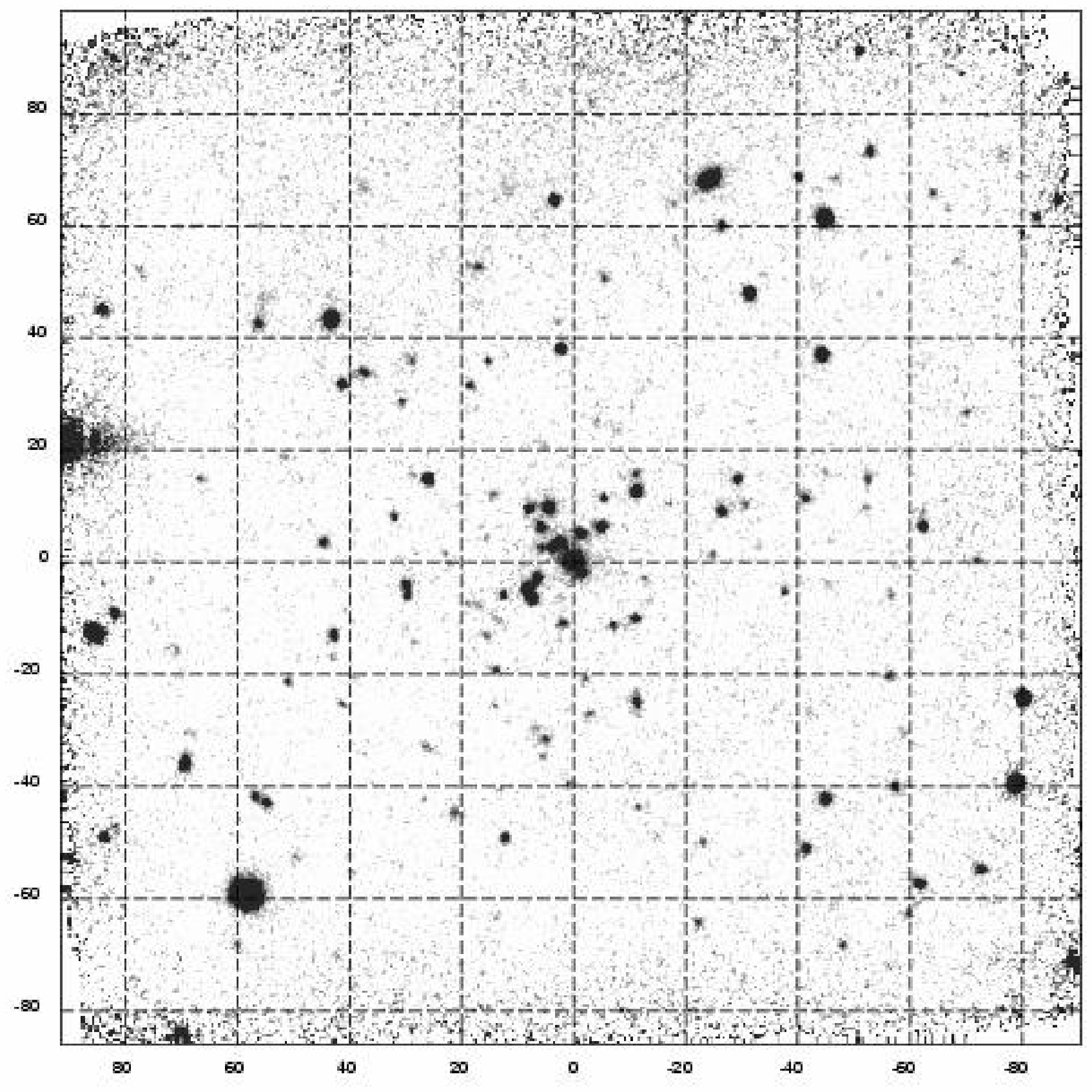}
\caption{$K$-band image of MS 1137.5+6625.  The coordinates are in relative arcsec from the
central brightest galaxy.  North is up and East to the left. }
\label{m1137k}
\end{figure}
\clearpage
\begin{figure}[p]
\plotone{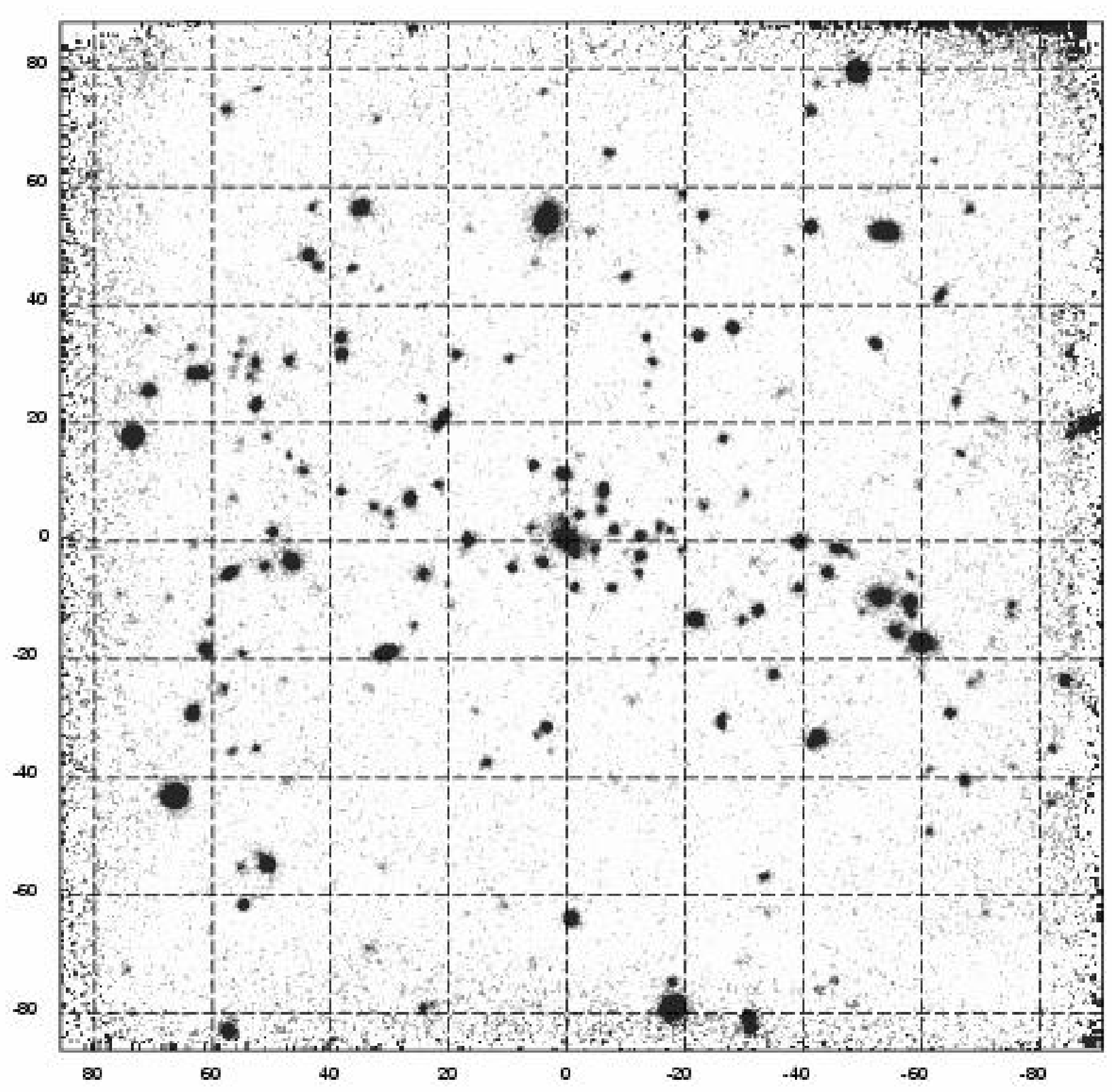}
\caption{$K$-band image of MS 1054.5-0320.  The coordinates are in relative arcsec from the
central brightest galaxy.  North is up and East to the left. }
\label{m1054k}
\end{figure}
\clearpage
\begin{figure}[p]
\plotone{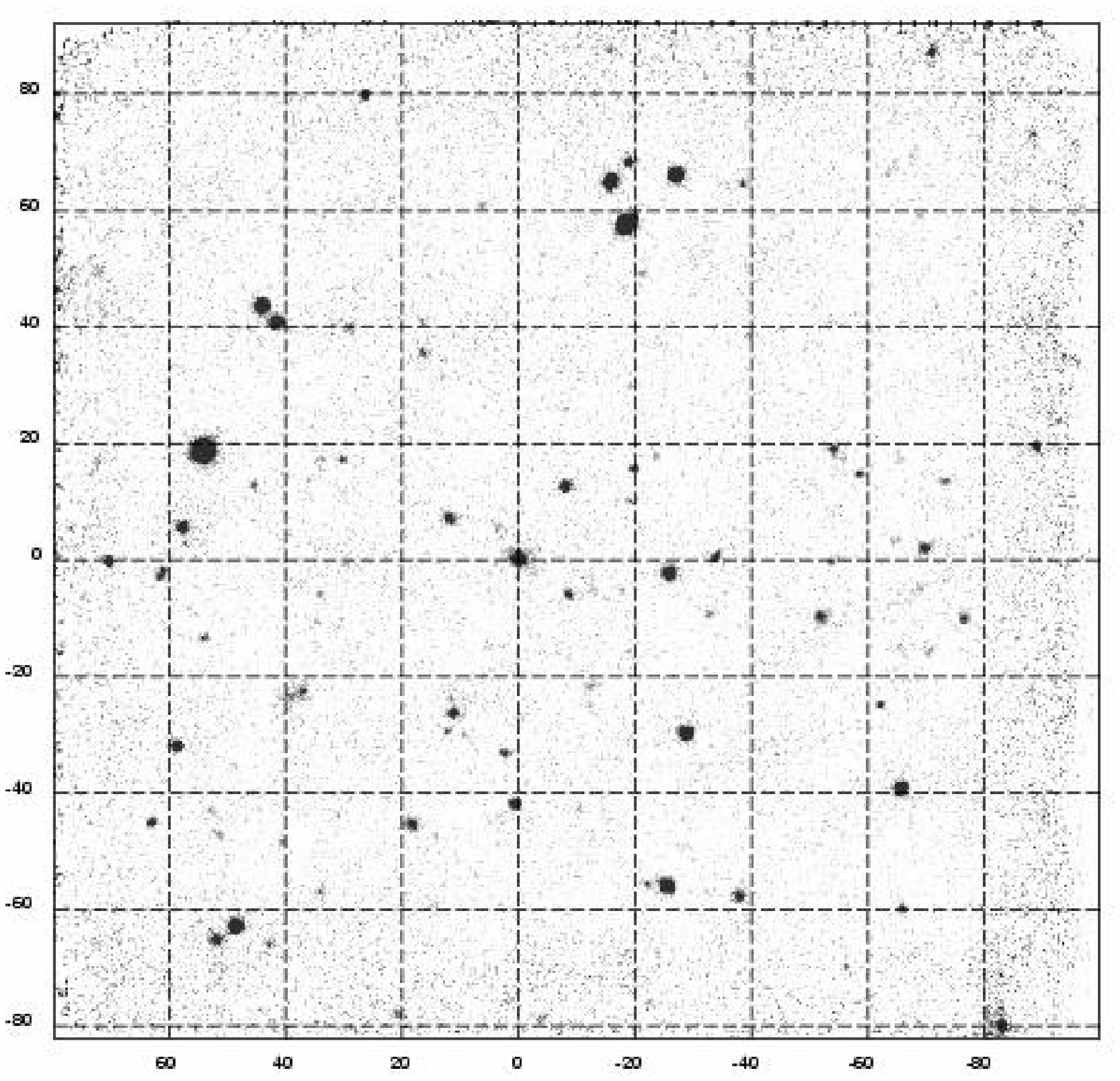}
\caption{$H$-band image of GHO~0021+0406.  The coordinates are in relative arcsec from the
central brightest galaxy.  North is up and East to the left. }
\label{g0021h}
\end{figure}
\clearpage
\begin{figure}[p]
\plotone{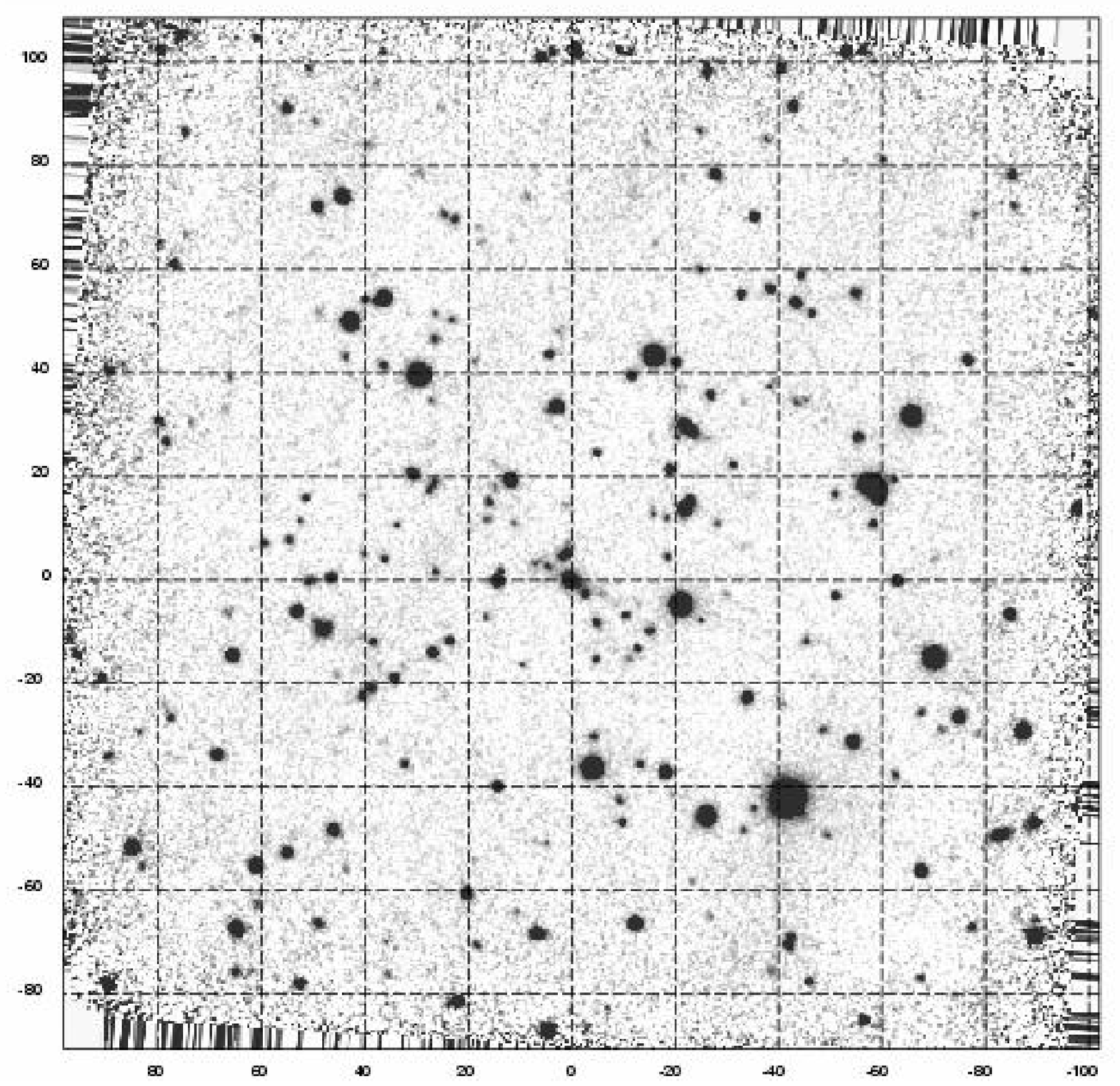}
\caption{$K$-band image of 3C~6.1.  The coordinates are in relative arcsec from the
central brightest galaxy.  North is up and East to the left. }
\label{3c6p1k}
\end{figure}
\clearpage
\begin{figure}[p]
\plotone{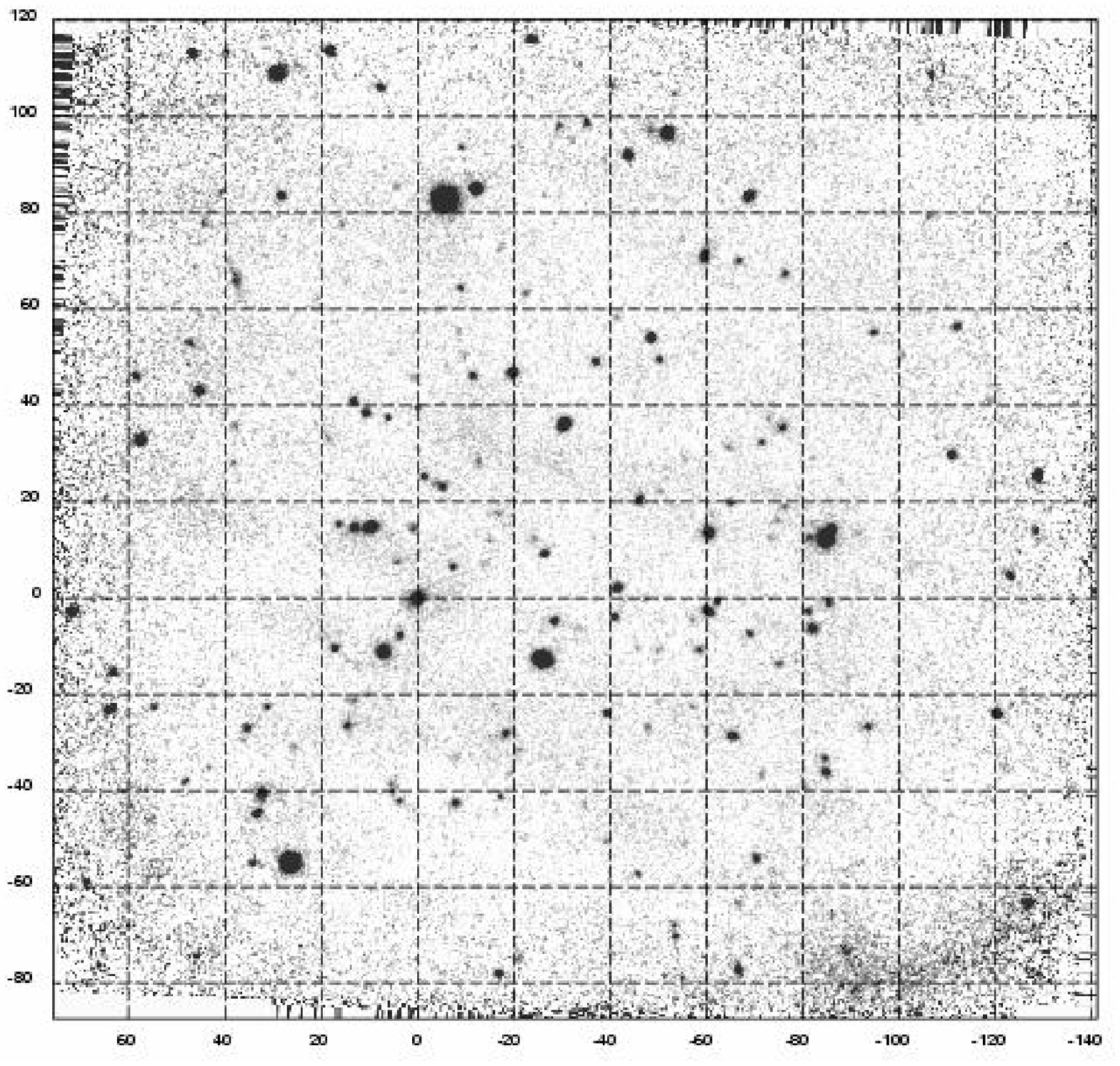}
\caption{$K$-band image of GHO~1603+4313.  The coordinates are in relative arcsec from the
central brightest galaxy.  North is up and East to the left. }
\label{g1603k}
\end{figure}
\clearpage
\begin{figure}[p]
\plotone{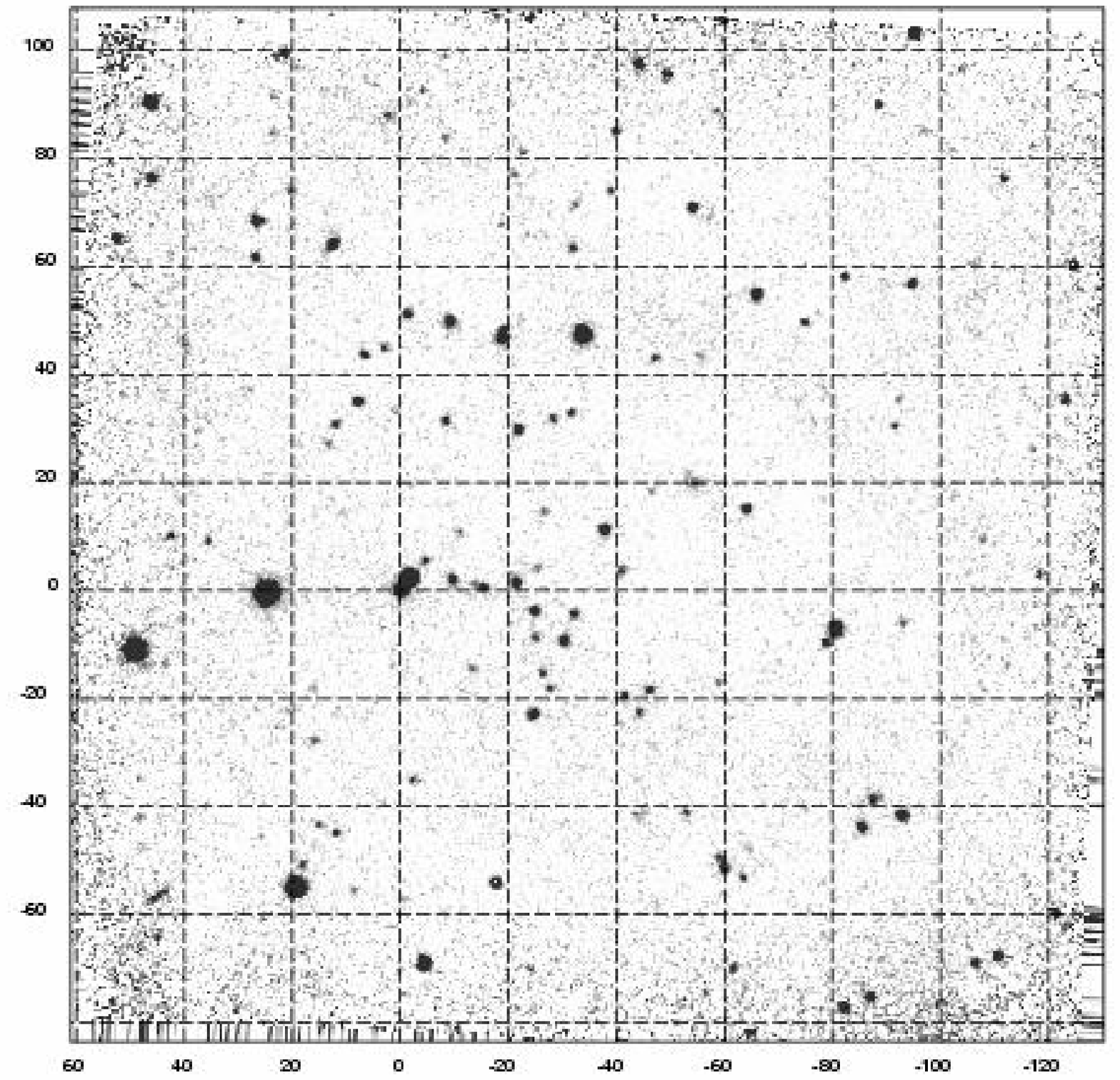}
\caption{$K$-band image of GHO~1604+4329.  The coordinates are in relative arcsec from the
central brightest galaxy.  North is up and East to the left. }
\label{g1604k}
\end{figure}
\clearpage
\begin{figure}[p]
\plotone{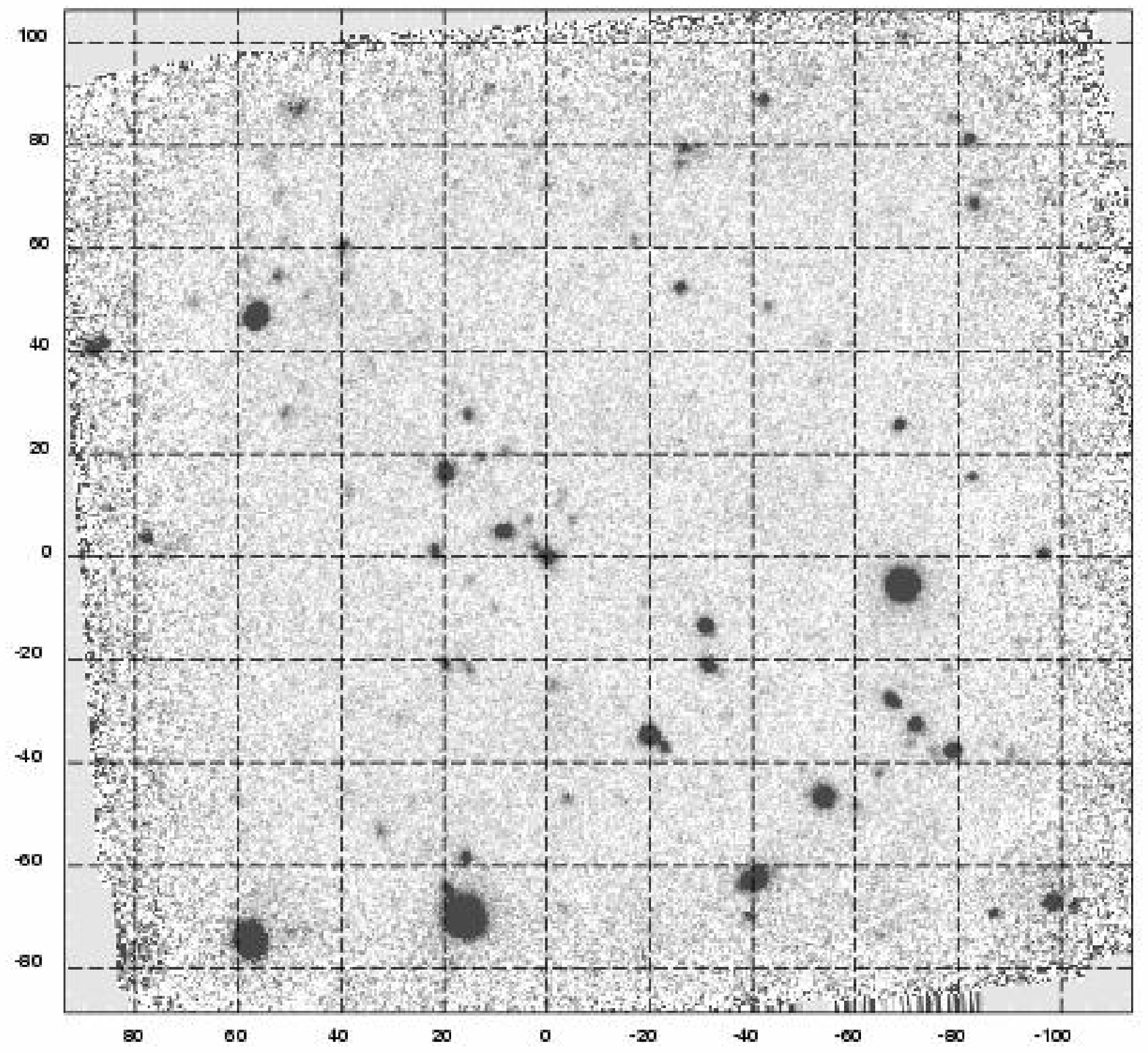}
\caption{$K$-band image of 3C~184.  The coordinates are in relative arcsec from the
central brightest galaxy.  North is up and East to the left. }
\label{3c184k}
\end{figure}
\clearpage
\begin{figure}[p]
\plotone{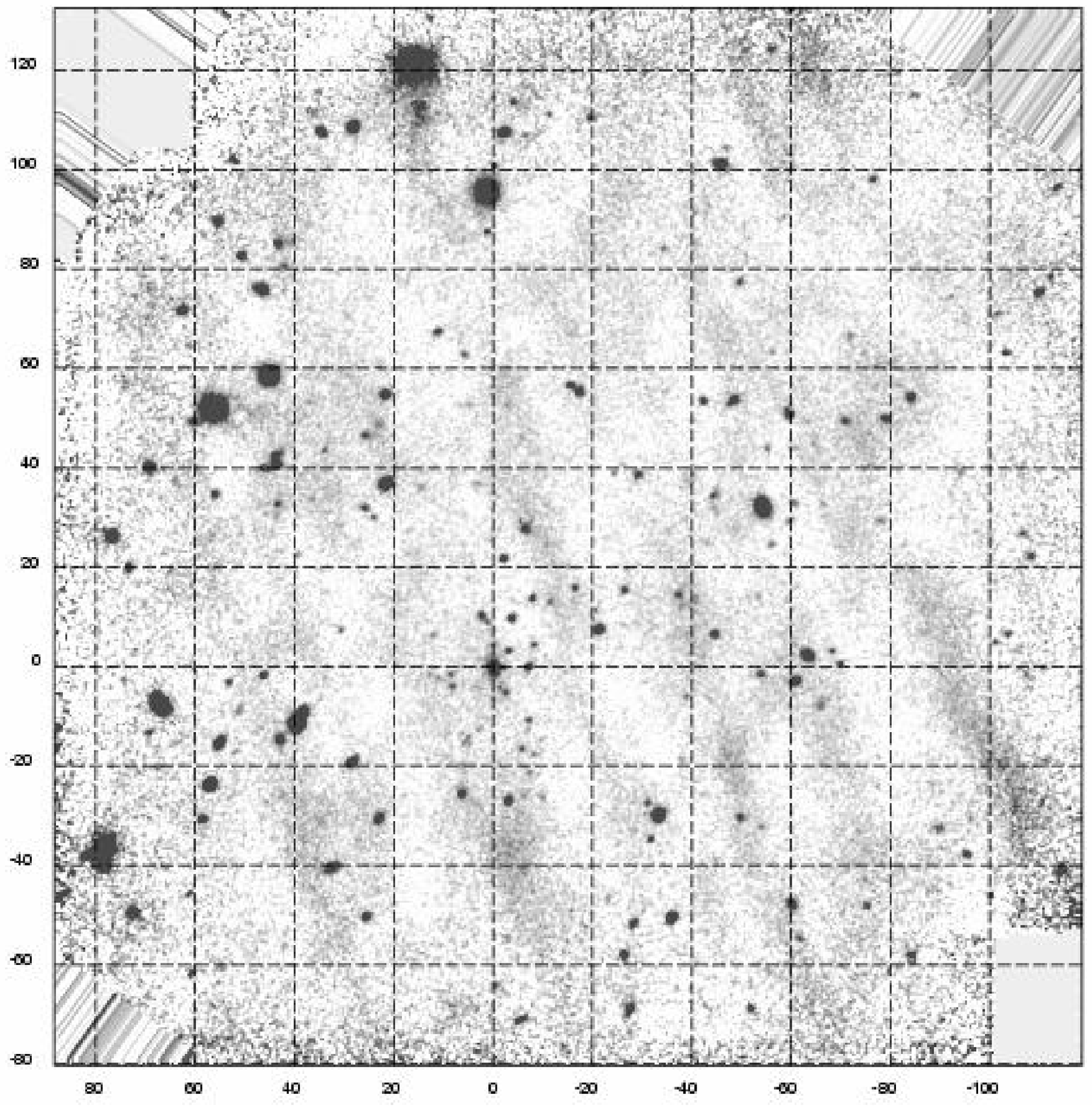}
\caption{$K$-band image of 3C~210.  The coordinates are in relative arcsec from the
central brightest galaxy.  North is up and East to the left. }
\label{3c210k}
\end{figure}
\clearpage
\begin{figure}[p]
\plotone{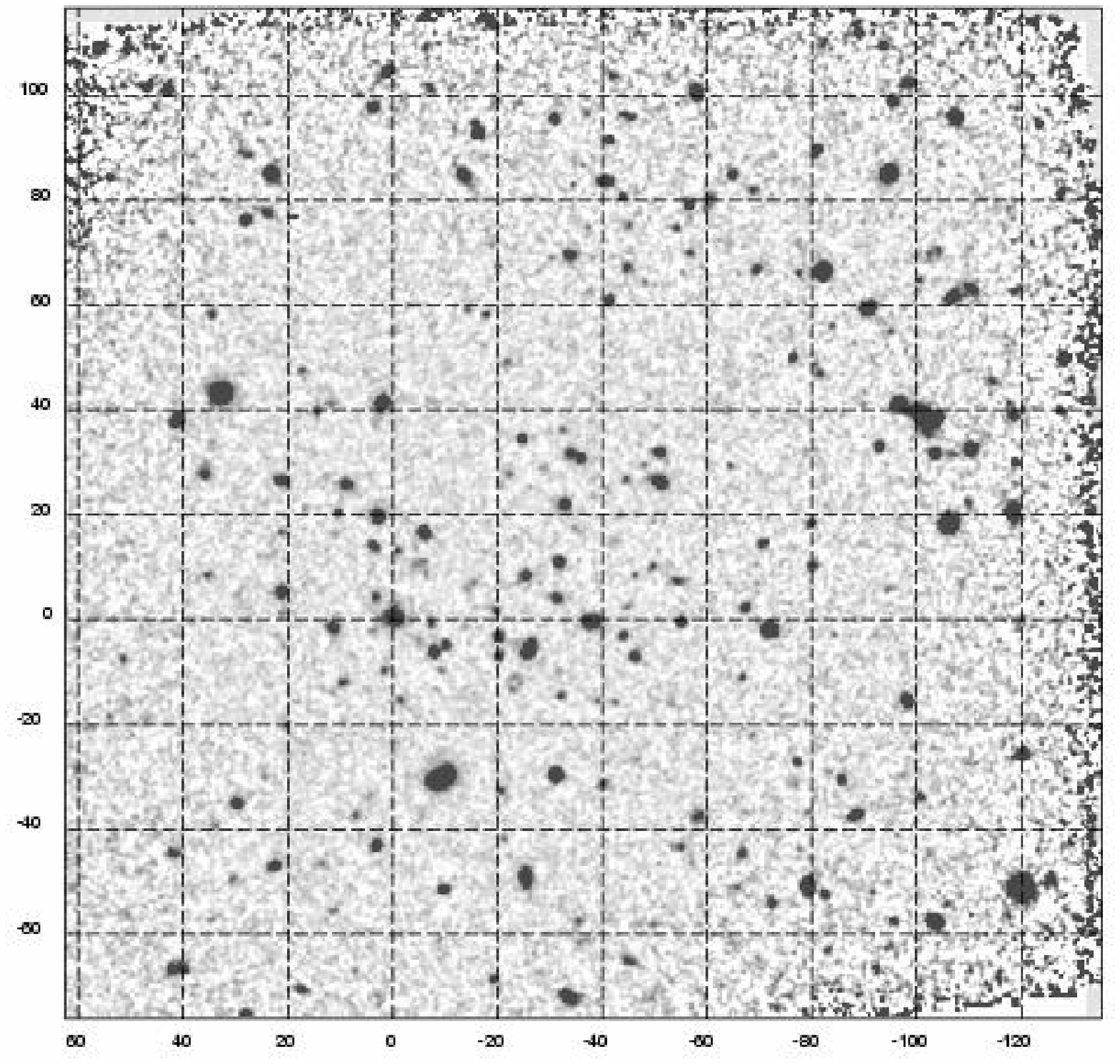}
\caption{$K$-band image of RDCS~0848+4453.  The coordinates are in relative arcsec from the
central brightest galaxy.  North is up and East to the left. }
\label{cl1k}
\end{figure}
\clearpage

\begin{deluxetable}{lcccccc}
\small
\tablecaption{Galaxy Cluster Sample}
\tablewidth{0pt}
\tablehead{
\colhead{Name} & \colhead{R.A.} & \colhead{Dec.} & \colhead{Field}&
\colhead{$z$} & \colhead{$K_{lim}$} & \colhead{L$_x$} 
\\
\colhead{} & \colhead{J2000} & \colhead{J2000} & \colhead{arcmin$^2$}
& \colhead{} & \colhead{mag} & \colhead{$10^{45}$ ergs $s^{-1}$}
}
\startdata
Abell 1146	& 11:01:14.48 & $-$22:43:53 & 59.3 & 0.142 & 16.5 & 3.89 \tablenotemark{b}	\\
Abell 3305	& 05:01:52.89 & $-$39:12:48 & 52.9 & 0.157 & 16.5 & 0.08 \tablenotemark{c} \\
MS 0906.5+1110	& 09:09:12.68 & $+$10:58:29 & 57.5 & 0.180 & 16.9 & 0.36 \tablenotemark{d} \\
Abell 1689	& 13:11:29.49 & $-$01:20:28 & 54.1 & 0.182 & 17.4 & 3.89 \tablenotemark{b} \\
Abell 1942	& 14:38:21.85 & $+$03:40:13 & 15.9 & 0.224 & 17.5 & 0.23 \tablenotemark{e} \\
MS 1253.9+0456	& 12:56:26.42 & $+$04:40:13 & 16.1 & 0.230 & 17.2 & 0.48 \tablenotemark{f} \\
Abell 1525	& 12:21:57.75 & $-$01:08:03 & 16.0 & 0.259 & 17.5 & \nodata \\
MS 1008.1-1224	& 10:10:32.27 & $-$12:39:52 & 17.2 & 0.301 & 17.9 & 0.70 \tablenotemark{b} \\
MS 1147.3+1103	& 11:49:52.68 & $+$10:46:37 & 14.9 & 0.303 & 17.9 & 0.32 \tablenotemark{b} \\
AC 118		& 00:14:20.68 & $-$30:24:00 & 24.4 & 0.308 & 18.5 & 4.78 \tablenotemark{g} \\
AC 103		& 20:57:01.00 & $-$64:39:48 & 23.8 & 0.311 & 18.2 & \nodata \\
AC 114		& 22:58:48.37 & $-$34:48:09 & 21.8 & 0.312 & 18.3 & 2.92 \tablenotemark{b} \\
MS 2137.3-234 	& 21:40:15.16 & $-$23:39:40 & 24.3 & 0.313 & 18.1 & 1.50 \tablenotemark{b} \\
Abell S0506	& 05:01:05.93 & $-$24:24:59 & 18.4 & 0.316 & 18.0 & 1.35 \tablenotemark{h} \\
MS 1358.1+6245	& 13:59:50.62 & $+$62:31:05 & 21.5 & 0.328 & 18.2 & 0.63 \tablenotemark{d}\\
Cl 2244-02 	& 22:47:13.17 & $-$02:05:39 & 21.5 & 0.330 & 18.8 & 0.23 \tablenotemark{b} \\
Abell 370\tablenotemark{a}&02:39:53.8&$-$01:34:24&22.4&0.374&18.1& 1.66 \tablenotemark{b} \\
Cl 0024+16 	& 00:26:35.72 & $+$17:09:45 & 21.4 & 0.391 & 18.8 & 0.34 \tablenotemark{b} \\ 
Abell 851\tablenotemark{a}&09:43:02.6&$+$46:58:37&21.5&0.405&18.7& 1.31 \tablenotemark{b} \\
GHO 0303+1706 	& 03:06:19.10 & $+$17:18:49 & 21.3 & 0.418 & 18.8 & 0.31 \tablenotemark{i} \\
3C~313		& 15:11:00.01 & $+$07:51:50 & 21.3 & 0.461 & 18.5 & \nodata \\
3C~295		& 14:11:20.50 & $+$52:12:10 & 21.9 & 0.461 & 18.8 & 2.18 \tablenotemark{b} \\	
F1557.19TC 	& 04:12:46.63 & $-$65:50:55 & 24.1 & 0.510 & 19.1 & 0.06 \tablenotemark{j} \\
Vidal 14	& 00:49:10.93 & $-$24:40:43 & 22.1 & 0.520 & 18.0 & \nodata \\ 
GHO 1601+4253 	& 16:03:12.21 & $+$42:45:25 & 21.6 & 0.539 & 19.2 & \nodata \\
MS 0451.6-0306 	& 04:54:10.87 & $-$03:00:57 &  6.6 & 0.539 & 19.2 & 4.68 \tablenotemark{b} \\
Cl 0016+16 	& 00:18:33.51 & $+$16:25:15 & 21.7 & 0.545 & 19.1 & 2.46 \tablenotemark{b} \\
J1888.16CL 	& 00:56:57.07 & $-$27:40:30 & 21.1 & 0.560 & 19.2 & 0.20 \tablenotemark{j} \\
MS 2053.7-0449	& 20:56:21.25 & $-$04:37:53 & 27.7 & 0.582 & 19.2 & 0.12 \tablenotemark{f} \\
GHO 0317+1521	& 03:20:01.40 & $+$15:32:00 &  7.4 & 0.583 & 19.2 & \nodata  \\
GHO 0229+0035   & 02:31:44.99 & $+$00:48:15 &  6.5 & 0.607 & 19.3 & \nodata  \\
3C 220.1	& 09:32:39.65 & $+$79:06:32 & 20.2 & 0.620 & 19.5 & \nodata \\
GHO 2201+0258	& 22:04:03.94 & $+$03:12:48 &  8.1 & 0.640 & 19.3 & \nodata \\
3C 34 		& 01:10:18.52 & $+$31:47:19 &  6.5 & 0.689 & 19.1 & \nodata \\ 
GHO 2155+0321   & 21:57:54.04 & $+$03:47:25 &  7.0 & 0.7\tablenotemark{o}& 19.4 & \nodata \\
GHO 1322+3027 	& 13:24:48.87 & $+$30:11:38 &  6.5 & 0.751 & 20.3 & 0.14 \tablenotemark{b} \\ 
MS 1137.5+6625	& 11:40:22.21 & $+$66:08:13 &  7.1 & 0.782 & 20.0 & 1.26 \tablenotemark{k} \\
MS 1054.5-032 	& 10:56:59.54 & $-$03:37:36 &  6.5 & 0.828 & 20.3 & 3.37 \tablenotemark{l} \\
GHO 0021+0406   & 00:23:54.47 & $+$04:23:13 &  6.5 & 0.832 & 20.0 & \nodata   \\
3C 6.1		& 00:16:31.23 & $+$79:16:49 &  6.3 & 0.840 & 19.8 & \nodata \\
GHO 1603+4313 	& 16:04:24.53 & $+$43:04:40 &  6.9 & 0.895 & 20.3 & 0.25 \tablenotemark{m} \\
GHO 1604+4329	& 16:04:35.98 & $+$43:21:06 &  7.2 & 0.920 & 20.1 & $<$0.7\tablenotemark{m} \\
3C~184   	& 07:39:24.31 & $+$70:23:15 &  6.2 & 0.996 & 20.3 & \nodata \\
3C~210   	& 08:58:09.90 & $+$27:50:52 &  6.2 & 1.169 & 20.5 & \nodata  \\
RDCS~0848+4453	& 08:48:35.93 & $+$44:53:37 &  6.2 & 1.273 & 20.5 & 0.07 \tablenotemark{n}\\	
\enddata
\tablenotetext{a}{Photometry previously reported in SED95}
\tablenotetext{b}{Wu, Xue, \& Fang(1999)}
\tablenotetext{c}{David, Forman, \& Jones (1999)}
\tablenotetext{d}{Lewis {et~al.}(1999)}
\tablenotetext{e}{Lea \& Henry (1988)}
\tablenotetext{f}{Gioia \& Luppino 1994}
\tablenotetext{g}{Wu, Xue, \& Fang(1999) - listed as Abell 2744}
\tablenotetext{h}{Wu, Xue, \& Fang(1999) - listed as CL 0500-24}
\tablenotetext{i}{Henry {et~al.}(1982)}
\tablenotetext{j}{Smail {et~al.}(1997)}
\tablenotetext{k}{Borgani {et~al.}(2001)}
\tablenotetext{l}{Jeltema {et~al.}(2001)}
\tablenotetext{m}{Castander {et~al.}(1994)}
\tablenotetext{n}{Stanford {et~al.}(2001)}
\tablenotetext{o}{The redshift and even the reality of this cluster is
in doubt; see Oke et al.\ 1998}
\label{sample}
\end{deluxetable}

\begin{deluxetable}{lllllllcllcll}
\tablenum{2}
\tableheadfrac{0.01}
\tabletypesize{\footnotesize}
\tablewidth{0pt}
\tablecaption{Observing Run Information}
\tablehead{
\colhead{Cluster} & \multicolumn{2}{c}{$K_s$} &  
\multicolumn{2}{c}{$H$} & \multicolumn{2}{c}{$J$} &
\multicolumn{3}{c}{Red} & \multicolumn{3}{c}{Blue}\\ 
\colhead{} & 
\colhead{tel.}&\colhead{date}&
\colhead{tel.}&\colhead{date}&
\colhead{tel.}&\colhead{date}&
\colhead{f}&\colhead{tel.}&\colhead{date}&
\colhead{f}&\colhead{tel.}&\colhead{date}
} 
\startdata
A1146  &C1.5&18Feb94&\nodata&\nodata&C1.5&18Feb94&$V$&C0.9&19Mar93&$B$&C0.9&19Mar93 \\
A3305  &C1.5&20Feb94&\nodata&\nodata&C1.5&20Feb94&$R$&C1.5&29Aug95&$B$&C1.5&29Aug95 \\
M0906  &C1.5&20Feb94&\nodata&\nodata&C1.5&20Feb94&$R$&P5  &02Feb95&$B$&P5  &02Feb95 \\
A1689  &C1.5&18Feb94&\nodata&\nodata&C1.5&17Feb94&$R$&K2.1&23May95&$B$&K2.1&23May95 \\
A1942  &C1.5&18Feb94&\nodata&\nodata&C1.5&18Feb94&$R$&K2.1&22May95&$B$&K2.1&22May95 \\
M1253  &C1.5&19Feb94&\nodata&\nodata&C1.5&19Feb94&$R$&K2.1&21May95&$B$&K2.1&21May95 \\
A1525  &C1.5&19Feb94&\nodata&\nodata&C1.5&19Feb94&$R$&K2.1&22May95&$B$&K2.1&22May95 \\
M1008  &C1.5&17Feb94&\nodata&\nodata&C1.5&17Feb94&$R$&K2.1&22May95&$g$&P5  &02Feb95 \\
M1147  &C1.5&20Feb94&\nodata&\nodata&C1.5&20Feb94&$R$&K2.1&21May95&$g$&P5  &02Feb95 \\
AC118  &C1.5&02Sep95&C1.5&02Sep95  &C1.5&01Sep95&$R$&C1.5&27Aug95&$g$&C1.5&27Aug95 \\
AC103  &C1.5&21Sep94&C1.5&22Sep94  &C1.5&22Sep94&$R$&C1.5&30Aug95&$g$&C1.5&30Aug95 \\
AC114  &C1.5&02Sep95&C1.5&02Sep95  &C1.5&02Sep95&$R$&C1.5&29Aug95&$g$&C1.5&30Aug95 \\
M2137  &C1.5&02Sep95&C1.5&01Sep95  &C1.5&01Sep95&$R$&C1.5&29Aug95&$g$&C1.5&29Aug95 \\
S0506  &C1.5&18Feb94&\nodata&\nodata&C1.5&17Feb94&$R$&C1.5&29Aug95&$g$&C1.5&29Aug95 \\
M1358  &K2.1&03Apr93&\nodata&\nodata&K2.1&03Apr93&$R$&K2.1&21May95&$g$&K2.1&25May95 \\
C2244  &K2.1&02Oct93&K2.1&02Oct93  &K2.1&02Oct93&$R$&K2.1&29Oct94&$g$&K2.1&29Oct94 \\
C0024  &K2.1&03Oct93&K2.1&03Oct93  &K2.1&03Oct93&$R$&K2.1&29Oct94&$g$&K2.1&31Oct94 \\
G0303  &K2.1&01Oct93&K2.1&01Oct93  &K2.1&01Oct93&$R$&K2.1&31Oct94&$g$&K2.1&01Nov94 \\
3C313  &K2.1&05Apr93&\nodata&\nodata&K2.1&05Apr93&$i$&K2.1&22May95&$V$&K2.1&22May95 \\
3C295  &K2.1&04Apr93&K2.1&04Apr93  &K2.1&04Apr93&$i$&K2.1&25May95&$V$&K2.1&23May95 \\
F1557  &C1.5&03Sep95&C1.5&04Sep95  &C1.5&03Sep95&$I$&C1.5&30Aug95&$V$&C1.5&30Aug95 \\
V14    &C1.5&22Sep94&\nodata&\nodata&C1.5&22Sep94&$I$&C1.5&27Aug95&$V$&C1.5&27Aug95\\
G1601  &K2.1&03Apr93&K2.1&04Apr93  &K2.1&04Apr93&$i$&K2.1&22May95&$V$&K2.1&20May95 \\
M0451  &K4  &14Dec94&C1.5&04Sep95  &C1.5&04Sep95&$i$&P5  &03Feb95&$V$&P5  &03Feb95 \\
C0016  &K2.1&01Oct93&K2.1&02Oct93  &K2.1&02Oct93&$I$&K2.1&31Oct94&$V$&K2.1&01Nov94 \\
J1888  &C1.5&03Sep95&C1.5&03Sep95  &C1.5&02Sep95&$I$&C1.5&30Aug95&$V$&C1.5&30Aug95 \\
M2053  &C1.5&02Sep95&\nodata&\nodata&C1.5&03Sep95&$I$&K2.1&29Oct94&$V$&K2.1&31Oct94 \\
G0317  &K4  &14Dec94&\nodata&\nodata&K4  &22Sep96&$I$&K2.1&07Oct96&$V$&K2.1&07Oct96 \\
G0229  &\nodata&\nodata&K4&23Sep96  &K4  &23Sep96&$I$&K2.1&07Oct96&$V$&K2.1&07Oct96 \\
3C220.1&K2.1&04Apr93&K2.1&04Apr93  &K2.1&04Apr93&$I\tablenotemark{a}$&HST&20Apr96&$V\tablenotemark{b}$&HST&20Apr96 \\
G2201  &K4  &16Dec94&\nodata&\nodata&K4  &23Sep96&$I$&K2.1&07Oct96&$V$&K2.1&08Oct96 \\
3C34   &K4  &15Dec94&K4  &22Sep96  &K4  &22Sep96&$i$&P5  &02Feb95&$V$&P5  &02Feb95 \\
G2155  &\nodata&\nodata&K4&23Sep96  &K4  &23Sep96&$I$&K2.1&07Oct96&$V$&K2.1&10Oct96 \\
G1322  &K4  &18Apr95&K4  &08May96  &K4  &18Apr95&$i$&P5  &02Feb95&$R$&K2.1&20May95 \\
M1137  &K4  &07May96&K4  &08May96  &K4  &09May96&$i$&K2.1&21May95&$R$&W3.5 &27May97 \\
M1054  &K4  &07May96&K4  &08May96  &K4  &09May96&$i$&P5  &02Feb95&$R$&K4  &18May96 \\
G0021  &\nodata&\nodata&K4&23Sep96  &K4  &23Sep96&$I$&K2.1&07Oct96&$R$&K2.1&07Oct96 \\
G1603  &K4  &14Apr95&K4  &03May96  &K4  &07May96&$i$&K2.1&20May95&$R$&K2.1&20May95 \\
G1604  &K4  &18Apr95&K4  &08May96  &K4  &07May96&$i$&K2.1&23May95&$R$&W3.5 &14Apr97 \\
3C~184 &K4  &12Dec94&\nodata  &\nodata  &K4  &16Dec94&$I\tablenotemark{a}$&HST&11Nov95&\nodata&\nodata &\nodata \\
3C~210 &K4  &13Dec94&\nodata  &\nodata  &K4  &13Dec94&$I\tablenotemark{a}$&HST&18May96&\nodata&\nodata &\nodata \\
R0848\tablenotemark{c} &K4  &21Feb97&K4  &24Feb97  &K4  &20Feb97&$I$&K4&18Mar97&$R$&K4 &13May94 \\

\enddata
\label{obslog}
\tablenotetext{a}{WFPC2/F814W} \tablenotetext{b}{WFPC2/F555W}
\tablenotetext{c}{$B$-band images also obtained at K4m on 02Feb95}

\end{deluxetable}

\begin{deluxetable}{rrrccccc} 
\tabletypesize{\footnotesize} 
\tablewidth{0pt}
\tableheadfrac{0.01}
\tablenum{3} 
\tablecaption{Abell 1146} 
\tablehead{ 
\colhead{ID}&\colhead{$\delta$x}&\colhead{$\delta$y}& 
\colhead{$Kt$}&
\colhead{$Ka$}&
\colhead{$J-K$}&
\colhead{$V-K$}&
\colhead{$B-K$}\\
\colhead{}&\colhead{arcsec}&\colhead{arcsec} & 
\colhead{}&\colhead{$r=3\arcsec$}&\colhead{$r=3\arcsec$}&
\colhead{$r=3\arcsec$}&\colhead{$r=3\arcsec$}
} 
\startdata 
  1 & 227.4 &   14.6 & 10.44$_{-0.01}^{+ 0.01}$ &10.78$_{-0.01}^{+ 0.01}$ &  0.20$_{-0.01}^{+ 0.01}$ & 2.21$_{-0.01}^{+ 0.01}$ & 2.65$_{-0.01}^{+ 0.01}$  \\ 
  2 &  99.9 &  -39.7 & 11.41$_{-0.01}^{+ 0.01}$ &11.69$_{-0.01}^{+ 0.01}$ &  0.56$_{-0.01}^{+ 0.01}$ & 2.48$_{-0.01}^{+ 0.01}$ & 3.74$_{-0.01}^{+ 0.01}$  \\ 
  3 &   0.0 &    0.0 & 12.10$_{-0.01}^{+ 0.01}$ &14.19$_{-0.01}^{+ 0.01}$ &  1.14$_{-0.01}^{+ 0.01}$ & 3.84$_{-0.01}^{+ 0.01}$ & 5.50$_{-0.01}^{+ 0.01}$  \\ 
  4 & 190.9 &  111.1 & 12.26$_{-0.01}^{+ 0.01}$ &12.42$_{-0.01}^{+ 0.01}$ &  0.56$_{-0.01}^{+ 0.01}$ & 2.33$_{-0.01}^{+ 0.01}$ & 3.48$_{-0.01}^{+ 0.01}$  \\ 
  5 & 173.5 &  -76.6 & 12.49$_{-0.01}^{+ 0.01}$ &12.69$_{-0.01}^{+ 0.01}$ &  0.61$_{-0.01}^{+ 0.01}$ & 2.76$_{-0.01}^{+ 0.01}$ & 3.97$_{-0.01}^{+ 0.01}$  \\ 
  6 &  83.2 & -115.9 & 12.79$_{-0.01}^{+ 0.01}$ &13.03$_{-0.01}^{+ 0.01}$ &  0.43$_{-0.01}^{+ 0.01}$ & 2.15$_{-0.01}^{+ 0.01}$ & 3.18$_{-0.01}^{+ 0.01}$  \\ 
  7 &-192.0 &  -77.1 & 12.96$_{-0.01}^{+ 0.01}$ &13.16$_{-0.01}^{+ 0.01}$ &  0.31$_{-0.01}^{+ 0.01}$ & 1.46$_{-0.01}^{+ 0.01}$ & 2.28$_{-0.01}^{+ 0.01}$  \\ 
  8 &-223.0 &  -77.9 & 13.03$_{-0.02}^{+ 0.02}$ &14.08$_{-0.01}^{+ 0.01}$ &  1.18$_{-0.02}^{+ 0.02}$ & 3.83$_{-0.02}^{+ 0.02}$ & 5.42$_{-0.02}^{+ 0.02}$  \\
\enddata 
\label{a1146} 
\end{deluxetable} 

\begin{deluxetable}{rrrccccc} 
\tabletypesize{\footnotesize} 
\tablewidth{0pt} 
\tablenum{4} 
\tablecaption{Abell 3305} 
\tablehead{ 
\colhead{ID}&\colhead{$\delta$x}&\colhead{$\delta$y}& 
\colhead{$Kt$}&
\colhead{$Ka$}&
\colhead{$J-K$}&
\colhead{$R-K$}&
\colhead{$B-K$}\\
\colhead{}&\colhead{arcsec}&\colhead{arcsec} & 
\colhead{}&\colhead{$r=1.42\arcsec$}&\colhead{$r=1.42\arcsec$}&
\colhead{$r=1.42\arcsec$}&\colhead{$r=1.42\arcsec$}
} 
\startdata 
  1 & 154.3 &  125.6  &  9.30$_{-0.01}^{+ 0.01}$ &10.06$_{-0.01}^{+ 0.01}$ & 99.00\phm{+0.0} & 2.66$_{-0.01}^{+ 0.01}$ & 2.55$_{-0.01}^{+ 0.01}$  \\ 
  2 & -23.8 &  197.7  & 11.81$_{-0.01}^{+ 0.01}$ &12.26$_{-0.01}^{+ 0.01}$ &  0.31$_{-0.01}^{+ 0.01}$ & 1.19$_{-0.01}^{+ 0.01}$ & 2.02$_{-0.01}^{+ 0.01}$  \\ 
  3 & 163.0 &  230.8  & 11.99$_{-0.01}^{+ 0.01}$ &12.38$_{-0.01}^{+ 0.01}$ &  0.38$_{-0.01}^{+ 0.01}$ & 1.27$_{-0.01}^{+ 0.01}$ & 2.55$_{-0.01}^{+ 0.01}$  \\ 
  4 &  33.2 & -122.3  & 12.23$_{-0.01}^{+ 0.01}$ &12.67$_{-0.01}^{+ 0.01}$ &  0.25$_{-0.01}^{+ 0.01}$ & 0.80$_{-0.01}^{+ 0.01}$ & 1.40$_{-0.01}^{+ 0.01}$  \\ 
  5 &-106.4 & -190.9  & 12.58$_{-0.01}^{+ 0.01}$ &13.01$_{-0.01}^{+ 0.01}$ &  0.62$_{-0.01}^{+ 0.01}$ & 1.85$_{-0.01}^{+ 0.01}$ & 3.80$_{-0.01}^{+ 0.01}$  \\ 
  6 &-138.5 &  155.6  & 13.17$_{-0.01}^{+ 0.01}$ &14.55$_{-0.01}^{+ 0.01}$ &  1.12$_{-0.02}^{+ 0.02}$ & 2.69$_{-0.02}^{+ 0.02}$ & 4.11$_{-0.02}^{+ 0.02}$  \\ 
  7 &   0.0 &    0.0  & 13.24$_{-0.01}^{+ 0.01}$ &14.48$_{-0.01}^{+ 0.01}$ &  1.09$_{-0.01}^{+ 0.01}$ & 2.90$_{-0.01}^{+ 0.01}$ & 5.27$_{-0.01}^{+ 0.01}$  \\ 
  8 & 176.5 &  252.1  & 13.29$_{-0.01}^{+ 0.01}$ &13.70$_{-0.01}^{+ 0.01}$ &  0.75$_{-0.01}^{+ 0.01}$ &-99.00\phm{+0.0} &-99.00\phm{+0.0}  \\
\enddata 
\label{a3305} 
\end{deluxetable} 

\begin{deluxetable}{rrrrrrrr} 
\tabletypesize{\footnotesize} 
\tablewidth{0pt} 
\tablenum{5} 
\tablecaption{MS0906+1110} 
\tablehead{ 
\colhead{ID}&\colhead{$\delta$x}&\colhead{$\delta$y}& 
\colhead{$Kt$}&
\colhead{$Ka$}&
\colhead{$J-K$}&
\colhead{$R-K$}&
\colhead{$B-K$}\\
\colhead{}&\colhead{arcsec}&\colhead{arcsec} & 
\colhead{}&\colhead{$r=3\arcsec$}&\colhead{$r=3\arcsec$}&
\colhead{$r=3\arcsec$}&\colhead{$r=3\arcsec$}
} 
\startdata 
  1 & 219.4 & -134.2 & 10.11$_{-0.01}^{+ 0.01}$ &10.42$_{-0.01}^{+ 0.01}$ &  0.31$_{-0.01}^{+ 0.01}$ & 7.58$_{-0.03}^{+ 0.03}$ & 6.11$_{-0.01}^{+ 0.01}$  \\ 
  2 &  90.6 & -160.0 & 12.01$_{-0.01}^{+ 0.01}$ &12.61$_{-0.01}^{+ 0.01}$ &  0.79$_{-0.01}^{+ 0.01}$ & 5.15$_{-0.01}^{+ 0.01}$ & 4.15$_{-0.01}^{+ 0.01}$  \\ 
  3 &-235.0 &  203.7 & 12.18$_{-0.01}^{+ 0.01}$ &12.58$_{-0.01}^{+ 0.01}$ &  0.19$_{-0.01}^{+ 0.01}$ & 5.24$_{-0.02}^{+ 0.02}$ & 3.94$_{-0.01}^{+ 0.01}$  \\ 
  4 & -51.6 &  190.6 & 12.74$_{-0.01}^{+ 0.01}$ &13.01$_{-0.01}^{+ 0.01}$ &  0.82$_{-0.01}^{+ 0.01}$ & 5.07$_{-0.01}^{+ 0.01}$ & 7.31$_{-0.02}^{+ 0.02}$  \\ 
  5 &   0.0 &    0.0 & 12.81$_{-0.01}^{+ 0.01}$ &14.67$_{-0.01}^{+ 0.01}$ &  1.25$_{-0.01}^{+ 0.01}$ & 3.27$_{-0.01}^{+ 0.01}$ & 5.69$_{-0.01}^{+ 0.01}$  \\ 
  6 & 150.2 &  172.4 & 12.89$_{-0.01}^{+ 0.01}$ &13.15$_{-0.01}^{+ 0.01}$ &  0.21$_{-0.01}^{+ 0.01}$ & 4.66$_{-0.02}^{+ 0.02}$ & 3.35$_{-0.01}^{+ 0.01}$  \\ 
  7 &-237.8 & -178.1 & 13.08$_{-0.01}^{+ 0.01}$ &13.44$_{-0.01}^{+ 0.01}$ &  0.55$_{-0.01}^{+ 0.01}$ & 4.28$_{-0.01}^{+ 0.01}$ & 3.32$_{-0.01}^{+ 0.01}$  \\ 
  8 & -85.4 &  -38.4 & 13.22$_{-0.01}^{+ 0.01}$ &14.25$_{-0.01}^{+ 0.01}$ &  1.38$_{-0.01}^{+ 0.01}$ & 3.57$_{-0.01}^{+ 0.01}$ & 5.67$_{-0.01}^{+ 0.01}$  \\
\enddata 
\label{m0906} 
\end{deluxetable} 

\begin{deluxetable}{rrrrrrrr} 
\tabletypesize{\footnotesize} 
\tablewidth{0pt} 
\tablenum{6} 
\tablecaption{Abell 1689} 
\tablehead{ 
\colhead{ID}&\colhead{$\delta$x}&\colhead{$\delta$y}& 
\colhead{$Kt$}&
\colhead{$Ka$}&
\colhead{$J-K$}&
\colhead{$R-K$}&
\colhead{$B-K$}\\
\colhead{}&\colhead{arcsec}&\colhead{arcsec} & 
\colhead{}&\colhead{$r=1.4\arcsec$}&\colhead{$r=1.4\arcsec$}&
\colhead{$r=1.4\arcsec$}&\colhead{$r=1.4\arcsec$}
} 
\startdata 
  1 & 214.3 & -166.2 & 10.88$_{-0.01}^{+ 0.01}$ &11.24$_{-0.01}^{+ 0.01}$ &  0.47$_{-0.01}^{+ 0.01}$ &-99.00\phm{+0.0} &-99.00\phm{+0.0}  \\ 
  2 & 185.8 & -108.6 & 11.04$_{-0.01}^{+ 0.01}$ &11.38$_{-0.01}^{+ 0.01}$ &  0.64$_{-0.01}^{+ 0.01}$ &-99.00\phm{+0.0} &-99.00\phm{+0.0}  \\ 
  3 & -17.7 & -122.7 & 11.37$_{-0.01}^{+ 0.01}$ &11.75$_{-0.01}^{+ 0.01}$ &  0.55$_{-0.01}^{+ 0.01}$ & 2.00$_{-0.01}^{+ 0.01}$ & 2.85$_{-0.01}^{+ 0.01}$  \\ 
  4 & -99.2 &  171.3 & 12.38$_{-0.01}^{+ 0.01}$ &12.73$_{-0.01}^{+ 0.01}$ &  0.83$_{-0.01}^{+ 0.01}$ &-99.00\phm{+0.0} &-99.00\phm{+0.0}  \\ 
  5 &-132.5 & -149.8 & 12.60$_{-0.01}^{+ 0.01}$ &13.00$_{-0.01}^{+ 0.01}$ &  0.64$_{-0.01}^{+ 0.01}$ & 3.69$_{-0.01}^{+ 0.01}$ & 6.41$_{-0.02}^{+ 0.02}$  \\ 
  6 &-135.9 &  -39.7 & 12.81$_{-0.01}^{+ 0.01}$ &13.20$_{-0.01}^{+ 0.01}$ &  0.46$_{-0.01}^{+ 0.01}$ &-99.00\phm{+0.0} & 3.44$_{-0.01}^{+ 0.01}$  \\ 
  7 & 135.0 &  -17.5 & 13.19$_{-0.01}^{+ 0.01}$ &13.63$_{-0.01}^{+ 0.01}$ &  0.63$_{-0.01}^{+ 0.01}$ & 2.00$_{-0.01}^{+ 0.01}$ & 3.84$_{-0.01}^{+ 0.01}$  \\ 
  8 &   0.0 &    0.0 & 13.26$_{-0.01}^{+ 0.01}$ &15.06$_{-0.01}^{+ 0.01}$ &  1.19$_{-0.01}^{+ 0.01}$ & 3.03$_{-0.01}^{+ 0.01}$ & 5.35$_{-0.01}^{+ 0.01}$  \\
\enddata 
\label{a1689} 
\end{deluxetable} 

\begin{deluxetable}{rrrrrrrr} 
\tabletypesize{\footnotesize} 
\tablewidth{0pt} 
\tablenum{7} 
\tablecaption{Abell 1942} 
\tablehead{ 
\colhead{ID}&\colhead{$\delta$x}&\colhead{$\delta$y}& 
\colhead{$Kt$}&
\colhead{$Ka$}&
\colhead{$J-K$}&
\colhead{$R-K$}&
\colhead{$B-K$}\\
\colhead{}&\colhead{arcsec}&\colhead{arcsec} & 
\colhead{}&\colhead{$r=1.54\arcsec$}&\colhead{$r=1.54\arcsec$}&
\colhead{$r=1.54\arcsec$}&\colhead{$r=1.54\arcsec$}
} 
\startdata 
  1 &   75.0 &   73.8 & 10.83$_{-0.01}^{+ 0.01}$ &11.13$_{-0.01}^{+ 0.01}$ &  0.67$_{-0.01}^{+ 0.01}$ & 2.92$_{-0.01}^{+ 0.01}$ & 3.95$_{-0.01}^{+ 0.01}$  \\ 
  2 &  136.0 &  -43.4 & 11.04$_{-0.01}^{+ 0.01}$ &11.34$_{-0.01}^{+ 0.01}$ &  0.41$_{-0.01}^{+ 0.01}$ & 2.61$_{-0.01}^{+ 0.01}$ & 3.35$_{-0.01}^{+ 0.01}$  \\ 
  3 &  -31.9 & -120.9 & 12.11$_{-0.01}^{+ 0.01}$ &12.51$_{-0.01}^{+ 0.01}$ &  0.37$_{-0.01}^{+ 0.01}$ & 1.65$_{-0.01}^{+ 0.01}$ & 2.36$_{-0.01}^{+ 0.01}$  \\ 
  4 &    0.0 &    0.0 & 13.15$_{-0.01}^{+ 0.01}$ &14.77$_{-0.01}^{+ 0.01}$ &  1.21$_{-0.01}^{+ 0.01}$ & 3.04$_{-0.01}^{+ 0.01}$ & 5.65$_{-0.01}^{+ 0.01}$  \\ 
  5 &  -43.7 & -111.4 & 13.16$_{-0.01}^{+ 0.01}$ &14.91$_{-0.01}^{+ 0.01}$ &  0.96$_{-0.01}^{+ 0.01}$ & 2.07$_{-0.01}^{+ 0.01}$ & 3.29$_{-0.01}^{+ 0.01}$  \\ 
  6 &  135.1 &   15.6 & 13.60$_{-0.01}^{+ 0.01}$ &13.84$_{-0.01}^{+ 0.01}$ &  0.40$_{-0.01}^{+ 0.01}$ & 1.51$_{-0.01}^{+ 0.01}$ & 3.33$_{-0.01}^{+ 0.01}$  \\ 
  7 &   67.1 &  -74.5 & 13.87$_{-0.01}^{+ 0.01}$ &15.04$_{-0.01}^{+ 0.01}$ &  1.33$_{-0.01}^{+ 0.01}$ & 3.09$_{-0.01}^{+ 0.01}$ & 5.75$_{-0.01}^{+ 0.01}$  \\ 
  8 &   14.5 &   17.0 & 14.33$_{-0.01}^{+ 0.01}$ &14.54$_{-0.01}^{+ 0.01}$ &  0.25$_{-0.01}^{+ 0.01}$ & 0.93$_{-0.01}^{+ 0.01}$ & 2.24$_{-0.01}^{+ 0.01}$  \\
\enddata 
\label{a1942} 
\end{deluxetable} 

\begin{deluxetable}{rrrrrrrr} 
\tabletypesize{\footnotesize} 
\tablewidth{0pt} 
\tablenum{8} 
\tablecaption{MS1253+0456} 
\tablehead{ 
\colhead{ID}&\colhead{$\delta$x}&\colhead{$\delta$y}& 
\colhead{$Kt$}&
\colhead{$Ka$}&
\colhead{$J-K$}&
\colhead{$R-K$}&
\colhead{$B-K$}\\
\colhead{}&\colhead{arcsec}&\colhead{arcsec} & 
\colhead{}&\colhead{$r=1.42\arcsec$}&\colhead{$r=1.4\arcsec$}&
\colhead{$r=1.42\arcsec$}&\colhead{$r=1.42\arcsec$}
} 
\startdata 
  1 & -87.5 &   88.9  & 11.86$_{-0.01}^{+ 0.01}$ &12.23$_{-0.01}^{+ 0.01}$ &  0.62$_{-0.01}^{+ 0.01}$ & 1.91$_{-0.01}^{+ 0.01}$ & 3.21$_{-0.01}^{+ 0.01}$  \\ 
  2 &  86.4 &  118.2  & 12.68$_{-0.01}^{+ 0.01}$ &12.97$_{-0.01}^{+ 0.01}$ &  0.45$_{-0.01}^{+ 0.01}$ & 1.34$_{-0.01}^{+ 0.01}$ & 2.35$_{-0.01}^{+ 0.01}$  \\ 
  3 &   0.0 &    0.0  & 13.56$_{-0.01}^{+ 0.01}$ &14.76$_{-0.01}^{+ 0.01}$ &  1.37$_{-0.01}^{+ 0.01}$ & 3.35$_{-0.01}^{+ 0.01}$ & 5.94$_{-0.01}^{+ 0.01}$  \\ 
  4 &-137.5 &  108.6  & 14.01$_{-0.02}^{+ 0.02}$ &14.98$_{-0.01}^{+ 0.01}$ &  1.34$_{-0.02}^{+ 0.02}$ & 3.40$_{-0.02}^{+ 0.02}$ & 5.75$_{-0.02}^{+ 0.02}$  \\ 
  5 &   9.9 &  -47.8  & 14.50$_{-0.02}^{+ 0.02}$ &14.78$_{-0.01}^{+ 0.01}$ &  0.62$_{-0.01}^{+ 0.01}$ & 1.41$_{-0.01}^{+ 0.01}$ & 2.55$_{-0.01}^{+ 0.01}$  \\ 
  6 &-142.1 &   57.1  & 14.58$_{-0.04}^{+ 0.04}$ &15.58$_{-0.03}^{+ 0.03}$ &  1.57$_{-0.04}^{+ 0.04}$ & 3.61$_{-0.03}^{+ 0.03}$ & 5.75$_{-0.03}^{+ 0.04}$  \\ 
  7 &  -1.7 &   31.9  & 14.72$_{-0.03}^{+ 0.03}$ &15.30$_{-0.01}^{+ 0.02}$ &  1.39$_{-0.02}^{+ 0.02}$ & 3.32$_{-0.02}^{+ 0.02}$ & 5.84$_{-0.02}^{+ 0.02}$  \\ 
  8 &-125.7 &   30.4  & 14.81$_{-0.04}^{+ 0.04}$ &15.81$_{-0.03}^{+ 0.03}$ &  1.56$_{-0.04}^{+ 0.04}$ & 3.34$_{-0.03}^{+ 0.04}$ & 5.76$_{-0.04}^{+ 0.04}$  \\
\enddata 
\label{m1253} 
\end{deluxetable} 

\begin{deluxetable}{rrrrrrrr} 
\tabletypesize{\footnotesize} 
\tablewidth{0pt} 
\tablenum{9} 
\tablecaption{Abell 1525} 
\tablehead{ 
\colhead{ID}&\colhead{$\delta$x}&\colhead{$\delta$y}& 
\colhead{$Kt$}&
\colhead{$Ka$}&
\colhead{$J-K$}&
\colhead{$R-K$}&
\colhead{$B-K$}\\
\colhead{}&\colhead{arcsec}&\colhead{arcsec} & 
\colhead{}&\colhead{$r=1.42\arcsec$}&\colhead{$r=1.42\arcsec$}&
\colhead{$r=1.42\arcsec$}&\colhead{$r=1.42\arcsec$}
} 
\startdata 
  1 &   0.0 &    0.0  & 14.25$_{-0.01}^{+ 0.01}$ &15.33$_{-0.01}^{+ 0.01}$ &  1.26$_{-0.01}^{+ 0.01}$ & 3.27$_{-0.01}^{+ 0.01}$ & 5.97$_{-0.01}^{+ 0.01}$  \\ 
  2 &  -7.0 &    4.5  & 14.31$_{-0.01}^{+ 0.01}$ &15.26$_{-0.01}^{+ 0.01}$ &  1.18$_{-0.01}^{+ 0.01}$ & 3.04$_{-0.01}^{+ 0.01}$ & 5.42$_{-0.01}^{+ 0.01}$  \\ 
  3 &  12.5 &   -1.5  & 14.62$_{-0.01}^{+ 0.01}$ &15.45$_{-0.01}^{+ 0.01}$ &  1.21$_{-0.01}^{+ 0.01}$ & 3.35$_{-0.01}^{+ 0.01}$ & 6.09$_{-0.02}^{+ 0.02}$  \\ 
  4 &  71.8 &  -20.9  & 14.64$_{-0.01}^{+ 0.01}$ &15.61$_{-0.01}^{+ 0.01}$ &  1.39$_{-0.02}^{+ 0.02}$ & 3.31$_{-0.02}^{+ 0.02}$ & 5.94$_{-0.02}^{+ 0.02}$  \\ 
  5 &  28.1 &    2.3  & 14.76$_{-0.01}^{+ 0.01}$ &15.68$_{-0.01}^{+ 0.01}$ &  1.19$_{-0.02}^{+ 0.02}$ & 3.27$_{-0.02}^{+ 0.02}$ & 6.03$_{-0.02}^{+ 0.02}$  \\ 
  6 & -72.6 &  -25.6  & 14.77$_{-0.03}^{+ 0.03}$ &15.63$_{-0.02}^{+ 0.02}$ &  1.35$_{-0.02}^{+ 0.02}$ & 3.35$_{-0.02}^{+ 0.02}$ & 6.00$_{-0.02}^{+ 0.02}$  \\ 
  7 & -89.0 &  -46.3  & 14.87$_{-0.03}^{+ 0.03}$ &15.73$_{-0.02}^{+ 0.02}$ &  1.37$_{-0.02}^{+ 0.02}$ & 3.25$_{-0.02}^{+ 0.02}$ & 5.97$_{-0.02}^{+ 0.02}$  \\ 
  8 &  26.1 &   18.1  & 14.95$_{-0.02}^{+ 0.02}$ &15.59$_{-0.01}^{+ 0.01}$ &  1.21$_{-0.01}^{+ 0.01}$ & 3.34$_{-0.01}^{+ 0.01}$ & 6.05$_{-0.02}^{+ 0.02}$  \\
\enddata 
\label{a1525} 
\end{deluxetable} 

\begin{deluxetable}{rrrrrrrr} 
\tabletypesize{\footnotesize} 
\tablewidth{0pt} 
\tablenum{10} 
\tablecaption{MS1008-1225} 
\tablehead{ 
\colhead{ID}&\colhead{$\delta$x}&\colhead{$\delta$y}& 
\colhead{$Kt$}&
\colhead{$Ka$}&
\colhead{$J-K$}&
\colhead{$R-K$}&
\colhead{$B-K$}\\
\colhead{}&\colhead{arcsec}&\colhead{arcsec} & 
\colhead{}&\colhead{$r=1.42\arcsec$}&\colhead{$r=1.42\arcsec$}&
\colhead{$r=1.42\arcsec$}&\colhead{$r=1.42\arcsec$}
} 
\startdata 
  1 & 122.1 &  123.2 &  9.58$_{-0.01}^{+ 0.01}$ &10.12$_{-0.01}^{+ 0.01}$ &  1.13$_{-0.01}^{+ 0.01}$ & 3.62$_{-0.01}^{+ 0.01}$ & 6.87$_{-0.01}^{+ 0.01}$  \\ 
  2 &-140.8 &   92.7 & 10.22$_{-0.01}^{+ 0.01}$ &10.72$_{-0.01}^{+ 0.01}$ &  0.49$_{-0.01}^{+ 0.01}$ & 3.09$_{-0.01}^{+ 0.01}$ & 6.27$_{-0.01}^{+ 0.01}$  \\ 
  3 & 121.6 &  116.5 & 12.02$_{-0.01}^{+ 0.01}$ &12.33$_{-0.01}^{+ 0.01}$ &  0.75$_{-0.01}^{+ 0.01}$ & 2.30$_{-0.01}^{+ 0.01}$ & 4.81$_{-0.01}^{+ 0.01}$  \\ 
  4 & -87.7 &  142.0 & 12.34$_{-0.01}^{+ 0.01}$ &12.82$_{-0.01}^{+ 0.01}$ &  0.61$_{-0.01}^{+ 0.01}$ & 1.96$_{-0.01}^{+ 0.01}$ & 4.40$_{-0.01}^{+ 0.01}$  \\ 
  5 &   0.0 &    0.0 & 13.63$_{-0.01}^{+ 0.01}$ &15.60$_{-0.01}^{+ 0.01}$ &  1.29$_{-0.01}^{+ 0.01}$ & 3.46$_{-0.01}^{+ 0.01}$ & 5.17$_{-0.01}^{+ 0.01}$  \\ 
  6 &  60.2 &   -5.1 & 13.88$_{-0.01}^{+ 0.01}$ &14.22$_{-0.01}^{+ 0.01}$ &  0.37$_{-0.01}^{+ 0.01}$ & 1.27$_{-0.01}^{+ 0.01}$ & 3.10$_{-0.01}^{+ 0.01}$  \\ 
  7 &   6.0 & -122.6 & 13.94$_{-0.01}^{+ 0.01}$ &14.34$_{-0.01}^{+ 0.01}$ &  0.30$_{-0.01}^{+ 0.01}$ & 1.33$_{-0.01}^{+ 0.01}$ & 3.00$_{-0.01}^{+ 0.01}$  \\ 
  8 &-133.5 &  -53.9 & 14.46$_{-0.01}^{+ 0.01}$ &15.33$_{-0.01}^{+ 0.01}$ &  1.35$_{-0.01}^{+ 0.01}$ & 3.34$_{-0.01}^{+ 0.01}$ & 4.89$_{-0.01}^{+ 0.01}$  \\
\enddata 
\label{m1008} 
\end{deluxetable} 

\begin{deluxetable}{rrrrrrrr} 
\tabletypesize{\footnotesize} 
\tablewidth{0pt} 
\tablenum{11} 
\tablecaption{MS1147+1103} 
\tablehead{ 
\colhead{ID}&\colhead{$\delta$x}&\colhead{$\delta$y}& 
\colhead{$Kt$}&
\colhead{$Ka$}&
\colhead{$J-K$}&
\colhead{$R-K$}&
\colhead{$B-K$}\\
\colhead{}&\colhead{arcsec}&\colhead{arcsec} & 
\colhead{}&\colhead{$r=1.65\arcsec$}&\colhead{$r=1.65\arcsec$}&
\colhead{$r=1.65\arcsec$}&\colhead{$r=1.65\arcsec$}
} 
\startdata 
  1 & 113.6 &  119.4  & 14.15$_{-0.01}^{+ 0.01}$ &15.10$_{-0.01}^{+ 0.01}$ &  1.25$_{-0.01}^{+ 0.01}$ & 2.94$_{-0.01}^{+ 0.01}$ & 4.22$_{-0.01}^{+ 0.01}$  \\ 
  2 &   0.2 &   -0.1  & 14.16$_{-0.01}^{+ 0.01}$ &15.47$_{-0.01}^{+ 0.01}$ &  1.34$_{-0.01}^{+ 0.02}$ & 3.28$_{-0.01}^{+ 0.01}$ & 4.98$_{-0.01}^{+ 0.01}$  \\ 
  3 & -50.1 &   12.1  & 14.28$_{-0.01}^{+ 0.01}$ &14.67$_{-0.01}^{+ 0.01}$ &  0.41$_{-0.01}^{+ 0.01}$ & 1.13$_{-0.01}^{+ 0.01}$ & 2.56$_{-0.01}^{+ 0.01}$  \\ 
  4 &  15.9 &   98.9  & 14.65$_{-0.01}^{+ 0.01}$ &15.00$_{-0.01}^{+ 0.01}$ &  0.54$_{-0.01}^{+ 0.01}$ & 1.61$_{-0.01}^{+ 0.01}$ & 2.76$_{-0.01}^{+ 0.01}$  \\ 
  5 &  56.2 &  106.3  & 14.70$_{-0.01}^{+ 0.01}$ &15.02$_{-0.01}^{+ 0.01}$ &  0.62$_{-0.01}^{+ 0.01}$ & 1.87$_{-0.01}^{+ 0.01}$ & 3.11$_{-0.01}^{+ 0.01}$  \\ 
  6 & -54.0 &   41.8  & 14.85$_{-0.02}^{+ 0.02}$ &15.17$_{-0.01}^{+ 0.01}$ &  0.53$_{-0.01}^{+ 0.01}$ & 1.40$_{-0.01}^{+ 0.01}$ & 2.51$_{-0.01}^{+ 0.01}$  \\ 
  7 &  46.2 &   57.9  & 14.90$_{-0.02}^{+ 0.02}$ &15.94$_{-0.02}^{+ 0.02}$ &  1.37$_{-0.02}^{+ 0.02}$ & 2.81$_{-0.02}^{+ 0.02}$ & 3.87$_{-0.02}^{+ 0.02}$  \\ 
  8 &  41.5 &   33.4  & 15.10$_{-0.02}^{+ 0.02}$ &15.65$_{-0.01}^{+ 0.01}$ &  1.35$_{-0.02}^{+ 0.02}$ & 3.18$_{-0.02}^{+ 0.02}$ & 4.81$_{-0.02}^{+ 0.02}$  \\
\enddata 
\label{m1147} 
\end{deluxetable} 

\begin{deluxetable}{rrrrrrrrrc} 
\tabletypesize{\footnotesize} 
\tablewidth{0pt} 
\tablenum{12} 
\tablecaption{AC 118} 
\tablehead{ 
\colhead{ID}&\colhead{$\delta$x}&\colhead{$\delta$y}& 
\colhead{$Kt$}&
\colhead{$Ka$}&
\colhead{$H-K$}&
\colhead{$J-K$}&
\colhead{$R-K$}&
\colhead{$g-K$}&
\colhead{type} \\ 
\colhead{}&\colhead{arcsec}&\colhead{arcsec} & 
\colhead{}&\colhead{$r=1.89\arcsec$}&\colhead{$r=1.89\arcsec$}&
\colhead{$r=1.89\arcsec$}& 
\colhead{$r=1.89\arcsec$}&\colhead{$r=1.89\arcsec$}
} 
\startdata 
  1 & -59.1 &  116.0 & 10.84$_{-0.01}^{+ 0.01}$ & 11.23$_{-0.01}^{+ 0.01}$ & 0.02$_{-0.01}^{+ 0.01}$ &  0.59$_{-0.01}^{+ 0.01}$ & 1.76$_{-0.01}^{+ 0.01}$ & 2.26$_{-0.01}^{+ 0.01}$ &      5  \\ 
  2 & -93.9 & -106.9 & 12.78$_{-0.01}^{+ 0.01}$ & 13.25$_{-0.01}^{+ 0.01}$ & 0.04$_{-0.01}^{+ 0.01}$ &  0.61$_{-0.01}^{+ 0.01}$ & 1.49$_{-0.01}^{+ 0.01}$ & 2.13$_{-0.01}^{+ 0.01}$ &      0  \\ 
  3 &  44.9 & -104.6 & 13.62$_{-0.01}^{+ 0.01}$ & 13.92$_{-0.01}^{+ 0.01}$ & 0.25$_{-0.01}^{+ 0.01}$ &  0.87$_{-0.01}^{+ 0.01}$ & 3.14$_{-0.01}^{+ 0.01}$ & 4.39$_{-0.01}^{+ 0.01}$ &      0  \\ 
  4 &   0.0 &    0.0 & 13.69$_{-0.01}^{+ 0.01}$ & 15.13$_{-0.01}^{+ 0.01}$ & 0.72$_{-0.01}^{+ 0.01}$ &  1.51$_{-0.01}^{+ 0.01}$ & 3.76$_{-0.01}^{+ 0.01}$ & 5.13$_{-0.01}^{+ 0.01}$ &      1  \\ 
  5 &-140.4 &   95.9 & 13.85$_{-0.01}^{+ 0.01}$ & 15.43$_{-0.01}^{+ 0.01}$ & 0.62$_{-0.01}^{+ 0.01}$ &  1.51$_{-0.01}^{+ 0.01}$ & 3.44$_{-0.01}^{+ 0.01}$ & 4.70$_{-0.02}^{+ 0.02}$ &      0  \\ 
  6 &-101.2 &   82.9 & 13.87$_{-0.01}^{+ 0.01}$ & 15.33$_{-0.01}^{+ 0.01}$ & 0.63$_{-0.01}^{+ 0.01}$ &  1.53$_{-0.01}^{+ 0.01}$ & 3.49$_{-0.01}^{+ 0.01}$ & 4.79$_{-0.01}^{+ 0.01}$ &      0  \\ 
  7 &  18.5 &  -19.8 & 13.91$_{-0.01}^{+ 0.01}$ & 15.14$_{-0.01}^{+ 0.01}$ & 0.74$_{-0.01}^{+ 0.01}$ &  1.48$_{-0.01}^{+ 0.01}$ & 3.77$_{-0.01}^{+ 0.01}$ & 5.19$_{-0.01}^{+ 0.01}$ &      1  \\ 
  8 & -27.9 &  154.5 & 14.00$_{-0.01}^{+ 0.01}$ & 15.30$_{-0.01}^{+ 0.01}$ & 0.67$_{-0.01}^{+ 0.01}$ &  1.55$_{-0.01}^{+ 0.01}$ & 3.50$_{-0.01}^{+ 0.01}$ & 4.74$_{-0.01}^{+ 0.01}$ &      0  \\
\enddata 
\label{ac118} 
\end{deluxetable} 

\begin{deluxetable}{rrrrrrrrrc} 
\tabletypesize{\footnotesize} 
\tablewidth{0pt} 
\tablenum{13} 
\tablecaption{AC 103} 
\tablehead{ 
\colhead{ID}&\colhead{$\delta$x}&\colhead{$\delta$y}& 
\colhead{$Kt$}&
\colhead{$Ka$}&
\colhead{$H-K$}&
\colhead{$J-K$}&
\colhead{$R-K$}&
\colhead{$g-K$}&
\colhead{type} \\ 
\colhead{}&\colhead{arcsec}&\colhead{arcsec} & 
\colhead{}&\colhead{$r=1.89\arcsec$}&\colhead{$r=1.89\arcsec$}&
\colhead{$r=1.89\arcsec$}& 
\colhead{$r=1.89\arcsec$}&\colhead{$r=1.89\arcsec$}
} 
\startdata 
  1 &   4.1 & -100.3  & 11.00$_{-0.01}^{+ 0.01}$ & 11.25$_{-0.01}^{+ 0.01}$ & 0.14$_{-0.01}^{+ 0.01}$ &  0.50$_{-0.01}^{+ 0.01}$ & 2.01$_{-0.01}^{+ 0.01}$ & 2.39$_{-0.01}^{+ 0.01}$ &      0  \\ 
  2 &   5.0 &  -69.6  & 11.81$_{-0.01}^{+ 0.01}$ & 12.08$_{-0.01}^{+ 0.01}$ & 0.21$_{-0.01}^{+ 0.01}$ &  0.86$_{-0.01}^{+ 0.01}$ & 3.06$_{-0.01}^{+ 0.01}$ & 4.34$_{-0.01}^{+ 0.01}$ &      0  \\ 
  3 &  73.6 &  115.8  & 11.85$_{-0.01}^{+ 0.01}$ & 12.26$_{-0.01}^{+ 0.01}$ & 0.12$_{-0.01}^{+ 0.01}$ &  0.43$_{-0.01}^{+ 0.01}$ & 1.23$_{-0.01}^{+ 0.01}$ & 1.63$_{-0.01}^{+ 0.01}$ &      5  \\ 
  4 & 130.5 &   89.3  & 12.06$_{-0.01}^{+ 0.01}$ & 12.48$_{-0.01}^{+ 0.01}$ & 0.15$_{-0.01}^{+ 0.01}$ &  0.54$_{-0.01}^{+ 0.01}$ & 1.51$_{-0.01}^{+ 0.01}$ & 2.13$_{-0.01}^{+ 0.01}$ &      0  \\ 
  5 & -33.9 &  -67.7  & 12.26$_{-0.01}^{+ 0.01}$ & 12.50$_{-0.01}^{+ 0.01}$ & 0.07$_{-0.01}^{+ 0.01}$ &  0.39$_{-0.01}^{+ 0.01}$ & 1.35$_{-0.01}^{+ 0.01}$ & 1.83$_{-0.01}^{+ 0.01}$ &      0  \\ 
  6 &  87.2 &  -54.5  & 12.30$_{-0.01}^{+ 0.01}$ & 12.53$_{-0.01}^{+ 0.01}$ & 0.25$_{-0.01}^{+ 0.01}$ &  0.87$_{-0.01}^{+ 0.01}$ & 3.17$_{-0.01}^{+ 0.01}$ & 4.47$_{-0.01}^{+ 0.01}$ &      0  \\ 
  7 & 116.9 &  -76.9  & 13.07$_{-0.01}^{+ 0.01}$ & 14.15$_{-0.01}^{+ 0.01}$ & 0.56$_{-0.01}^{+ 0.01}$ &  1.33$_{-0.01}^{+ 0.01}$ & 3.52$_{-0.01}^{+ 0.01}$ & 4.50$_{-0.01}^{+ 0.01}$ &      0  \\ 
  8 &  40.3 &  -53.3  & 13.22$_{-0.01}^{+ 0.01}$ & 13.43$_{-0.01}^{+ 0.01}$ & 0.14$_{-0.01}^{+ 0.01}$ &  0.62$_{-0.01}^{+ 0.01}$ & 1.85$_{-0.01}^{+ 0.01}$ & 2.58$_{-0.01}^{+ 0.01}$ &      0  \\
\enddata 
\label{ac103} 
\end{deluxetable} 

\begin{deluxetable}{rrrrrrrr} 
\tabletypesize{\footnotesize} 
\tablewidth{0pt} 
\tablenum{14} 
\tablecaption{AC 114} 
\tablehead{ 
\colhead{ID}&\colhead{$\delta$x}&\colhead{$\delta$y}& 
\colhead{$Kt$}&
\colhead{$Ka$}&
\colhead{$J-K$}&
\colhead{$R-K$}&
\colhead{$g-K$}\\
\colhead{}&\colhead{arcsec}&\colhead{arcsec} & 
\colhead{}&\colhead{$r=1.83\arcsec$}&\colhead{$r=1.83\arcsec$}&
\colhead{$r=1.83\arcsec$}&\colhead{$r=1.83\arcsec$}
} 
\startdata 
  1 & -87.6 &   37.3  & 12.88$_{-0.01}^{+ 0.01}$ &13.23$_{-0.01}^{+ 0.01}$ &  0.67$_{-0.01}^{+ 0.01}$ & 1.77$_{-0.01}^{+ 0.01}$ & 2.45$_{-0.01}^{+ 0.01}$  \\ 
  2 &  -2.6 &   86.8  & 13.07$_{-0.01}^{+ 0.01}$ &13.31$_{-0.01}^{+ 0.01}$ &  0.85$_{-0.01}^{+ 0.01}$ & 3.93$_{-0.01}^{+ 0.01}$ & 5.28$_{-0.01}^{+ 0.01}$  \\ 
  3 & -84.9 &   43.6  & 13.31$_{-0.01}^{+ 0.01}$ &13.64$_{-0.01}^{+ 0.01}$ &  0.59$_{-0.01}^{+ 0.01}$ & 1.48$_{-0.01}^{+ 0.01}$ & 2.10$_{-0.01}^{+ 0.01}$  \\ 
  4 &   0.0 &    0.0  & 13.38$_{-0.01}^{+ 0.01}$ &14.95$_{-0.01}^{+ 0.01}$ &  1.48$_{-0.01}^{+ 0.01}$ & 3.68$_{-0.01}^{+ 0.01}$ & 5.02$_{-0.01}^{+ 0.01}$  \\ 
  5 &-139.0 &   33.8  & 13.67$_{-0.01}^{+ 0.01}$ &14.23$_{-0.01}^{+ 0.01}$ &  0.61$_{-0.01}^{+ 0.01}$ & 1.32$_{-0.01}^{+ 0.01}$ & 1.96$_{-0.01}^{+ 0.01}$  \\ 
  6 & 120.0 & -120.0  & 13.82$_{-0.01}^{+ 0.01}$ &14.05$_{-0.01}^{+ 0.01}$ &  0.67$_{-0.01}^{+ 0.01}$ & 1.98$_{-0.01}^{+ 0.01}$ & 2.82$_{-0.01}^{+ 0.01}$  \\ 
  7 & 119.6 &  141.6  & 14.06$_{-0.01}^{+ 0.01}$ &14.34$_{-0.01}^{+ 0.01}$ &  0.70$_{-0.01}^{+ 0.01}$ & 2.87$_{-0.01}^{+ 0.01}$ & 4.09$_{-0.01}^{+ 0.01}$  \\ 
  8 & -77.0 &   22.9  & 14.20$_{-0.01}^{+ 0.01}$ &15.15$_{-0.01}^{+ 0.01}$ &  1.48$_{-0.01}^{+ 0.01}$ & 3.47$_{-0.01}^{+ 0.01}$ & 4.75$_{-0.01}^{+ 0.01}$  \\
\enddata 
\label{ac114} 
\end{deluxetable} 

\begin{deluxetable}{rrrrrrrrrc} 
\tabletypesize{\footnotesize} 
\tablewidth{0pt} 
\tablenum{15} 
\tablecaption{MS2137-2339} 
\tablehead{ 
\colhead{ID}&\colhead{$\delta$x}&\colhead{$\delta$y}& 
\colhead{$Kt$}&
\colhead{$Ka$}&
\colhead{$H-K$}&
\colhead{$J-K$}&
\colhead{$R-K$}&
\colhead{$g-K$}&
\colhead{type} \\ 
\colhead{}&\colhead{arcsec}&\colhead{arcsec} & 
\colhead{}&\colhead{$r=1.94\arcsec$}&\colhead{$r=1.94\arcsec$}&
\colhead{$r=1.94\arcsec$}& 
\colhead{$r=1.94\arcsec$}&\colhead{$r=1.94\arcsec$}
} 
\startdata 
  1 & -49.5 &  103.3  & 10.20$_{-0.01}^{+ 0.01}$ & 10.50$_{-0.01}^{+ 0.01}$ & 0.20$_{-0.01}^{+ 0.01}$ &  0.80$_{-0.01}^{+ 0.01}$ & 2.43$_{-0.01}^{+ 0.01}$ & 2.76$_{-0.01}^{+ 0.01}$ &      5  \\ 
  2 &  34.6 &   40.6  & 11.46$_{-0.01}^{+ 0.01}$ & 11.69$_{-0.01}^{+ 0.01}$ & 0.03$_{-0.01}^{+ 0.01}$ &  0.37$_{-0.01}^{+ 0.01}$ & 1.57$_{-0.01}^{+ 0.01}$ & 1.88$_{-0.01}^{+ 0.01}$ &      0  \\ 
  3 &  57.8 &  135.9  & 12.67$_{-0.01}^{+ 0.01}$ & 12.91$_{-0.01}^{+ 0.01}$ & 0.08$_{-0.01}^{+ 0.01}$ &  0.54$_{-0.01}^{+ 0.01}$ & 1.52$_{-0.01}^{+ 0.01}$ & 2.13$_{-0.01}^{+ 0.01}$ &      0  \\ 
  4 & 139.4 &   11.7  & 12.90$_{-0.01}^{+ 0.01}$ & 13.14$_{-0.01}^{+ 0.01}$ & 0.02$_{-0.01}^{+ 0.01}$ &  0.32$_{-0.01}^{+ 0.01}$ & 1.23$_{-0.01}^{+ 0.01}$ & 1.73$_{-0.01}^{+ 0.01}$ &      0  \\ 
  5 &   0.0 &    0.0  & 13.41$_{-0.01}^{+ 0.01}$ & 14.82$_{-0.01}^{+ 0.01}$ & 0.71$_{-0.01}^{+ 0.01}$ &  1.44$_{-0.01}^{+ 0.01}$ & 3.52$_{-0.01}^{+ 0.01}$ & 4.70$_{-0.01}^{+ 0.01}$ &      1  \\ 
  6 &  33.8 & -111.9  & 14.32$_{-0.01}^{+ 0.01}$ & 14.56$_{-0.01}^{+ 0.01}$ & 0.14$_{-0.01}^{+ 0.01}$ &  0.64$_{-0.01}^{+ 0.01}$ & 1.78$_{-0.01}^{+ 0.01}$ & 2.49$_{-0.01}^{+ 0.01}$ &      0  \\ 
  7 & 122.2 &   14.3  & 14.35$_{-0.01}^{+ 0.01}$ & 15.20$_{-0.01}^{+ 0.01}$ & 0.60$_{-0.01}^{+ 0.01}$ &  1.38$_{-0.01}^{+ 0.01}$ & 3.36$_{-0.01}^{+ 0.01}$ & 3.98$_{-0.01}^{+ 0.01}$ &      0  \\ 
  8 &-112.2 &  -91.6  & 14.39$_{-0.01}^{+ 0.01}$ & 14.87$_{-0.01}^{+ 0.01}$ & 0.17$_{-0.01}^{+ 0.01}$ &  0.93$_{-0.01}^{+ 0.01}$ & 2.61$_{-0.01}^{+ 0.01}$ & 3.80$_{-0.01}^{+ 0.01}$ &      0  \\
\enddata 
\label{m2137} 
\end{deluxetable} 

\begin{deluxetable}{rrrrrrrr} 
\tabletypesize{\footnotesize} 
\tablewidth{0pt} 
\tablenum{16} 
\tablecaption{Abell S0506} 
\tablehead{ 
\colhead{ID}&\colhead{$\delta$x}&\colhead{$\delta$y}& 
\colhead{$Kt$}&
\colhead{$Ka$}&
\colhead{$J-K$}&
\colhead{$R-K$}&
\colhead{$g-K$}\\
\colhead{}&\colhead{arcsec}&\colhead{arcsec} & 
\colhead{}&\colhead{$r=1.42\arcsec$}&\colhead{$r=1.42\arcsec$}&
\colhead{$r=1.42\arcsec$}&\colhead{$r=1.42\arcsec$}
} 
\startdata 
  1  & -84.9 &   96.5  & 10.06$_{-0.01}^{+ 0.01}$ &10.61$_{-0.01}^{+ 0.01}$ &  0.67$_{-0.01}^{+ 0.01}$ & 2.54$_{-0.01}^{+ 0.01}$ & 2.71$_{-0.01}^{+ 0.01}$  \\ 
  2  &  99.9 &  -30.6  & 12.73$_{-0.01}^{+ 0.01}$ &13.02$_{-0.01}^{+ 0.01}$ &  0.84$_{-0.01}^{+ 0.01}$ & 2.54$_{-0.01}^{+ 0.01}$ & 3.92$_{-0.01}^{+ 0.01}$  \\ 
  3  & -29.2 &   75.7  & 13.71$_{-0.01}^{+ 0.01}$ &14.11$_{-0.01}^{+ 0.01}$ &  0.40$_{-0.01}^{+ 0.01}$ & 1.06$_{-0.01}^{+ 0.01}$ & 1.83$_{-0.01}^{+ 0.01}$  \\ 
  4  &-100.1 &   67.1  & 14.54$_{-0.01}^{+ 0.01}$ &15.07$_{-0.01}^{+ 0.01}$ &  0.30$_{-0.01}^{+ 0.01}$ & 0.73$_{-0.01}^{+ 0.01}$ & 1.38$_{-0.01}^{+ 0.01}$  \\ 
  5  & -42.0 &  -48.2  & 14.55$_{-0.01}^{+ 0.01}$ &14.82$_{-0.01}^{+ 0.01}$ &  0.44$_{-0.01}^{+ 0.01}$ & 1.48$_{-0.01}^{+ 0.01}$ & 2.45$_{-0.01}^{+ 0.01}$  \\ 
  6  & -65.8 &  -42.2  & 14.57$_{-0.01}^{+ 0.01}$ &14.97$_{-0.01}^{+ 0.01}$ &  0.75$_{-0.01}^{+ 0.01}$ & 2.82$_{-0.01}^{+ 0.01}$ & 4.31$_{-0.01}^{+ 0.01}$  \\ 
  7  & 133.9 & -108.6  & 14.64$_{-0.01}^{+ 0.01}$ &14.99$_{-0.01}^{+ 0.01}$ &  0.58$_{-0.01}^{+ 0.01}$ & 2.81$_{-0.01}^{+ 0.01}$ & 4.30$_{-0.01}^{+ 0.01}$  \\ 
  8  &  26.8 &  -15.5  & 14.74$_{-0.01}^{+ 0.01}$ &15.84$_{-0.01}^{+ 0.01}$ &  1.38$_{-0.01}^{+ 0.02}$ & 3.37$_{-0.01}^{+ 0.01}$ & 4.94$_{-0.01}^{+ 0.01}$  \\
\enddata 
\label{s0506} 
\end{deluxetable} 

\begin{deluxetable}{rrrrrrrr} 
\tabletypesize{\footnotesize} 
\tablewidth{0pt} 
\tablenum{17} 
\tablecaption{MS1358+6245} 
\tablehead{ 
\colhead{ID}&\colhead{$\delta$x}&\colhead{$\delta$y}& 
\colhead{$Kt$}&
\colhead{$Ka$}&
\colhead{$J-K$}&
\colhead{$R-K$}&
\colhead{$g-K$}\\
\colhead{}&\colhead{arcsec}&\colhead{arcsec} & 
\colhead{}&\colhead{$r=1.37\arcsec$}&\colhead{$r=1.37\arcsec$}&
\colhead{$r=1.37\arcsec$}&\colhead{$r=1.37\arcsec$}
} 
\startdata 
  1 &   4.7 & -116.6  & 11.90$_{-0.01}^{+ 0.01}$ &12.26$_{-0.01}^{+ 0.01}$ &  0.90$_{-0.01}^{+ 0.01}$ & 2.16$_{-0.01}^{+ 0.01}$ & 2.67$_{-0.01}^{+ 0.01}$  \\ 
  2 &   0.0 &    0.0  & 13.99$_{-0.01}^{+ 0.01}$ &15.57$_{-0.01}^{+ 0.01}$ &  1.52$_{-0.01}^{+ 0.01}$ & 3.73$_{-0.01}^{+ 0.01}$ & 5.47$_{-0.01}^{+ 0.01}$  \\ 
  3 &-117.7 &   12.7  & 14.52$_{-0.01}^{+ 0.01}$ &14.89$_{-0.01}^{+ 0.01}$ &  0.59$_{-0.01}^{+ 0.01}$ & 4.21$_{-0.01}^{+ 0.01}$ & 6.03$_{-0.01}^{+ 0.01}$  \\ 
  4 &  52.9 &  -88.4  & 14.65$_{-0.01}^{+ 0.01}$ &15.53$_{-0.01}^{+ 0.01}$ &  1.43$_{-0.01}^{+ 0.01}$ & 3.65$_{-0.01}^{+ 0.01}$ & 5.33$_{-0.01}^{+ 0.01}$  \\ 
  5 &  17.7 &  -77.3  & 14.68$_{-0.01}^{+ 0.01}$ &15.52$_{-0.01}^{+ 0.01}$ &  1.52$_{-0.01}^{+ 0.01}$ & 3.76$_{-0.01}^{+ 0.01}$ & 5.52$_{-0.01}^{+ 0.01}$  \\ 
  6 &  -7.6 &  -17.9  & 14.99$_{-0.01}^{+ 0.01}$ &15.95$_{-0.01}^{+ 0.01}$ &  1.52$_{-0.01}^{+ 0.01}$ & 3.65$_{-0.01}^{+ 0.01}$ & 5.41$_{-0.01}^{+ 0.01}$  \\ 
  7 &  20.5 &  -59.2  & 15.28$_{-0.01}^{+ 0.01}$ &15.86$_{-0.01}^{+ 0.01}$ &  1.52$_{-0.01}^{+ 0.01}$ & 3.73$_{-0.01}^{+ 0.01}$ & 5.51$_{-0.01}^{+ 0.01}$  \\ 
  8 &-131.5 &   82.0  & 15.29$_{-0.02}^{+ 0.02}$ &16.34$_{-0.01}^{+ 0.01}$ &  1.36$_{-0.02}^{+ 0.02}$ & 3.35$_{-0.02}^{+ 0.02}$ & 5.00$_{-0.02}^{+ 0.02}$  \\
\enddata 
\label{m1358} 
\end{deluxetable} 

\begin{deluxetable}{rrrrrrrrrc} 
\tabletypesize{\footnotesize} 
\tablewidth{0pt} 
\tablenum{18} 
\tablecaption{Cl2244-0205} 
\tablehead{ 
\colhead{ID}&\colhead{$\delta$x}&\colhead{$\delta$y}& 
\colhead{$Kt$}&
\colhead{$Ka$}&
\colhead{$H-K$}&
\colhead{$J-K$}&
\colhead{$R-K$}&
\colhead{$g-K$}&
\colhead{type} \\ 
\colhead{}&\colhead{arcsec}&\colhead{arcsec} & 
\colhead{}&\colhead{$r=1.64\arcsec$}&\colhead{$r=1.64\arcsec$}&
\colhead{$r=1.64\arcsec$}& 
\colhead{$r=1.64\arcsec$}&\colhead{$r=1.64\arcsec$}
} 
\startdata 
  1 &  83.6 &  111.2  & 13.78$_{-0.01}^{+ 0.01}$ & 14.10$_{-0.01}^{+ 0.01}$ & 0.09$_{-0.01}^{+ 0.01}$ &  0.64$_{-0.01}^{+ 0.01}$ & 2.21$_{-0.01}^{+ 0.01}$ & 3.42$_{-0.01}^{+ 0.01}$ &      0  \\ 
  2 &-132.3 &  130.5  & 14.13$_{-0.01}^{+ 0.01}$ & 14.55$_{-0.01}^{+ 0.01}$ & 1.69$_{-0.01}^{+ 0.01}$ &  1.10$_{-0.01}^{+ 0.01}$ & 2.51$_{-0.01}^{+ 0.01}$ & 3.74$_{-0.01}^{+ 0.01}$ &      0  \\ 
  3 & -60.8 &  -38.9  & 14.52$_{-0.01}^{+ 0.01}$ & 14.84$_{-0.01}^{+ 0.01}$ & 0.18$_{-0.01}^{+ 0.01}$ &  0.85$_{-0.01}^{+ 0.01}$ & 2.87$_{-0.01}^{+ 0.01}$ & 4.18$_{-0.01}^{+ 0.01}$ &      5  \\ 
  4 &   0.0 &    0.0  & 14.58$_{-0.01}^{+ 0.01}$ & 15.49$_{-0.01}^{+ 0.01}$ & 0.70$_{-0.01}^{+ 0.01}$ &  1.47$_{-0.01}^{+ 0.01}$ & 3.56$_{-0.01}^{+ 0.01}$ & 5.23$_{-0.04}^{+ 0.04}$ &      1  \\ 
  5 &  37.7 &   94.6  & 14.60$_{-0.01}^{+ 0.01}$ & 15.52$_{-0.01}^{+ 0.01}$ & 0.66$_{-0.01}^{+ 0.01}$ &  1.43$_{-0.01}^{+ 0.01}$ & 3.44$_{-0.01}^{+ 0.01}$ & 5.13$_{-0.04}^{+ 0.04}$ &      0  \\ 
  6 & -13.4 &    1.8  & 14.62$_{-0.01}^{+ 0.01}$ & 15.86$_{-0.01}^{+ 0.01}$ & 0.72$_{-0.01}^{+ 0.01}$ &  1.48$_{-0.01}^{+ 0.01}$ & 3.70$_{-0.01}^{+ 0.01}$ & 5.20$_{-0.05}^{+ 0.05}$ &      1  \\ 
  7 & 107.4 &  -80.8  & 14.66$_{-0.01}^{+ 0.01}$ & 15.03$_{-0.01}^{+ 0.01}$ & 0.28$_{-0.01}^{+ 0.01}$ &  0.71$_{-0.01}^{+ 0.01}$ & 2.80$_{-0.01}^{+ 0.01}$ & 4.24$_{-0.01}^{+ 0.01}$ &      0  \\ 
  8 &   6.6 &    4.2  & 14.69$_{-0.01}^{+ 0.01}$ & 15.62$_{-0.01}^{+ 0.01}$ & 0.70$_{-0.01}^{+ 0.01}$ &  1.43$_{-0.01}^{+ 0.01}$ & 3.50$_{-0.01}^{+ 0.01}$ & 5.15$_{-0.04}^{+ 0.04}$ &      1  \\
\enddata 
\label{c2244} 
\end{deluxetable} 

\begin{deluxetable}{rrrrrrrrrc} 
\tabletypesize{\footnotesize} 
\tablewidth{0pt} 
\tablenum{19} 
\tablecaption{Cl0024+16} 
\tablehead{ 
\colhead{ID}&\colhead{$\delta$x}&\colhead{$\delta$y}& 
\colhead{$Kt$}&
\colhead{$Ka$}&
\colhead{$H-K$}&
\colhead{$J-K$}&
\colhead{$R-K$}&
\colhead{$g-K$}&
\colhead{type} \\ 
\colhead{}&\colhead{arcsec}&\colhead{arcsec} & 
\colhead{}&\colhead{$r=1.91\arcsec$}&\colhead{$r=1.91\arcsec$}&
\colhead{$r=1.91\arcsec$}& 
\colhead{$r=1.91\arcsec$}&\colhead{$r=1.91\arcsec$}
} 
\startdata 
  1 &  15.3 & -119.7  & 11.73$_{-0.01}^{+ 0.01}$ & 11.90$_{-0.01}^{+ 0.01}$ & 0.11$_{-0.01}^{+ 0.01}$ &  0.84$_{-0.01}^{+ 0.01}$ & 2.59$_{-0.01}^{+ 0.01}$ & 2.89$_{-0.01}^{+ 0.01}$ &      0  \\ 
  2 &-121.9 &  -33.3  & 12.12$_{-0.01}^{+ 0.01}$ & 12.26$_{-0.01}^{+ 0.01}$ & 0.28$_{-0.01}^{+ 0.01}$ &  0.95$_{-0.01}^{+ 0.01}$ & 3.87$_{-0.01}^{+ 0.01}$ & 5.21$_{-0.01}^{+ 0.01}$ &      0  \\ 
  3 &   0.0 &    0.0  & 14.23$_{-0.01}^{+ 0.01}$ & 15.53$_{-0.01}^{+ 0.01}$ & 0.77$_{-0.01}^{+ 0.01}$ &  1.63$_{-0.01}^{+ 0.01}$ & 3.71$_{-0.01}^{+ 0.01}$ & 5.44$_{-0.01}^{+ 0.01}$ &      1  \\ 
  4 &   5.3 &    2.6  & 14.38$_{-0.01}^{+ 0.01}$ & 15.63$_{-0.01}^{+ 0.01}$ & 0.76$_{-0.01}^{+ 0.01}$ &  1.59$_{-0.01}^{+ 0.01}$ & 3.68$_{-0.01}^{+ 0.01}$ & 5.42$_{-0.01}^{+ 0.01}$ &      1  \\ 
  5 &  -8.9 &   -4.7  & 14.45$_{-0.01}^{+ 0.01}$ & 15.68$_{-0.01}^{+ 0.01}$ & 0.74$_{-0.01}^{+ 0.01}$ &  1.60$_{-0.01}^{+ 0.01}$ & 3.68$_{-0.01}^{+ 0.01}$ & 5.45$_{-0.01}^{+ 0.01}$ &      1  \\ 
  6 &  10.8 &   79.1  & 14.65$_{-0.01}^{+ 0.01}$ & 15.02$_{-0.01}^{+ 0.01}$ &99.00\phm{+0.0} &  0.41$_{-0.01}^{+ 0.01}$ & 3.59$_{-0.01}^{+ 0.01}$ & 4.98$_{-0.01}^{+ 0.01}$ &      0  \\ 
  7 & -61.7 &   71.5  & 14.81$_{-0.01}^{+ 0.01}$ & 15.76$_{-0.01}^{+ 0.01}$ & 5.92$_{-0.01}^{+ 0.00}$ &  6.57$_{-0.01}^{+ 0.00}$ & 3.66$_{-0.61}^{+ 0.00}$ &12.48$_{-0.01}^{+ 0.00}$ &      0  \\ 
  8 & -11.4 &  -24.6  & 14.84$_{-0.01}^{+ 0.01}$ & 15.90$_{-0.01}^{+ 0.01}$ & 0.75$_{-0.01}^{+ 0.01}$ &  1.59$_{-0.01}^{+ 0.01}$ & 3.68$_{-0.01}^{+ 0.01}$ & 5.30$_{-0.01}^{+ 0.01}$ &      1  \\
\enddata 
\label{c0024} 
\end{deluxetable} 

\begin{deluxetable}{rrrrrrrrrc} 
\tabletypesize{\footnotesize} 
\tablewidth{0pt} 
\tablenum{20} 
\tablecaption{GHO~0303+1706} 
\tablehead{ 
\colhead{ID}&\colhead{$\delta$x}&\colhead{$\delta$y}& 
\colhead{$Kt$}&
\colhead{$Ka$}&
\colhead{$H-K$}&
\colhead{$J-K$}&
\colhead{$R-K$}&
\colhead{$g-K$}&
\colhead{type} \\ 
\colhead{}&\colhead{arcsec}&\colhead{arcsec} & 
\colhead{}&\colhead{$r=1.64\arcsec$}&\colhead{$r=1.64\arcsec$}&
\colhead{$r=1.64\arcsec$}& 
\colhead{$r=1.64\arcsec$}&\colhead{$r=1.64\arcsec$}
} 
\startdata 
  1 &  20.1 &  178.4  & 12.52$_{-0.01}^{+ 0.01}$ & 12.74$_{-0.01}^{+ 0.01}$ & 0.04$_{-0.01}^{+ 0.01}$ &  0.45$_{-0.01}^{+ 0.01}$ & 1.23$_{-0.01}^{+ 0.01}$ & 2.05$_{-0.01}^{+ 0.01}$ &      0  \\ 
  2 &  77.7 &  177.3  & 13.38$_{-0.01}^{+ 0.01}$ & 13.60$_{-0.01}^{+ 0.01}$ & 0.18$_{-0.01}^{+ 0.01}$ &  0.90$_{-0.01}^{+ 0.01}$ & 2.98$_{-0.01}^{+ 0.01}$ & 4.16$_{-0.01}^{+ 0.01}$ &      0  \\ 
  3 &  39.6 &  -49.4  & 13.96$_{-0.01}^{+ 0.01}$ & 15.15$_{-0.01}^{+ 0.01}$ & 0.75$_{-0.01}^{+ 0.01}$ &  1.67$_{-0.01}^{+ 0.01}$ & 4.14$_{-0.01}^{+ 0.01}$ & 5.68$_{-0.01}^{+ 0.01}$ &      0  \\ 
  4 &  -4.8 &  157.8  & 14.34$_{-0.01}^{+ 0.01}$ & 14.53$_{-0.01}^{+ 0.01}$ & 0.22$_{-0.01}^{+ 0.01}$ &  0.89$_{-0.01}^{+ 0.01}$ & 3.13$_{-0.01}^{+ 0.01}$ & 4.31$_{-0.01}^{+ 0.01}$ &      0  \\ 
  5 &  57.3 &   64.7  & 14.50$_{-0.01}^{+ 0.01}$ & 15.38$_{-0.01}^{+ 0.01}$ & 0.61$_{-0.01}^{+ 0.01}$ &  1.38$_{-0.01}^{+ 0.01}$ & 3.45$_{-0.01}^{+ 0.01}$ & 4.50$_{-0.01}^{+ 0.01}$ &      0  \\ 
  6 &   0.0 &    0.0  & 14.54$_{-0.01}^{+ 0.01}$ & 15.74$_{-0.01}^{+ 0.01}$ & 0.79$_{-0.01}^{+ 0.01}$ &  1.71$_{-0.01}^{+ 0.01}$ & 4.02$_{-0.01}^{+ 0.01}$ & 5.62$_{-0.01}^{+ 0.01}$ &      1  \\ 
  7 & -87.8 &   77.3  & 14.79$_{-0.01}^{+ 0.01}$ & 15.88$_{-0.01}^{+ 0.01}$ & 0.75$_{-0.01}^{+ 0.01}$ &  1.68$_{-0.01}^{+ 0.01}$ & 3.99$_{-0.01}^{+ 0.01}$ & 5.56$_{-0.01}^{+ 0.01}$ &      1  \\ 
  8 &   4.0 &  132.0  & 15.15$_{-0.01}^{+ 0.01}$ & 15.51$_{-0.01}^{+ 0.01}$ & 0.27$_{-0.01}^{+ 0.01}$ &  0.88$_{-0.01}^{+ 0.01}$ & 3.82$_{-0.01}^{+ 0.01}$ & 5.00$_{-0.01}^{+ 0.01}$ &      0  \\
\enddata 
\label{g0303} 
\end{deluxetable} 

\begin{deluxetable}{rrrrrrrr} 
\tabletypesize{\footnotesize} 
\tablewidth{0pt} 
\tablenum{21} 
\tablecaption{3C 313} 
\tablehead{ 
\colhead{ID}&\colhead{$\delta$x}&\colhead{$\delta$y}& 
\colhead{$Kt$}&
\colhead{$Ka$}&
\colhead{$J-K$}&
\colhead{$I-K$}&
\colhead{$V-K$}\\
\colhead{}&\colhead{arcsec}&\colhead{arcsec} & 
\colhead{}&\colhead{$r=1.64\arcsec$}&\colhead{$r=1.64\arcsec$}&
\colhead{$r=1.64\arcsec$}&\colhead{$r=1.64\arcsec$}
} 
\startdata 
  1 & -31.2 & -135.7  & 13.16$_{-0.01}^{+ 0.01}$ &13.35$_{-0.01}^{+ 0.01}$ &  0.50$_{-0.01}^{+ 0.01}$ & 1.16$_{-0.01}^{+ 0.01}$ & 1.90$_{-0.01}^{+ 0.01}$  \\ 
  2 & 116.2 &   71.3  & 14.19$_{-0.01}^{+ 0.01}$ &14.48$_{-0.01}^{+ 0.01}$ &  0.34$_{-0.01}^{+ 0.01}$ & 0.86$_{-0.01}^{+ 0.01}$ & 1.73$_{-0.01}^{+ 0.01}$  \\ 
  3 &  64.9 &   84.1  & 14.21$_{-0.01}^{+ 0.01}$ &14.45$_{-0.01}^{+ 0.01}$ &  0.74$_{-0.01}^{+ 0.01}$ & 1.97$_{-0.01}^{+ 0.01}$ & 4.42$_{-0.01}^{+ 0.01}$  \\ 
  4 & 108.1 &  -65.0  & 14.60$_{-0.01}^{+ 0.01}$ &14.83$_{-0.01}^{+ 0.01}$ &  0.83$_{-0.01}^{+ 0.01}$ & 1.94$_{-0.01}^{+ 0.01}$ & 4.19$_{-0.01}^{+ 0.01}$  \\ 
  5 & -10.7 &   -9.8  & 14.76$_{-0.01}^{+ 0.01}$ &14.95$_{-0.01}^{+ 0.01}$ &  0.79$_{-0.01}^{+ 0.01}$ & 1.90$_{-0.01}^{+ 0.01}$ & 4.18$_{-0.01}^{+ 0.01}$  \\ 
  6 &   0.0 &    0.0  & 14.85$_{-0.01}^{+ 0.01}$ &15.85$_{-0.01}^{+ 0.01}$ &  1.56$_{-0.02}^{+ 0.02}$ & 2.83$_{-0.01}^{+ 0.01}$ & 5.02$_{-0.01}^{+ 0.01}$  \\ 
  7 & 112.0 &   77.0  & 14.92$_{-0.01}^{+ 0.01}$ &15.71$_{-0.01}^{+ 0.01}$ &  1.41$_{-0.02}^{+ 0.02}$ & 2.55$_{-0.01}^{+ 0.01}$ & 4.26$_{-0.01}^{+ 0.01}$  \\ 
  8 & -94.3 &   -6.1  & 15.16$_{-0.01}^{+ 0.01}$ &16.05$_{-0.01}^{+ 0.01}$ &  1.74$_{-0.03}^{+ 0.03}$ & 2.99$_{-0.01}^{+ 0.01}$ & 4.83$_{-0.01}^{+ 0.01}$  \\
\enddata 
\label{3c313} 
\end{deluxetable} 

\begin{deluxetable}{rrrrrrrrrc} 
\tabletypesize{\footnotesize} 
\tablewidth{0pt} 
\tablenum{22} 
\tablecaption{3C 295} 
\tablehead{ 
\colhead{ID}&\colhead{$\delta$x}&\colhead{$\delta$y}& 
\colhead{$Kt$}&
\colhead{$Ka$}&
\colhead{$H-K$}&
\colhead{$J-K$}&
\colhead{$I-K$}&
\colhead{$V-K$}&
\colhead{type} \\ 
\colhead{}&\colhead{arcsec}&\colhead{arcsec} & 
\colhead{}&\colhead{$r=1.77\arcsec$}&\colhead{$r=1.77\arcsec$}&
\colhead{$r=1.77\arcsec$}& 
\colhead{$r=1.77\arcsec$}&\colhead{$r=1.77\arcsec$}
} 
\startdata 
  1 &  -1.1 &  120.9  & 10.39$_{-0.01}^{+ 0.01}$ & 10.75$_{-0.01}^{+ 0.01}$ & 0.61$_{-0.01}^{+ 0.01}$ &  1.40$_{-0.01}^{+ 0.01}$ & 3.07$_{-0.01}^{+ 0.01}$ & 4.08$_{-0.01}^{+ 0.01}$ &      0  \\ 
  2 &  90.2 &  135.4  & 11.37$_{-0.01}^{+ 0.01}$ & 11.61$_{-0.01}^{+ 0.01}$ & 0.40$_{-0.01}^{+ 0.01}$ &  1.05$_{-0.01}^{+ 0.01}$ & 3.93$_{-0.01}^{+ 0.01}$ & 3.24$_{-0.01}^{+ 0.01}$ &      0  \\ 
  3 & -11.4 &  -61.0  & 11.56$_{-0.01}^{+ 0.01}$ & 11.95$_{-0.01}^{+ 0.01}$ & 0.25$_{-0.01}^{+ 0.01}$ &  0.98$_{-0.01}^{+ 0.01}$ & 2.10$_{-0.01}^{+ 0.01}$ & 3.27$_{-0.01}^{+ 0.01}$ &      5  \\ 
  4 & 100.2 &  -10.5  & 11.92$_{-0.01}^{+ 0.01}$ & 12.14$_{-0.01}^{+ 0.01}$ & 0.25$_{-0.01}^{+ 0.01}$ &  1.00$_{-0.01}^{+ 0.01}$ & 1.85$_{-0.01}^{+ 0.01}$ & 2.83$_{-0.01}^{+ 0.01}$ &      0  \\ 
  5 &-143.2 &  -28.8  & 12.06$_{-0.01}^{+ 0.01}$ & 12.26$_{-0.01}^{+ 0.01}$ & 0.21$_{-0.01}^{+ 0.01}$ &  0.70$_{-0.01}^{+ 0.01}$ & 1.74$_{-0.01}^{+ 0.01}$ & 2.63$_{-0.01}^{+ 0.01}$ &      0  \\ 
  6 &-110.6 &  129.9  & 12.86$_{-0.01}^{+ 0.01}$ & 13.09$_{-0.01}^{+ 0.01}$ & 0.10$_{-0.01}^{+ 0.01}$ &  0.44$_{-0.01}^{+ 0.01}$ & 1.19$_{-0.01}^{+ 0.01}$ & 2.10$_{-0.01}^{+ 0.01}$ &      0  \\ 
  7 &-132.5 &  -55.6  & 13.88$_{-0.01}^{+ 0.01}$ & 14.06$_{-0.01}^{+ 0.01}$ & 0.04$_{-0.01}^{+ 0.01}$ &  0.50$_{-0.01}^{+ 0.01}$ & 0.80$_{-0.01}^{+ 0.01}$ & 1.61$_{-0.01}^{+ 0.01}$ &      0  \\ 
  8 & -27.5 &  -72.4  & 14.03$_{-0.01}^{+ 0.01}$ & 14.21$_{-0.01}^{+ 0.01}$ & 0.02$_{-0.01}^{+ 0.01}$ &  0.46$_{-0.01}^{+ 0.01}$ & 0.82$_{-0.01}^{+ 0.01}$ & 1.63$_{-0.01}^{+ 0.01}$ &      5  \\
\enddata 
\label{3c295} 
\end{deluxetable} 

\begin{deluxetable}{rrrrrrrrrc} 
\tabletypesize{\footnotesize} 
\tablewidth{0pt} 
\tablenum{23} 
\tablecaption{F1557.19TC} 
\tablehead{ 
\colhead{ID}&\colhead{$\delta$x}&\colhead{$\delta$y}& 
\colhead{$Kt$}&
\colhead{$Ka$}&
\colhead{$H-K$}&
\colhead{$J-K$}&
\colhead{$I-K$}&
\colhead{$V-K$}&
\colhead{type} \\ 
\colhead{}&\colhead{arcsec}&\colhead{arcsec} & 
\colhead{}&\colhead{$r=1.68\arcsec$}&\colhead{$r=1.68\arcsec$}&
\colhead{$r=1.68\arcsec$}& 
\colhead{$r=1.68\arcsec$}&\colhead{$r=1.68\arcsec$}
} 
\startdata 
  1 &  28.5 &   46.9  & 11.20$_{-0.01}^{+ 0.01}$ & 11.44$_{-0.01}^{+ 0.01}$ & 0.07$_{-0.01}^{+ 0.01}$ &  0.37$_{-0.01}^{+ 0.01}$ & 1.51$_{-0.01}^{+ 0.01}$ & 2.49$_{-0.01}^{+ 0.01}$ &      5  \\ 
  2 &  11.1 &  -90.5  & 11.95$_{-0.01}^{+ 0.01}$ & 12.31$_{-0.01}^{+ 0.01}$ & 0.13$_{-0.01}^{+ 0.01}$ &  0.35$_{-0.01}^{+ 0.01}$ & 1.06$_{-0.01}^{+ 0.01}$ & 1.86$_{-0.01}^{+ 0.01}$ &      0  \\ 
  3 &  97.8 &   23.7  & 12.20$_{-0.01}^{+ 0.01}$ & 12.44$_{-0.01}^{+ 0.01}$ & 0.05$_{-0.01}^{+ 0.01}$ &  0.39$_{-0.01}^{+ 0.01}$ & 1.13$_{-0.01}^{+ 0.01}$ & 1.97$_{-0.01}^{+ 0.01}$ &      5  \\ 
  4 & -98.3 &  -27.2  & 12.57$_{-0.01}^{+ 0.01}$ & 13.18$_{-0.01}^{+ 0.01}$ & 0.17$_{-0.01}^{+ 0.01}$ &  0.66$_{-0.01}^{+ 0.01}$ & 1.12$_{-0.01}^{+ 0.01}$ & 2.18$_{-0.01}^{+ 0.01}$ &      0  \\ 
  5 &  59.4 &  178.2  & 12.88$_{-0.01}^{+ 0.01}$ & 13.15$_{-0.01}^{+ 0.01}$ & 0.21$_{-0.01}^{+ 0.01}$ &  0.84$_{-0.01}^{+ 0.01}$ & 1.93$_{-0.01}^{+ 0.01}$ & 4.08$_{-0.01}^{+ 0.01}$ &      0  \\ 
  6 & -76.1 &  -35.1  & 12.99$_{-0.01}^{+ 0.01}$ & 13.49$_{-0.01}^{+ 0.01}$ & 0.24$_{-0.01}^{+ 0.01}$ &  0.83$_{-0.01}^{+ 0.01}$ & 1.58$_{-0.01}^{+ 0.01}$ & 3.11$_{-0.01}^{+ 0.01}$ &      0  \\ 
  7 & -82.8 &  122.2  & 14.04$_{-0.01}^{+ 0.01}$ & 14.56$_{-0.01}^{+ 0.01}$ & 0.15$_{-0.01}^{+ 0.01}$ &  0.88$_{-0.01}^{+ 0.01}$ & 1.51$_{-0.01}^{+ 0.01}$ & 2.85$_{-0.01}^{+ 0.01}$ &      0  \\ 
  8 & -22.4 &   26.4  & 14.06$_{-0.01}^{+ 0.01}$ & 15.26$_{-0.01}^{+ 0.01}$ & 0.45$_{-0.01}^{+ 0.01}$ &  1.16$_{-0.01}^{+ 0.01}$ & 2.36$_{-0.01}^{+ 0.01}$ & 3.63$_{-0.01}^{+ 0.01}$ &      2  \\
\enddata 
\label{f1557} 
\end{deluxetable} 

\begin{deluxetable}{rrrrrrrr} 
\tabletypesize{\footnotesize} 
\tablewidth{0pt} 
\tablenum{24} 
\tablecaption{Vidal 14} 
\tablehead{ 
\colhead{ID}&\colhead{$\delta$x}&\colhead{$\delta$y}& 
\colhead{$Kt$}&
\colhead{$Ka$}&
\colhead{$J-K$}&
\colhead{$I-K$}&
\colhead{$V-K$}\\
\colhead{}&\colhead{arcsec}&\colhead{arcsec} & 
\colhead{}&\colhead{$r=1.74\arcsec$}&\colhead{$r=1.74\arcsec$}&
\colhead{$r=1.74\arcsec$}&\colhead{$r=1.74\arcsec$}
} 
\startdata 
  1 &  112.3 &  -85.4 & 13.08$_{-0.01}^{+ 0.01}$ &13.37$_{-0.01}^{+ 0.01}$ &  0.74$_{-0.01}^{+ 0.01}$ & 2.13$_{-0.01}^{+ 0.01}$ & 4.35$_{-0.01}^{+ 0.01}$  \\ 
  2 &  -71.6 &   52.7 & 13.35$_{-0.01}^{+ 0.01}$ &13.91$_{-0.01}^{+ 0.01}$ &  0.14$_{-0.01}^{+ 0.01}$ & 0.46$_{-0.01}^{+ 0.01}$ & 1.42$_{-0.01}^{+ 0.01}$  \\ 
  3 &  -10.3 &  -51.9 & 13.39$_{-0.01}^{+ 0.01}$ &13.68$_{-0.01}^{+ 0.01}$ &  0.44$_{-0.01}^{+ 0.01}$ & 1.07$_{-0.01}^{+ 0.01}$ & 2.34$_{-0.01}^{+ 0.01}$  \\ 
  4 &   46.9 &   18.8 & 13.88$_{-0.01}^{+ 0.01}$ &14.20$_{-0.01}^{+ 0.01}$ &  0.57$_{-0.01}^{+ 0.01}$ & 1.50$_{-0.01}^{+ 0.01}$ & 2.77$_{-0.01}^{+ 0.01}$  \\ 
  5 &   37.7 &   36.8 & 14.00$_{-0.01}^{+ 0.01}$ &14.39$_{-0.01}^{+ 0.01}$ &  0.38$_{-0.01}^{+ 0.01}$ & 1.09$_{-0.01}^{+ 0.01}$ & 2.22$_{-0.01}^{+ 0.01}$  \\ 
  6 &    0.0 &    0.0 & 14.50$_{-0.01}^{+ 0.01}$ &15.68$_{-0.01}^{+ 0.01}$ &  1.55$_{-0.02}^{+ 0.02}$ & 2.90$_{-0.02}^{+ 0.02}$ & 5.69$_{-0.02}^{+ 0.02}$  \\ 
  7 &  -65.3 & -103.0 & 14.94$_{-0.02}^{+ 0.02}$ &15.64$_{-0.01}^{+ 0.01}$ &  0.48$_{-0.02}^{+ 0.02}$ & 1.57$_{-0.02}^{+ 0.02}$ & 4.66$_{-0.02}^{+ 0.02}$  \\ 
  8 & -102.1 &   92.0 & 15.10$_{-0.03}^{+ 0.04}$ &16.21$_{-0.02}^{+ 0.02}$ &  1.36$_{-0.02}^{+ 0.02}$ & 2.16$_{-0.02}^{+ 0.02}$ & 4.07$_{-0.02}^{+ 0.02}$  \\
\enddata 
\label{v14} 
\end{deluxetable} 

\begin{deluxetable}{rrrrrrrrrc} 
\tabletypesize{\footnotesize} 
\tablewidth{0pt} 
\tablenum{25} 
\tablecaption{GHO 1601+4253} 
\tablehead{ 
\colhead{ID}&\colhead{$\delta$x}&\colhead{$\delta$y}& 
\colhead{$Kt$}&
\colhead{$Ka$}&
\colhead{$H-K$}&
\colhead{$J-K$}&
\colhead{$I-K$}&
\colhead{$V-K$}&
\colhead{type} \\ 
\colhead{}&\colhead{arcsec}&\colhead{arcsec} & 
\colhead{}&\colhead{$r=1.77\arcsec$}&\colhead{$r=1.77\arcsec$}&
\colhead{$r=1.77\arcsec$}& 
\colhead{$r=1.77\arcsec$}&\colhead{$r=1.77\arcsec$}
} 
\startdata 
  1 & -58.5 & -127.8  & 12.77$_{-0.01}^{+ 0.01}$ & 12.93$_{-0.01}^{+ 0.01}$ & 0.21$_{-0.01}^{+ 0.01}$ &  0.60$_{-0.01}^{+ 0.01}$ & 1.13$_{-0.01}^{+ 0.01}$ & 2.17$_{-0.01}^{+ 0.01}$ &      0  \\ 
  2 &  59.3 & -140.2  & 12.80$_{-0.01}^{+ 0.01}$ & 12.95$_{-0.01}^{+ 0.01}$ & 0.23$_{-0.01}^{+ 0.01}$ &  0.78$_{-0.01}^{+ 0.01}$ & 1.39$_{-0.01}^{+ 0.01}$ & 2.61$_{-0.01}^{+ 0.01}$ &      0  \\ 
  3 &  82.4 &    5.7  & 14.66$_{-0.01}^{+ 0.01}$ & 14.86$_{-0.01}^{+ 0.01}$ & 0.07$_{-0.01}^{+ 0.01}$ &  0.50$_{-0.01}^{+ 0.01}$ & 1.26$_{-0.01}^{+ 0.01}$ & 2.51$_{-0.01}^{+ 0.01}$ &      0  \\ 
  4 &   0.0 &    0.0  & 14.99$_{-0.01}^{+ 0.01}$ & 16.08$_{-0.01}^{+ 0.01}$ & 0.62$_{-0.01}^{+ 0.01}$ &  1.49$_{-0.01}^{+ 0.01}$ & 2.89$_{-0.01}^{+ 0.01}$ & 5.53$_{-0.01}^{+ 0.01}$ &      1  \\ 
  5 &  66.9 &   93.3  & 15.39$_{-0.01}^{+ 0.01}$ & 15.57$_{-0.01}^{+ 0.01}$ & 0.03$_{-0.01}^{+ 0.01}$ &  0.44$_{-0.01}^{+ 0.01}$ & 1.13$_{-0.01}^{+ 0.01}$ & 2.32$_{-0.01}^{+ 0.01}$ &      0  \\ 
  6 &   9.5 &   28.5  & 15.46$_{-0.01}^{+ 0.01}$ & 16.28$_{-0.01}^{+ 0.01}$ & 0.67$_{-0.01}^{+ 0.01}$ &  1.52$_{-0.01}^{+ 0.01}$ & 2.88$_{-0.01}^{+ 0.01}$ & 5.50$_{-0.01}^{+ 0.01}$ &      1  \\ 
  7 & -34.7 &  -65.8  & 15.52$_{-0.01}^{+ 0.01}$ & 15.78$_{-0.01}^{+ 0.01}$ & 1.13$_{-0.01}^{+ 0.01}$ &  2.05$_{-0.01}^{+ 0.01}$ & 2.86$_{-0.01}^{+ 0.01}$ & 4.12$_{-0.01}^{+ 0.01}$ &      4  \\ 
  8 &  33.7 & -105.6  & 15.77$_{-0.01}^{+ 0.01}$ & 15.99$_{-0.01}^{+ 0.01}$ & 0.33$_{-0.01}^{+ 0.01}$ &  0.87$_{-0.01}^{+ 0.01}$ & 2.21$_{-0.01}^{+ 0.01}$ & 5.32$_{-0.01}^{+ 0.01}$ &      0  \\
\enddata 
\label{g1601} 
\end{deluxetable} 

\begin{deluxetable}{rrrrrrrrrc} 
\tabletypesize{\footnotesize} 
\tablewidth{0pt} 
\tablenum{26} 
\tablecaption{MS 0451.6-0306} 
\tablehead{ 
\colhead{ID}&\colhead{$\delta$x}&\colhead{$\delta$y}& 
\colhead{$Kt$}&
\colhead{$Ka$}&
\colhead{$H-K$}&
\colhead{$J-K$}&
\colhead{$I-K$}&
\colhead{$V-K$}&
\colhead{type} \\ 
\colhead{}&\colhead{arcsec}&\colhead{arcsec} & 
\colhead{}&\colhead{$r=1.95\arcsec$}&\colhead{$r=1.95\arcsec$}&
\colhead{$r=1.95\arcsec$}& 
\colhead{$r=1.95\arcsec$}&\colhead{$r=1.95\arcsec$}
} 
\startdata 
  1 & -33.9 &   44.5  & 13.38$_{-0.01}^{+ 0.01}$ & 13.66$_{-0.01}^{+ 0.01}$ & 0.15$_{-0.01}^{+ 0.01}$ &  0.45$_{-0.01}^{+ 0.01}$ & 1.00$_{-0.01}^{+ 0.01}$ & 1.51$_{-0.01}^{+ 0.01}$ &      5  \\ 
  2 & -32.7 &  -33.3  & 14.54$_{-0.01}^{+ 0.01}$ & 14.70$_{-0.01}^{+ 0.01}$ & 0.13$_{-0.01}^{+ 0.01}$ &  0.35$_{-0.01}^{+ 0.01}$ & 0.70$_{-0.01}^{+ 0.01}$ & 1.21$_{-0.01}^{+ 0.01}$ &      5  \\ 
  3 &   0.0 &    0.0  & 14.91$_{-0.01}^{+ 0.01}$ & 16.02$_{-0.01}^{+ 0.01}$ & 0.75$_{-0.01}^{+ 0.01}$ &  1.58$_{-0.02}^{+ 0.02}$ & 3.00$_{-0.01}^{+ 0.01}$ & 5.08$_{-0.01}^{+ 0.01}$ &      1  \\ 
  4 &  73.0 &   43.3  & 14.99$_{-0.01}^{+ 0.01}$ & 15.25$_{-0.01}^{+ 0.01}$ & 0.28$_{-0.01}^{+ 0.01}$ &  0.84$_{-0.01}^{+ 0.01}$ & 2.06$_{-0.01}^{+ 0.01}$ & 4.33$_{-0.01}^{+ 0.01}$ &      5  \\ 
  5 &  14.4 &   74.3  & 15.00$_{-0.01}^{+ 0.01}$ & 15.17$_{-0.01}^{+ 0.01}$ & 0.08$_{-0.01}^{+ 0.01}$ &  0.40$_{-0.01}^{+ 0.01}$ & 0.82$_{-0.01}^{+ 0.01}$ & 1.42$_{-0.01}^{+ 0.01}$ &      0  \\ 
  6 & -65.2 &  -30.6  & 15.02$_{-0.01}^{+ 0.01}$ & 15.16$_{-0.01}^{+ 0.01}$ & 0.34$_{-0.01}^{+ 0.01}$ &  0.93$_{-0.01}^{+ 0.01}$ & 1.96$_{-0.01}^{+ 0.01}$ & 3.93$_{-0.01}^{+ 0.01}$ &      0  \\ 
  7 &  -3.5 &  -55.8  & 15.92$_{-0.01}^{+ 0.01}$ & 16.11$_{-0.01}^{+ 0.01}$ & 0.30$_{-0.01}^{+ 0.01}$ &  0.82$_{-0.01}^{+ 0.01}$ & 2.25$_{-0.01}^{+ 0.01}$ & 4.92$_{-0.01}^{+ 0.01}$ &      0  \\ 
  8 &  -6.6 &    4.7  & 16.06$_{-0.01}^{+ 0.01}$ & 16.61$_{-0.01}^{+ 0.01}$ & 0.74$_{-0.02}^{+ 0.02}$ &  1.68$_{-0.03}^{+ 0.03}$ & 3.05$_{-0.01}^{+ 0.01}$ & 4.97$_{-0.01}^{+ 0.01}$ &      4  \\
\enddata 
\label{m0451} 
\end{deluxetable} 

\begin{deluxetable}{rrrrrrrrrc} 
\tabletypesize{\footnotesize} 
\tablewidth{0pt} 
\tablenum{27} 
\tablecaption{C0016+16} 
\tablehead{ 
\colhead{ID}&\colhead{$\delta$x}&\colhead{$\delta$y}& 
\colhead{$Kt$}&
\colhead{$Ka$}&
\colhead{$H-K$}&
\colhead{$J-K$}&
\colhead{$I-K$}&
\colhead{$V-K$}&
\colhead{type} \\ 
\colhead{}&\colhead{arcsec}&\colhead{arcsec} & 
\colhead{}&\colhead{$r=1.77\arcsec$}&\colhead{$r=1.77\arcsec$}&
\colhead{$r=1.77\arcsec$}& 
\colhead{$r=1.77\arcsec$}&\colhead{$r=1.77\arcsec$}
} 
\startdata 
  1 &-127.3 &   37.8  & 13.11$_{-0.01}^{+ 0.01}$ & 14.10$_{-0.01}^{+ 0.01}$ & 0.38$_{-0.01}^{+ 0.01}$ &  1.03$_{-0.01}^{+ 0.01}$ & 2.18$_{-0.01}^{+ 0.01}$ & 3.56$_{-0.01}^{+ 0.01}$ &      0  \\ 
  2 & -70.8 &  -45.7  & 13.32$_{-0.01}^{+ 0.01}$ & 13.48$_{-0.01}^{+ 0.01}$ & 0.23$_{-0.01}^{+ 0.01}$ &  0.78$_{-0.01}^{+ 0.01}$ & 2.06$_{-0.01}^{+ 0.01}$ & 4.50$_{-0.01}^{+ 0.01}$ &      5  \\ 
  3 &-112.6 &   65.5  & 13.55$_{-0.01}^{+ 0.01}$ & 14.34$_{-0.01}^{+ 0.01}$ & 0.34$_{-0.01}^{+ 0.01}$ &  1.03$_{-0.01}^{+ 0.01}$ & 2.17$_{-0.01}^{+ 0.01}$ & 3.45$_{-0.01}^{+ 0.01}$ &      0  \\ 
  4 &-107.0 &   57.8  & 14.38$_{-0.01}^{+ 0.01}$ & 14.74$_{-0.01}^{+ 0.01}$ & 0.12$_{-0.01}^{+ 0.01}$ &  0.72$_{-0.01}^{+ 0.01}$ & 1.65$_{-0.01}^{+ 0.01}$ & 3.15$_{-0.01}^{+ 0.01}$ &      0  \\ 
  5 & -19.8 & -132.7  & 14.75$_{-0.01}^{+ 0.01}$ & 14.97$_{-0.01}^{+ 0.01}$ & 0.08$_{-0.01}^{+ 0.01}$ &  0.70$_{-0.01}^{+ 0.01}$ & 1.58$_{-0.01}^{+ 0.01}$ & 2.96$_{-0.01}^{+ 0.01}$ &      0  \\ 
  6 &-116.1 &   60.5  & 14.81$_{-0.01}^{+ 0.01}$ & 15.49$_{-0.01}^{+ 0.01}$ & 0.38$_{-0.01}^{+ 0.01}$ &  1.05$_{-0.01}^{+ 0.01}$ & 2.17$_{-0.01}^{+ 0.01}$ & 3.47$_{-0.01}^{+ 0.01}$ &      0  \\ 
  7 & -18.7 &  -33.4  & 14.89$_{-0.01}^{+ 0.01}$ & 15.72$_{-0.01}^{+ 0.01}$ & 0.80$_{-0.01}^{+ 0.01}$ &  1.77$_{-0.01}^{+ 0.01}$ & 3.16$_{-0.01}^{+ 0.01}$ & 4.93$_{-0.01}^{+ 0.01}$ &      4  \\ 
  8 &   4.9 &    3.2  & 15.06$_{-0.01}^{+ 0.01}$ & 16.16$_{-0.01}^{+ 0.01}$ & 0.76$_{-0.01}^{+ 0.01}$ &  1.65$_{-0.01}^{+ 0.01}$ & 3.04$_{-0.01}^{+ 0.01}$ & 5.50$_{-0.01}^{+ 0.01}$ &      1  \\
\enddata 
\label{c0016} 
\end{deluxetable} 

\input{tab28}

\begin{deluxetable}{rrrrrrrr} 
\tabletypesize{\footnotesize} 
\tablewidth{0pt} 
\tablenum{29} 
\tablecaption{MS 2053.7-0449} 
\tablehead{ 
\colhead{ID}&\colhead{$\delta$x}&\colhead{$\delta$y}& 
\colhead{$Kt$}&
\colhead{$Ka$}&
\colhead{$J-K$}&
\colhead{$I-K$}&
\colhead{$V-K$}\\
\colhead{}&\colhead{arcsec}&\colhead{arcsec} & 
\colhead{}&\colhead{$r=1.74\arcsec$}&\colhead{$r=1.74\arcsec$}&
\colhead{$r=1.74\arcsec$}&\colhead{$r=1.74\arcsec$}
} 
\startdata 
  1 & 128.9 &  116.4  &  9.48$_{-0.01}^{+ 0.01}$ & 9.91$_{-0.01}^{+ 0.01}$ &  1.35$_{-0.01}^{+ 0.01}$ & 4.22$_{-0.01}^{+ 0.01}$ & 4.86$_{-0.01}^{+ 0.01}$  \\ 
  2 &  80.1 & -100.4  & 11.22$_{-0.01}^{+ 0.01}$ &11.80$_{-0.01}^{+ 0.01}$ &  0.78$_{-0.01}^{+ 0.01}$ & 3.01$_{-0.01}^{+ 0.01}$ & 5.45$_{-0.01}^{+ 0.01}$  \\ 
  3 & -96.1 &   91.2  & 12.87$_{-0.01}^{+ 0.01}$ &13.31$_{-0.01}^{+ 0.01}$ &  0.73$_{-0.01}^{+ 0.01}$ & 2.06$_{-0.01}^{+ 0.01}$ & 3.30$_{-0.01}^{+ 0.01}$  \\ 
  4 & -52.2 &  -48.5  & 13.39$_{-0.01}^{+ 0.01}$ &13.83$_{-0.01}^{+ 0.01}$ &  0.51$_{-0.01}^{+ 0.01}$ & 1.55$_{-0.01}^{+ 0.01}$ & 2.46$_{-0.01}^{+ 0.01}$  \\ 
  5 & -47.6 & -109.8  & 13.41$_{-0.01}^{+ 0.01}$ &13.82$_{-0.01}^{+ 0.01}$ &  0.36$_{-0.01}^{+ 0.01}$ & 1.45$_{-0.01}^{+ 0.01}$ & 2.15$_{-0.01}^{+ 0.01}$  \\ 
  6 &-129.7 & -113.2  & 13.45$_{-0.01}^{+ 0.01}$ &14.07$_{-0.01}^{+ 0.01}$ &  0.56$_{-0.01}^{+ 0.01}$ & 1.42$_{-0.01}^{+ 0.01}$ & 2.40$_{-0.01}^{+ 0.01}$  \\ 
  7 & -73.1 &   35.3  & 13.50$_{-0.01}^{+ 0.01}$ &13.93$_{-0.01}^{+ 0.01}$ &  0.35$_{-0.01}^{+ 0.01}$ & 1.46$_{-0.01}^{+ 0.01}$ & 2.13$_{-0.01}^{+ 0.01}$  \\ 
  8 & -88.3 & -109.2  & 13.68$_{-0.01}^{+ 0.01}$ &14.18$_{-0.01}^{+ 0.01}$ &  0.36$_{-0.01}^{+ 0.01}$ & 1.21$_{-0.01}^{+ 0.01}$ & 1.91$_{-0.01}^{+ 0.01}$  \\
\enddata 
\label{m2053} 
\end{deluxetable} 

\input{tab30}

\input{tab31}

\input{tab32}

\input{tab33}

\input{tab34}

\input{tab35}

\input{tab36}

\input{tab37}

\input{tab38}

\input{tab39}

\input{tab40}

\input{tab41}

\input{tab42}

\input{tab43}

\input{tab44}

\input{tab45}

\end{document}